\documentclass{article}
\usepackage{cite}
\usepackage{amsmath,amssymb,amsfonts}
\usepackage{algorithmic}
\usepackage[hyphens]{url} 
\usepackage{graphicx}
\usepackage{textcomp}
\usepackage{xcolor}
\usepackage[a4paper, total={184mm,239mm}]{geometry}
\usepackage{fancyhdr}
\usepackage{lastpage}
\usepackage{datetime}

\def\BibTeX{{\rm B\kern-.05em{\sc i\kern-.025em b}\kern-.08em
    T\kern-.1667em\lower.7ex\hbox{E}\kern-.125emX}}

\newdateformat{monthyeardate}{%
	\monthname[\THEMONTH], \THEYEAR}

\newdateformat{yeardate}{%
    \THEYEAR}

\numberwithin{figure}{section}
\numberwithin{table}{section}

\usepackage[section]{placeins}

\begin{document}

\begin{center}
	
	\vspace*{8cm}	
	{\Huge {Towards a Domain Specific Solution for a New Generation of Wireless Modems: \\Technical Report  \yeardate\today\par}}
	\vspace{100px}
	{\Large {Prepared by Futurewei Wireless Research Center\\ \today \\}
	{Contributors: Alan Gatherer, Ashish Shrivastava, Hao Luan, Asheesh Kashyap, Zhenguo Gu, Miguel Dajer \\}
	{alan.gatherer, ashish.shrivastava, hao.luan, asheesh.kashyap, zhenguo.gu, miguel.dajer\}@futurewei.com \\}}
	\includegraphics[scale=0.5]{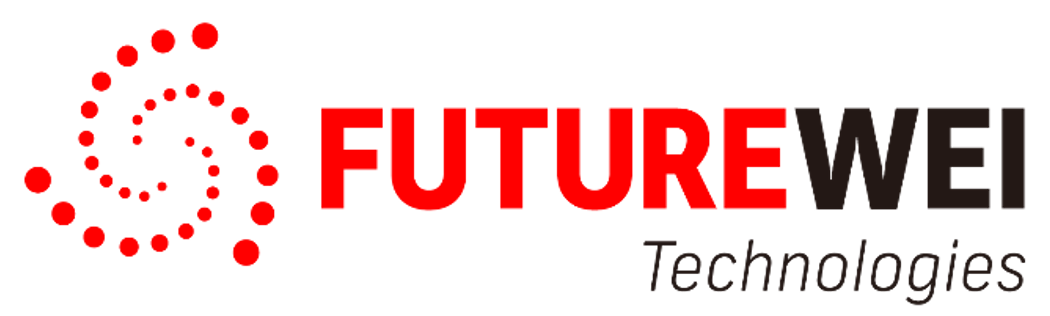}
\end{center}
\pagebreak

\section{\label{sec:intro}Introduction}
At the end of Moore’s Law, we find ourselves with an unwavering demand for more processing power. This is especially true in the realm of wireless communications, which is experiencing unprecedented growth in the areas of data rates, devices, and services. Fifth generation (5G) wireless technologies require a massive number of antennas and sophisticated signal processing to improve bandwidth and spectral efficiency \cite{Larsson2014}. The Internet of Things (IoT) is causing an enormous proliferation in the number of connected devices, predicted to exceed 30 billion devices by 2025 \cite{IoT2018}. And new service categories, such as ultra-reliable low latency communications (uRLLC), will produce new use cases, such as self-driving cars, robotic factories, and remote surgery \cite{5gAmericas}. Addressing these challenges in the “post Moore’s Law” era requires a novel approach to computer design. Architects can no longer rely on faster cores, or even more silicon, to save the day. An effective design requires an understanding of the dataflow and processing requirements of a particular application. It appears that we have entered the era of Domain Specific Architectures (DSA) \cite{HP2019}.

\subsection{A Brief Taxonomy of Domain Specific Languages and Architectures}
DSAs are far from a new concept. In the last decade, the emergence of Domain Specific Languages (DSLs) paired with DSAs has produced a new revolution in architecture design \cite{HP2019}, but even this has some much earlier examples. In what follows we will recognize the strong connection by using the term DSL/DSA to describe such holistic architectures. For DSAs, some of the earliest and best-known examples are the fixed function pipelines of early graphics subsystems \cite{Clark1982}. These specialized systems often required a DSL such as Iris GL and later OpenGL \cite{Khronos} to facilitate programming. With the help of Moore’s Law, these rigid systems were eventually displaced by more general and flexible graphics processing units (GPUs) \cite{Macedonia2003}. However, widespread adoption of GPUs for general purpose computing did not occur until the arrival of a more flexible DSL called CUDA \cite{Lindholm2008}. GPU’s eventually found use in a wide variety of applications, including a new “killer app” called deep convolutional neural networks \cite{Hinton2012}. While GPU’s proved to be an excellent platform for neural network training, they lacked the speed and efficiency for neural network inference. This led to a new, more specialized class of processors called TPU’s (tensor processing units) \cite{Jouppi2018}, which could be easily programmed with a specialized DSL (Tensorflow) \cite{tensorflow}. In all of these cases, a DSL abstracted the complexity of the underlying hardware and alleviated the burden of optimizing performance.

While GPUs and TPUs may be the best-known examples of DSAs, there are numerous other examples which may give us inspiration. In the field of molecular simulation, the Anton 2 supercomputer outperforms general-purpose hardware by two orders of magnitude \cite{Shaw2014}. This is achieved through a combination of specialized processing elements, compute tiles, and interconnection networks. Another excellent example is the Eyeriss 2 flexible accelerator for neural networks on mobile devices \cite{Sze2019}. Mobile neural network models tend to be sparse and have irregular access patterns. In addition, mobile devices tend to have extremely constrained power budgets, so energy efficiency is of paramount importance. In order to address these challenges, Eyeriss 2 uses specialized processing elements, minimizes data movement, and implements a flexible hierarchical mesh network for different operating modes. A third example, similar in many aspects to the wireless processing domain, is the Barefoot Tofino domain specific processor for networking \cite{Bosshart2018}. This packet processor contains configurable match and action pipelines that can be programmed using a specialized DSL called P4 \cite{p4language}. By eschewing the von Neumann processing model, minimizing control area, and maintaining a regular structure, the Tofino processor can achieve performance comparable to fixed function switches. In each of these examples, tradeoffs were made at every level of the hierarchy with a thorough understanding of the dataflow and processing requirements for that domain.

\subsection{Wireless Baseband as a "Dataflow Domain Specific Architecture"}
In the domain of wireless communications, and especially in the area of wireless baseband processing, there are certain concepts that help us define the architectural requirements. One of the most important concepts is that of a service. From the network operator’s perspective, a service refers to a particular use case scenario, with well-defined bandwidth and quality of service (QoS) requirements. However, from a modem developer’s perspective, a service implies a predefined set of dataflows with well understood, or at least well bounded, processing, storage and timing requirements. In this context, dataflows refer to the abstract dataflow model described in \cite{Arvind2003}, where nodes denote operations, and arcs denote data dependencies between operations. Dataflows may be static (known before run time) or dynamic (known at run time). Many variants of dataflow models exist, including synchronous dataflow \cite{Lee1987}, dataflow process networks \cite{Lee1995}, and scenario aware dataflow \cite{Stuijk2011}, with each model having its own mathematical tools for analysis. \\
In wireless communications, flows may be considered semistatic, where the dataflow patterns are known in advance and remain stable for a reasonable length of time. Additionally, 5G wireless communication systems have different real-time constraints for different services. The real-time requirements for enhanced mobile broadband (eMBB) are certainly different from the requirements for uRLLC or massive machine type communication (mMTC) \cite{Campos2017}. The complexity of wireless communications processing combined with the multiplicity of services and the varying goals of stakeholders makes it a most challenging area for architecture exploration. We would argue that an efficient solution would be difficult, if not impossible, to achieve without a deep understanding of the application domain. \\
At the highest level, wireless base station modem consists of an uplink flow and a downlink flow \cite{5G_book}. To minimize power and cost, in a cellular system multiple spectrum channels of this flow are implemented simultaneously on a baseband L1 System on Chip (SoC). This provides some unique challenges to the system designer.
\begin{enumerate}
	\item Each channel is a dataflow system with strong dependencies along a flow of data from the antenna to the layer 2 processing in uplink and in the reverse in downlink.
	\item They are Firm Real Time (FRT) Systems in that there is no point in continuing to process a flow once it becomes apparent it will not make its deadline. But a flow can be occasionally dropped (leading to an occasional loss of a user packet or a slightly degraded estimate) without harm to  the performance of the SoC as long as it does not lead to an avalanche effect through the system \cite{Hahn2015}. So flows must maintain real time isolation.
	\item Streams have different QoS levels both because of the different types of channels in the modem (such as PUSCH, PDSCH, PRACH \cite{5G_book}) and also because different use cases require different data rates and QoS (such as video streaming versus Industrial control).
	\item Channel density requirements are very aggressive so high levels of parallelism are required on the SoC and hence the dataflow must be represented in a Model of Computation (MoC) that allows for parallelism.
	\item Every transmission  slot the Layer 2 spectral management algorithm will choose a new set of users to transmit and receive, changing the dataflow requirements. Slots can be sub millisecond in size. This leads to a challenging runtime scheduling problem in time and across resources.
	\item The SoC must be highly reliable and withstand Heisenbugs in the field at a rate of less than one crash in the lifetime of the SoC \cite{Gatherer2020}.
\end{enumerate}
5th Generation cellular wireless (5G) exacerbates these problems with an even more varied and flexible set of potential modes of operation at higher data rates with more independent users on SoC. As a result the development of 5G baseband software has become tremendously difficult and error prone if done by hand. 
\subsection{Achieving Automation with Efficiency in 5G Modems}
Currently, the modem SoC used in 3G, 4G and now 5G systems are highly optimized SoC with high potential capacity and many accelerators \cite{Marvell/Octeon_CNF73xx} as well as specialized Digital Signal Processors (DSPs) optimized for fixed point and floating point, and with specialized instructions for modem operations. Essentially the embedded system space, and modem development in particular, has been developing DSA for many years now. The missing link between modem development and the new philosophy of DSL/DSA is that the current modems remain hard to program with much detailed hardware architecture knowledge required by the programmer. Once a modem has been programmed and tested the process of upgrading or modifying is difficult and time consuming and this limits innovation. So a DSL could be used to simplify this process. But to develop a successful DSL for 5G modem design the DSA must also be modified to be an efficient target for the tool chain supporting the DSL. Therefore the development of a successful DSL/DSA solution requires a holistic approach to the DSL/DSA that supports
\begin{enumerate}
	\item the DSL that describes the requirements in a suitably abstracted and hardware independent manner.
	\item the toolchain that automates translation of those requirements in an efficient and robust manner, provides formal guarantees against Heisenbugs, and recovers rapidly from dropped dataflows while continuing to achieve the real time requirements of most flows.
	\item the DSA that is the target that supports the toolchain output with specialized processing, data management and control features with high performance and low power.
\end{enumerate}

\begin{figure}[h]
	\includegraphics[width=0.8\linewidth]{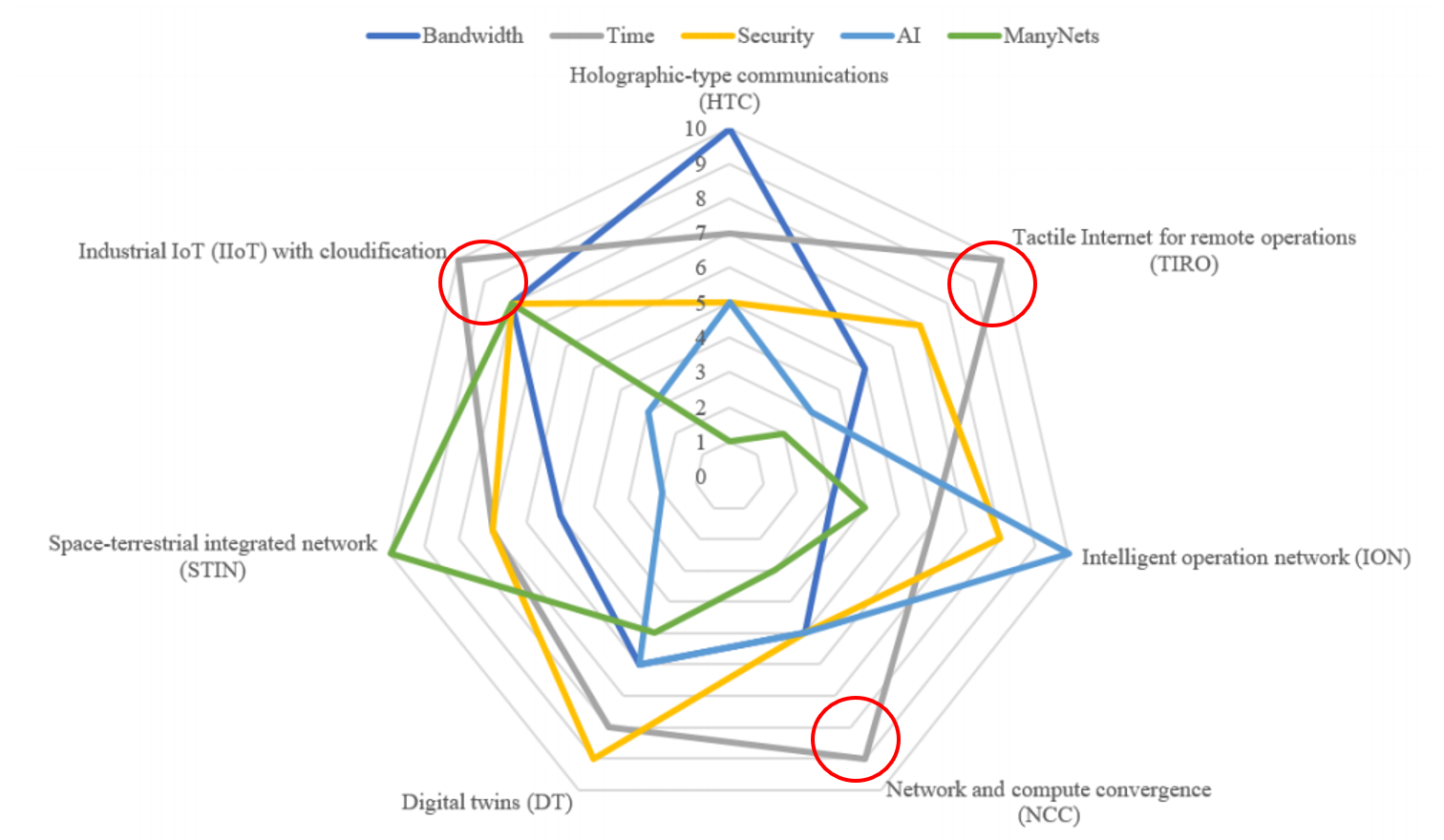}
	\centering
	\caption{5G to 6G requirements change highlighting latency and security as issues (Source \cite{ITU_T_012020})}
	\label{6Gspider} 	
\end{figure}

One of our chief concerns in development of this DSA is to make sure the architecture can be formally checked to remove Heisenbugs that can occur when real time constraints interact with limited memory and functional run time uncertainty. Uncertainty in the data flow is the result of many practical issues: 
\begin{itemize}
	\item Limited memory is a direct side effect of power and latency requirements that force data to be stored on the SoC in physical locations. Virtualization of memory is often not possible due to the dramatic latency hit that will occur in the event of a page miss. 
	\item Real time constraints force data to be dumped in memory by a certain time so the resources can be reused for another part of the data flow but if the memory is full at that time there will be a real time failure or a write before read error that can be hard to debug in a lab environment.
	\item Run time uncertainty occurs because the functions have many parameters that directly impact the complexity of the computation or the size of buffers to be processed. To create unique dataflow for every parameter combination is impossible so there will be run time uncertainty as well as data bufer size uncertainty associated with any dataflow node in the system.
	\item Competition for resources among data flow is a necessity to ensure efficient use of resources in the SoC. But when combined with run time uncertainty and local scheduling (a necessity to keep control overhead under control) it leads to uncertainty in the temporal ordering of dataflow node processing. Scheduling is generally an NP complete problem and so only a heuristic solution is possible.
\end{itemize}
This issues will worsen for 5G due to tighter latencies and a much larger set of requirements and use cases and then become much worse for 6G as can be seen in Figure \ref{6Gspider}.
\subsection{What is wrong with current SoC solutions}
Today the SoC solutions for baseband modems are dominated by OEM developed SoC and there is little information made public as to how they are programmed. However the public information that we have reveals that they are all built on what could be called a "Semi-coherent Global Virtual Memory" domain specific architecture which for brevity we shall call the GCM-DSA. An example of such an SoC is shown in \cite{Marvell/Octeon_CNF73xx}. SoC development in OEMs relies heavily on existing IP including memory controllers, crossbars and caches. These are hard to specify and verify and the IP provided by 3rd parties is configurable to allow for a wide variety of solutions.  3rd party code development and debug tools are also used and the expertise of the engineers tends to be in the development and maintenance of code in a global memory mapped system. Engineers will worry that debug will be difficult unless every register is exposed for peek and poke. 

All of these practical constraints push the system designer toward a classic three layer coherently cached system out to DDR. But this leads to all sorts of timing uncertainty in turn leading to performance losses and tricky Heisenbugs.  Maintaining latency constraints is then dealt with by exposing the L2 memory in the memory map or even exposing the L1 memory and hand managing the movement of data directly to and from the compute elements. Indeed for hardware accelerators caching is often not an option at all. This will lead to a complicated mixture of uncached compute elements, compute elements that are cached to and exposed L2 and some that cache all the way to DDR. Data movement is a mixture of hand managed using DMA and caching. ARM, the leader in embedded system SoC IP, has extended its coherent bus \lq\lq Corelink" specification to support a smorgasbord of cached and uncached transactions \cite{ARM_AMBA4ACE} but turning these into a workable system requires experience and care. \\

In summary, modem SoC are built as a GCM-DSA using IP that is not intended for the modem use case and it leads to a many practical difficulties in programming these devices:
\begin{itemize}
	\item Cached systems lead to timing uncertainty and Heisenbugs.
	\item Mixtures of cached, uncached, coherent, noncoherent regions are left to the programmer to manage for correctness
	\item Global memory exposure leads opens the system up to tricky Heisenbug problems due to read before write and write before read errors and buffer overflow errors.
	\item maintaining latency is done in an informal way by individual programmers and can break down when the system is integrated.
	\item hand management of data movement using a "distant" control processor in a cached region can lead to significant control overhead and latency.
\end{itemize}
In \cite{Gatherer2020} we outlined the \lq\lq Russian Doll" methodology often employed to manage the SoC but this \lq\lq super set of worst case" design approach has become highly inefficient because there are so many new use cases that use the system in significantly different ways.
\subsection{Summary of the goals and flow of this report}
In this report we describe the progress towards a Domain Specific Solution for modems in 5G and beyond. We use Dataflow as the basis to develop a DSL for the modem and show that an accelerator architecture can be developed to efficiently support this new language.  That is, we develop a DSA suitable for our DSL. We then show that this DSL/DSA combination can be formally checked for the kinds of difficult bugs we described in this introduction. 

We start in Section \ref{sec:dataflows} with a description of how to apply dataflow methodology as a DSL for baseband modems introducing the concepts of DataFlow (DF), DataFlow Fragments (DFF) and MicroFlows (MF). In Section \ref{sec:arch_tool_overview} we provide a top level description of our DSA and the overall tool flow that we are heading towards. In Section \ref{sec:SRE-Arch} we describe the DSA in detail. In Section \ref{sec:FV} we describe the formal checking aspects of the toolflow with detailed examples of how we modeled the DSA/DSL and applied formal checks to the dataflows.  We conclude and describe the work currently being planned and in progress in Section \ref{sec:Conclude_NextSteps}.

The report has two Appendices giving more detail of a detailed simulation model in the Mirabilis tool and details of the Message Formats used to manage dataflows in the DSA. 
\section{\label{sec:dataflows}Wireless Baseband Dataflows}

\subsection{\label{sec:dflow_graphs}Dataflow Graphs}

\begin{figure}[h]
	\includegraphics[width=0.6\linewidth, height=3cm]{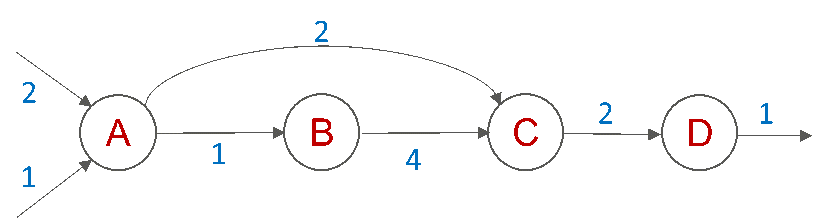}
	\centering
	\caption{Dataflow Graph}
	\label{fig:dataflow_graph} 
\end{figure}

Wireless baseband processing, like many other real-time workloads, can be modeled as a dataflow graph. In a dataflow graph, processing is modeled by “actors”, which are represented by circles, and communication is modeled by “edges” or lines. Actors produce and consume “tokens”, which are units of granularity of data along an edge. An example of a dataflow graph is shown in Figure ~\ref{fig:dataflow_graph}.

\begin{figure}[h]
	\includegraphics[width=1.0\linewidth, height=8cm]{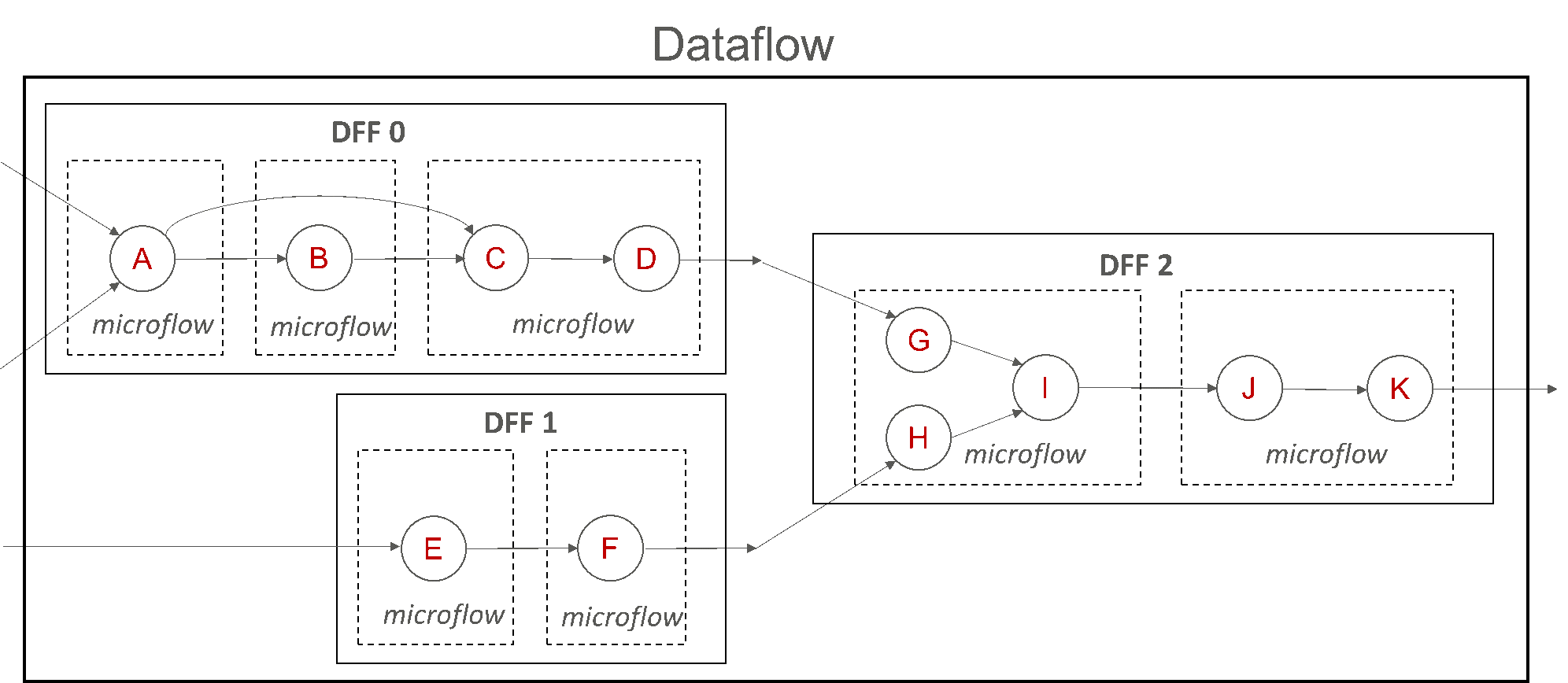}
	\centering
	\caption{Dataflow Decomposition}
	\label{fig:dataflow_decomp} 
\end{figure}

In more complicated dataflows, such as those required for wireless baseband processing, tasks may be decomposed into large granularity subtasks, which we refer to as dataflow fragments DFFs. Dataflow fragments may, in turn, be further decomposed into kernels, or tightly coupled collection of kernels, which we refer to as microflows. An example of a hierarchical decomposition of a dataflow graph into dataflow fragments, and subsequently into microflows is shown in Figure ~\ref{fig:dataflow_decomp}.

The reason for the introduction of new terminology is to provide a precise mapping between an abstract level of hierarchy in a dataflow graph, and a concrete level of hierarchy in the actual hardware. In our architecture, a DFF is an actor at the top level of the dataflow and executes on a dataflow accelerator called a Service Resource Element (SRE), which will be described in detail in Section \ref{sec:SRE-Arch}. A microflow is an actor within a DFF and executes on a single stage of the SRE. A stage is both a logical and physical partitioning of the SRE that contains the actual kernels and computational units needed to execute a microflow. The dataflow to hardware mapping is summarized in Table ~\ref{table:hardware_mapping}.

\begin{table}[h]
	\begin{center}
		\begin{tabular}{ |l|l| }
	    \hline 
		\textbf{Dataflow Abstraction Level} & 
		\textbf{Hardware Mapping} \\
		\hline\hline
		Dataflow Graph & Single or Multiple SREs\\
		\hline 
		Dataflow Fragment (DFF) & Single SRE \\
		\hline
		Microflow & Single SRE Stage \\
		\hline
	    \end{tabular}
    \caption{Dataflow to Hardware Mapping}
    \label{table:hardware_mapping}
    \end{center}
\end{table}

The following section gives an example of how an essential wireless baseband function, channel estimation, can be mapped to our architecture.

\subsection{\label{sec:channel_est}Channel Estimation Decomposition}

Channel estimation is a fundamental operation in wireless communications. Before a message can be transmitted over a channel, the properties of the channel must be determined. In 5G communications, the transmitter sends out known reference signals over multiple antennas at various frequencies and time intervals. The receiver can compare the actual received reference signals with the known transmitted reference signals and use this information to construct a channel model. A detailed discussion of channel estimation can be found in \cite{5G_book}.

\begin{figure}[h]
	\includegraphics[width=1.0\linewidth, height=3cm]{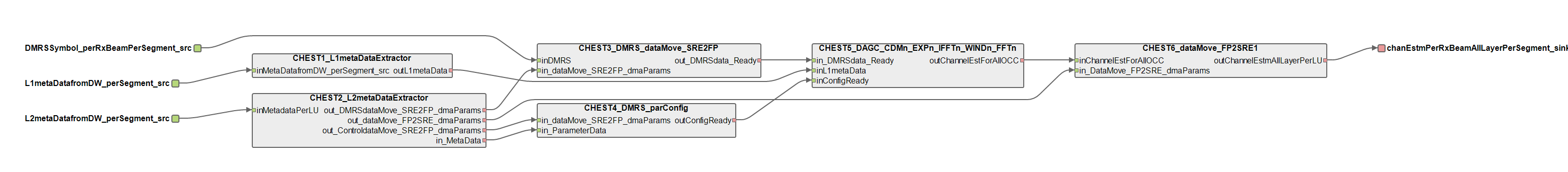}
	\centering
	\caption{Channel Estimation Dataflow Fragment}
	\label{fig:chest_dff} 
\end{figure}

The channel estimation function is represented in our architecture as a dataflow fragment (DFF), as shown in Figure ~\ref{fig:chest_dff}. A DFF is a directed, acyclic graph that defines the connectivity between microflows. The microflows are the actual kernels that execute on compute elements.

\begin{figure}[h]
	\includegraphics[width=1.0\linewidth, height=3cm]{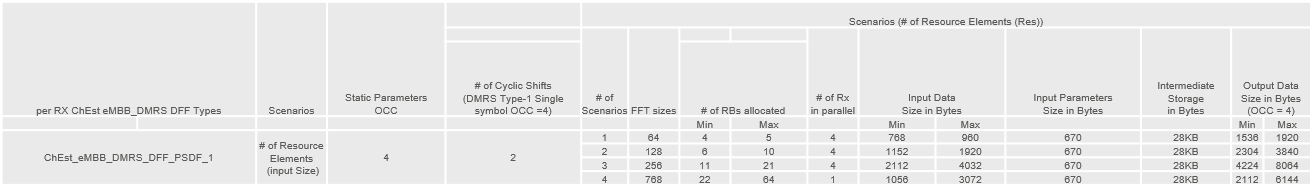}
	\centering
	\caption{Channel Estimation Scenarios}
	\label{fig:chest_scenarios} 
\end{figure}

\begin{figure}[h]
	\includegraphics[width=1.0\linewidth, height=3cm]{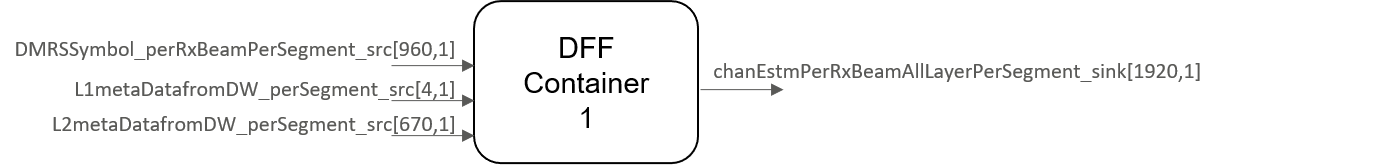}
	\centering
	\caption{Container for Scenario 1}
	\label{fig:container} 
\end{figure}

DFFs are created statically (offline) for  different scenarios. Because it is impractical to precompute all scenarios of a DFF, a number of commonly used scenarios are defined. We refer to a specific DFF scenario as a "container". In order to run a particular DFF, we must select the appropriate container in which our DFF fits. Figure ~\ref{fig:chest_scenarios} shows four different scenarios for channel estimation, while Figure ~\ref{fig:container} shows the corresponding container for scenario 1. In scenario 1, The input data size varies from 768 to 960 bytes, and the output data varies from 1536 to 1920 bytes. Hence, for scenario 1, we define a container that processes the worst case input size of 960 bytes and the worst case output size of 1920 bytes. With careful selection of containers, we can minimize the over allocation of resources. The advantage of containers is that we can accurately predict the resource requirements of a new DFF request, and whether it will interfere with DFFs that are already running on the machine. We can choose to reject a new DFF if it will cause other DFFs to miss their deadlines, thus providing isolation between DFFs.

Once a DFF as been accepted into the SRE, its microflows are loaded from memory. Each microflow contains producer and consumer information, the kernel to be executed, and the stage on which the microflow runs. All microflows belonging to the same DFF inherit a global tag from the DFF, which uniquely identifies each DFF instance in the system. Additionally, each microflow within a DFF has a unique local tag which has been precomputed. The (global tag, local tag) tuple uniquely identifies any microflow within the system.

\begin{figure}[h]
	\includegraphics[width=1.0\linewidth, height=8cm]{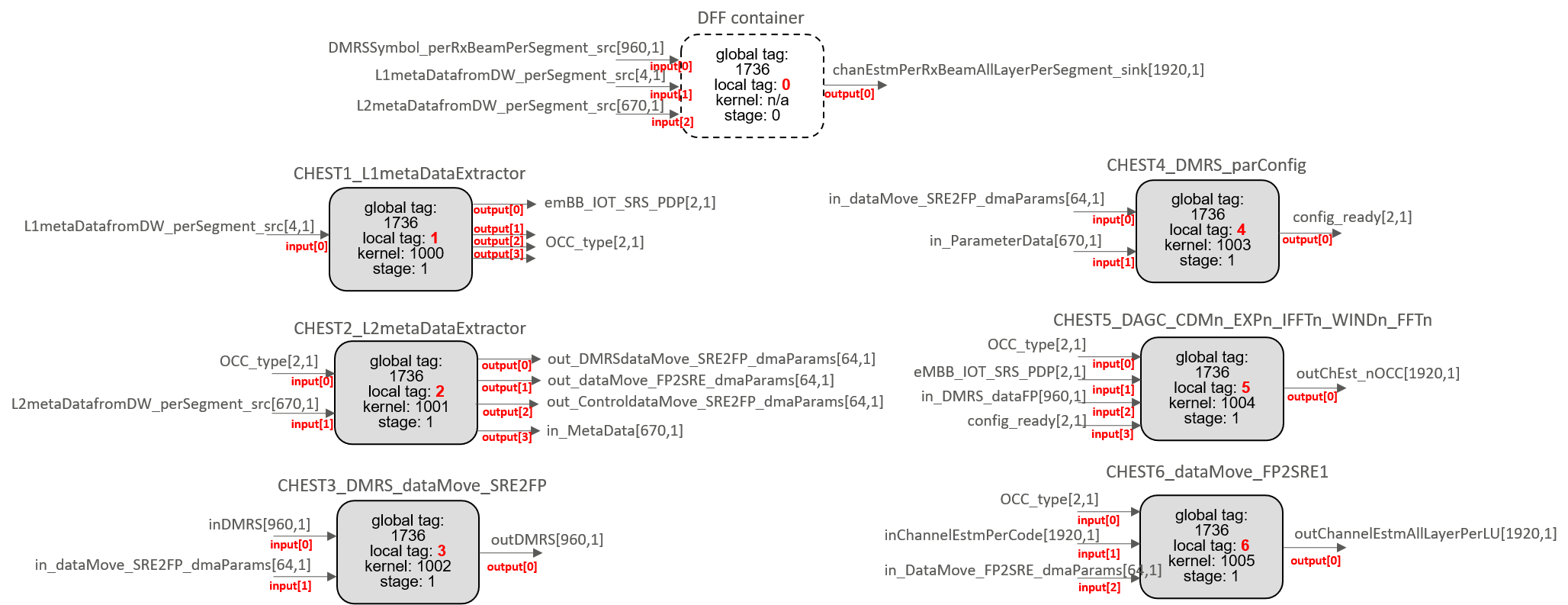}
	\centering
	\caption{Microflow Decomposition for Scenario 1}
	\label{fig:chest_microflows} 
\end{figure}

For the channel estimation example in Scenario 1, the microflow decomposition is shown in Figure ~\ref{fig:chest_microflows}. Notice that microflow 0 (i.e., the microflow with local tag 0) is always reserved to represent the DFF container. Hence, one can think of all microflows within a DFF as living inside the container boundary of microflow 0. This creates a simple numbering system for any microflow to communicate with the top level ports of the DFF. For instance, microflow 1, CHEST1\_L1metaDataExtractor, gets its input, L1metaDatafromDW\_perSegment\_src, from input port 1 on microflow 0. Similarly, microflow 6, CHEST6\_dataMove\_FP2SRE1, sends its output, {outChannelEstmAllLayerPerLU}, to output port 0 on microflow 0. A microflow will fire (begin execution) when all of its inputs are ready, and will retire when all of its outputs have been consumed.

In this example we have shown how channel estimation can be represented as a dataflow fragment (DFF), which can be subsequently decomposed into microflows. We believe this methodology can be extended to many different functions in wireless baseband processing, as well as other application domains.
\section{\label{sec:arch_tool_overview}A Novel Architecture and Toolchain Approach to Modem DSS}
In order to deal with these issues outlined in Section \ref{sec:intro} we propose a general philosophy of:
\begin{itemize}
	\item "Dataflow Native” hardware that 
	\begin{itemize}
		\item supports complete DFFs efficiently mixing in shared memory and compute resources.
		\item enforces stateless representation of flow, with all records erased locally on completion.
		\item simplies Heisenbug management through stateless DFF management.
		\item simplifies security via containerization of DFF (memory management stateless operation)
		\item maximizes functional flexibility with minimal control overhead using a hardened “OS”
	\end{itemize}
	\item “Firm Real Time Native” hardware that uses a policy based “pause, run, drop” strategy to removed the avalanche effect if a flow fails real time.
	\item “Energy Management Native” hardware that
	\begin{itemize}
		\item supports tooling for formal checking of performance.
		\item manages the flow of data thru memory as first class citizen.
		\item formally checks flows to manage energy rather than using speculation
		\item maximizes processing near/in memory
	\end{itemize}
\end{itemize}
\subsection{Hierarchical SoC decomposition and Formal Analysis}
\begin{figure}[h]
	\includegraphics[width=0.6\linewidth]{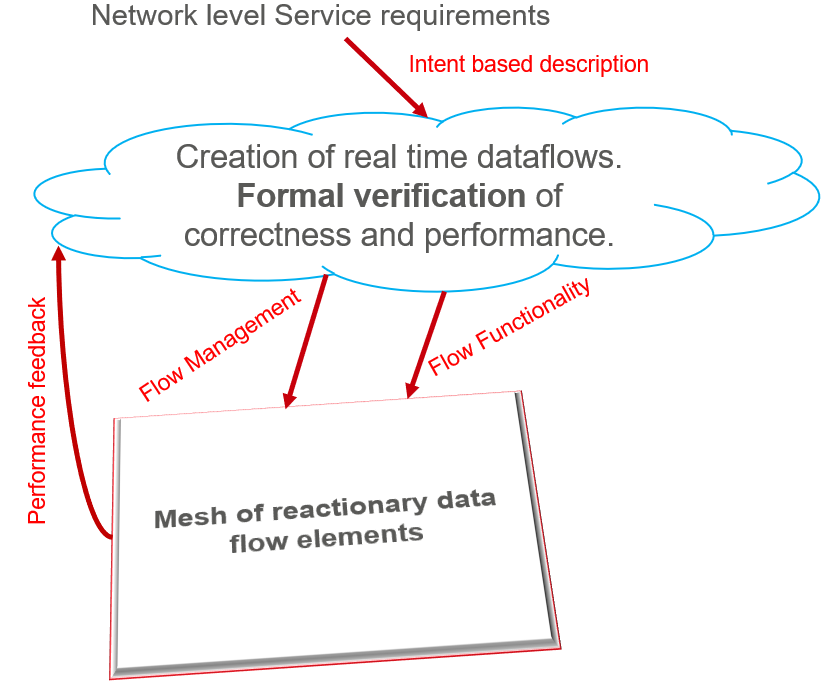}
	\centering
	\caption{Cloud Enhanced SoC Concept}
	\label{SoC_cloud} 	
\end{figure}

In order to allow a formal check of the SoC and also to make each layer as stateless as possible we propose a hierarchical decomposition. The top level of the SoC architecture is the cloud. Recognizing that all devices will be connected to the cloud we propose to use the cloud for long term control and configuration of the SoC for energy efficiency and formal verification of dataflow mixtures as part of the admission process for the SoC. Unlike previous generations, to our knowledge, the SoC is "dead in the water" without cloud connectivity with the cloud playing a critical low rate but continuous role in the reconfiguration of the SoC as shown in Figure \ref{SoC_cloud}.
The cloud will set parameters related to performance such as policy and scheduling (to be described in next Technical Report) that are used over a long configuration period during which time a certain mixture of DFF arriving within a know range of patterns is present in the SoC. Performance feedback at this level is also possible and can be used to tune policy. 

The SoC itself is an array of loosely connected DFF processing elements called Service Resource Elements (SRE) that are inspired by the Channel Elements from CDMA. The operation of the SRE is a critical component of our overall SoC strategy and we focus on it in this report. Its most salient top level features, seen from the SoC level are:
\begin{itemize}
	\item The SRE receive DFF control packets and then DFF data inputs, which they process to produce data outputs for the next DFFs. 
	\item Each DFF is a self contained DAG and is processed to "stateless" completion leaving only its outputs are result.
	\item There is a SoC level model for the arrival pattern of DFFs of each DFF flow. A flow of DFFs simply being a continuous arrival of DFFs at an SRE that is grouped for analysis convenience.
	\item  The SRE is not allowed to back pressure a flow. This means that the arrival pattern is maintained and can be formally analyzed.
	\item The SRE must also produce output data from a flow within the bounds of an SoC level departure pattern. SRE will be formally checked to make sure it achieves the pattern.
\end{itemize}

\begin{figure}[h]
	\includegraphics[width=1.0\linewidth]{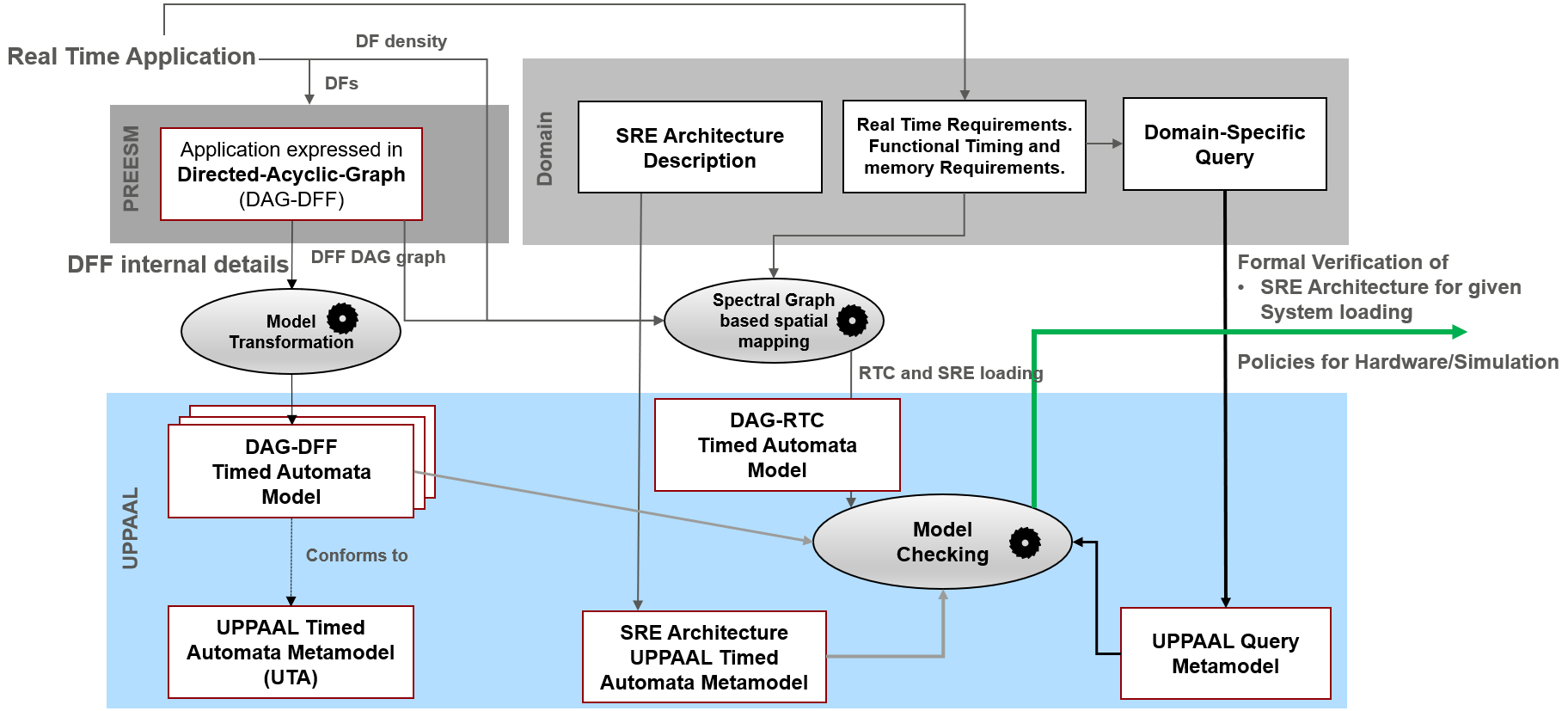}
	\centering
	\caption{Top Level Diagram of Formal Toolchain}
	\label{formal_toolchain} 	
\end{figure}

Once the input and output patterns are formally verified for an SRE it becomes a black box for SoC level checking of dataflow. The three concepts or input and output flow patterns and formal checking of each SRE to these patterns allows a hierarchical decomposition of the behavior of the SoC. The toolchain we present is therefore a critical aspect of this decomposition as it provides for a formal checking of the SRE when it experiences a set of flow patterns. The Complete formal toolchain is shown in Figure \ref{formal_toolchain}.
The flow is as follows:
\begin{enumerate}
	\item The Application engineer provides a set of DF along with the pattern of creation for each DF type. 
	\item The DF are partitioned into DFF using a dataflow tool such as the open source tool PREESM (see Section \ref{sec:FV}). This produces DFF with derived creation patterns. 
	\item Based on the DFF patterns each DFF type is statically mapped to one or more SRE. We have used spectral graph theory and an edge weight based heuristic to perform this and get intuitive results and have also used SAT and ILP. The static mapping allows us to develop patterns of DFF arrival for each DFF type for each SRE, though we have not automated this process yet.
	\item The DFF DAG along with its data packet and runtime ranges is converted to a model in a formal analysis tool. We use a Timed Automata (TA) metamodel converting from the xml output from PREESM to a format suitable for the UPPAAL TA analysis tool using a Python program (see section \ref{sec:FV}).
	\item The patterns of arrival and the DFF DAG are combined with a TA model of the SRE architecture and the complete SRE model is formally checked using a set of queries.
\end{enumerate}
If all queries pass the DFF loading on the SRE is considered to be formally feasible. If it fails the query(s) that fail will pinpoint the reason for failure and human intervention will be required to adjust the input real time application requirements. How to automate the process of relating the query failure to suggestions for how to modify the requirements is a topic for future research.

\subsection{The Service Resource Element}
\begin{figure}[h]
	\includegraphics[width=1.0\linewidth, height=6cm]{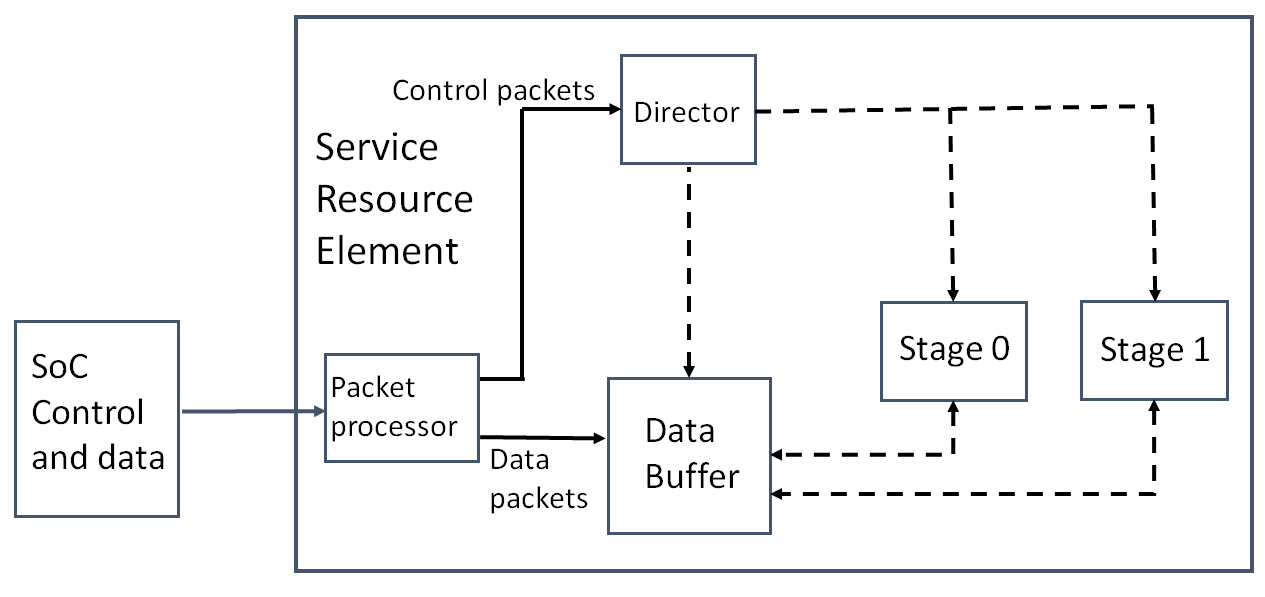}
	\centering
	\caption{Service Resource Element Architecture}
	\label{SRE_toplevel} 
\end{figure}

The service resource element (SRE) is the fundamental building block of our domain specific architecture. It was designed from the ground up to address the challenges encountered with current modem implementations. It was architected to work specifically with our tool flow and our application development philosophy. A high level view of the SRE and its internal structure is shown in Figure \ref{SRE_toplevel}.

\subsubsection{SRE basic structure}
In an SRE-based modem design, multiple SREs will be connected together using a network-on-chip (NoC). Each SRE is a computing cluster that can process several dataflow fragments (DFFs). An SRE is further partitioned into stages, where each stage processes individual microflows (refer to Section \ref{sec:dataflows} for a more detailed description of microflows).
\subsubsection{SRE interface to processing flows, Policy and Control} 
When a DFF needs to be processed, an external host, which could be another SRE, sends the SRE a control packet to request the necessary resources. The control packet generates a request to the director (the top level controller of the SRE), which applies a \emph{policy} to determine if the request will be rejected or accepted. \\
The policy is a set of rules that determines if any predetermined resource allocation strategies, which we refer to as ``Tetris" blocks, can be safely deployed given the current system load. \\
If the request is rejected, the director tells the packet processor to discard the corresponding data packets and to inform the initiator that the request was denied. However, if the request is accepted, the director informs the packet processor that the corresponding data packets may be stored in the data buffer. 
\subsubsection{SRE DFF decomposition and Microflow processing}
As the data arrives, the director dispatches \emph{microflow descriptors} to one or more stages. A microflow descriptor tells a stage everything it needs to know about how to execute a particular microflow, including where to get its input data, what kernel(s) to use, and where to send the output data. \\
During execution, microflows in a DFF consume, process, and produce tokens. A microflow fires when all of its inputs are ready, and retires when all of its outputs have been consumed. Microflows may send and receive tokens within the same stage, or may be required to send and receive tokens across stages. However, microflows that produce final output tokens (i.e, tokens that are the outputs of the DFF itself) send their output tokens back to the data buffer. Mulitple microflows may be in flight at the same time on the same stage and a DFF may spread across multiple stages.
\subsubsection{SRE DFF completion}
Once the packet processor has received all output tokens for a DFF, it sends the output data onto the next assigned processing block (this could be the same SRE, a different SRE or a SoC level control block) as one or more data packets. In keeping with our robust design philosophy, any microflow that fails to meet its deadline, or malfunctions in some other way, should not cause unrelated microflows to fail. If a microflow times out or fails, it deallocates its resources, cleans up its state, and generates an error code (see Section \ref{sec:error_codes}). 
\\

The following sections describe the SRE architecture and formal toolchain in more detail.
\section{\label{sec:SRE-Arch}The SoC Architecture of an SRE}
\begin{figure}
	\includegraphics[width=\linewidth, height=10cm]{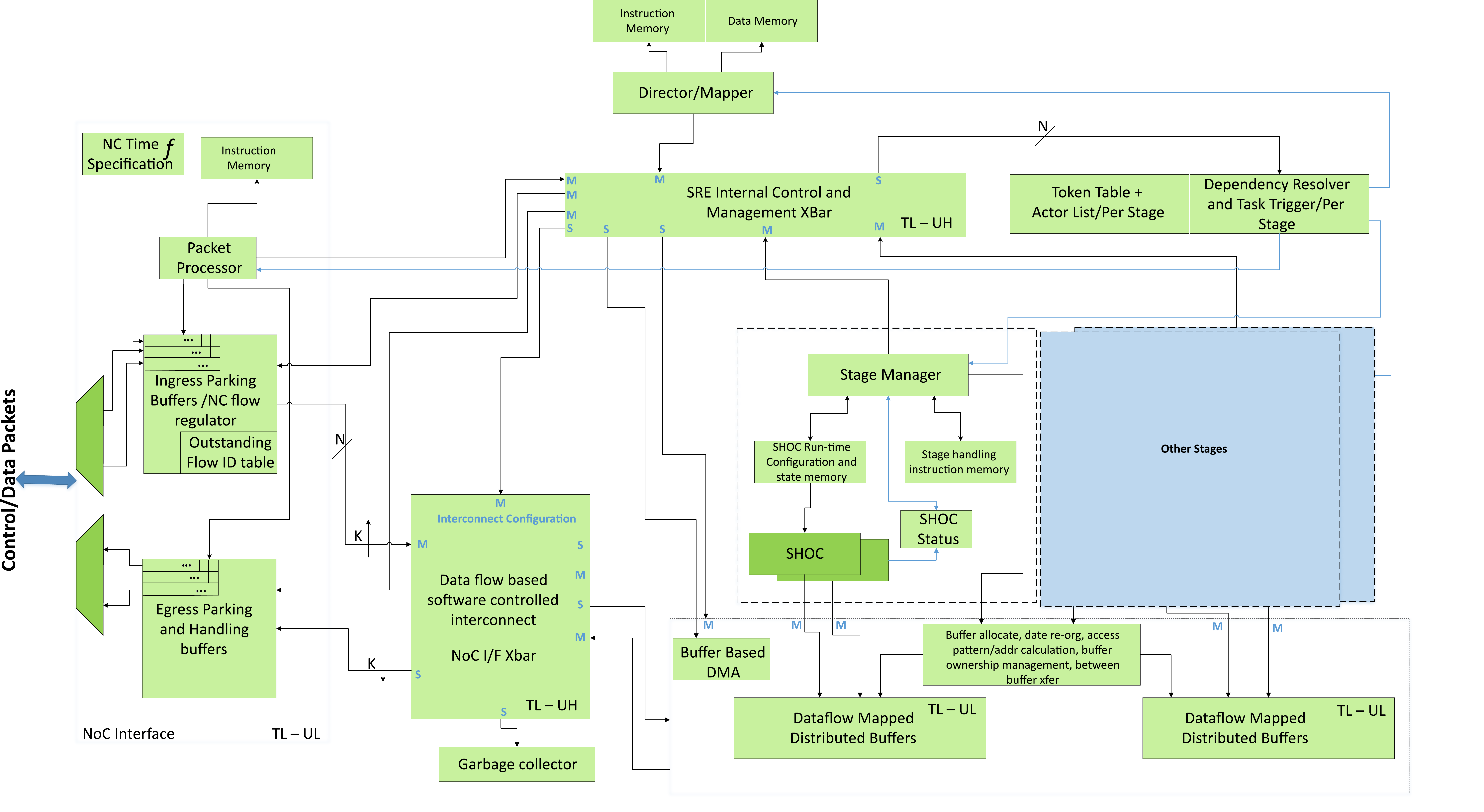}
	\centering
	\caption{SRE SoC Architectural Block Diagram}
	\label{fig:SRE-top-level-view} 
\end{figure}

Fig. \ref{fig:SRE-top-level-view} is the SoC block diagram of an SRE. The SRE has been architected to be highly configurable driven by the real system requirements imposed by the application, in our case wireless baseband modem. The SRE can be reconfigured during two different points in time.  The first one is during integration time where we can configure it with static system specifications such as number of sectors or number of antennas, these usually represent a typical base setting of base stations; these static configuration requirements and can be obtained once an SoC specification is derived. \\

The second one is the run-time reconfiguration enabled by the run-time selection of the number of stages,  and reconfiguration of the processing engine called SHOC, SHOC will be covered in great details in section \ref{subpara:SHOC}. \\

Though we have architected the SRE to be highly adaptive and flexible, it is also architected to support temporal composability without losing the need to satisfy the real-time QoS (Quality of Service) requirements mandated by wireless baseband processing. By its intrinsic nature a PRET (PREcision Time system)\cite{5757922} tries to improve overall resources utilization to further reduce the costs. It is therefore treated as a template for the SoC integration and development, which consists of the following key elements: \\
\begin{itemize}
	\item \textbf {The Director and Packet Processing Subsystem}: This subsystem is responsible for the control and data interfaces between an SRE and the rest of the SoC. It receives and interprets control and data packets, and generates, populates and terminates DFF information to the stages assigned. It also co-ordinates data packets' admission, sets up necessary data triggering information to the assigned stages that have ingress and egress interfaces, and orchestrates data movement once a DFF flow is completed. 
	\item \textbf {The Flexible Stage Subsystem}:  This subsystem consists of multiple identical collections of resources as shown inside the dash line plus the dependency resolver with the token table and actor list shown in Fig. \ref{fig:SRE-top-level-view}. It is the working engine of the overall architecture and it can be dynamically triggered based on relationships defined by data flow graphs with the combination usage of the data dependency resolver, the actor list and token table inside of a stage. Also, the SHOC(s) is(are) can be re-configured during run-time by the stage manager and depends on the functions defined by the data flow fragments assigned for a specific stage.  SHOC (Scalable Hybrid and Organizational Computing) element will be described in greater detail in a later section and consists of heterogeneous components that meet the high PPA efficiency required by the baseband processing system.
	\item \textbf {The Control and Data Infrastructure}: This subsystem consists of three switching interconnects: one is the control and management crossbar located in the top of the figure and is dedicated to control information exchange inside of an SRE; the second one is the data-flow based interconnect reserved for the NoC interface; The third one is located at bottom right of the figure and works together with the shared buffer and memory subsystem.
	\item \textbf {The Shared Buffer and memory Subsystem}:  This subsystem takes the lion share of the area inside of an SRE. It is a distributed and shared memory subsystem \cite{9241708}, and it basically has interfaces to all the other architectural components for data exchange purposes.
	\item \textbf {The DMA Subsystem}:  This subsystem is under the supervision of the stage manager(s), the director and packet processor inside an SRE so to efficiently move data across multiple stages and to the rest of SoC.
\end{itemize}

\subsection{\label{subsec:SRE-func-des} Functional and Operation Description of an SRE}
\subsubsection{\label{subsec:Director} The Director and Packet Processing Subsystem}
\begin{figure}[!h]
	\includegraphics[width=\linewidth, height=8cm]{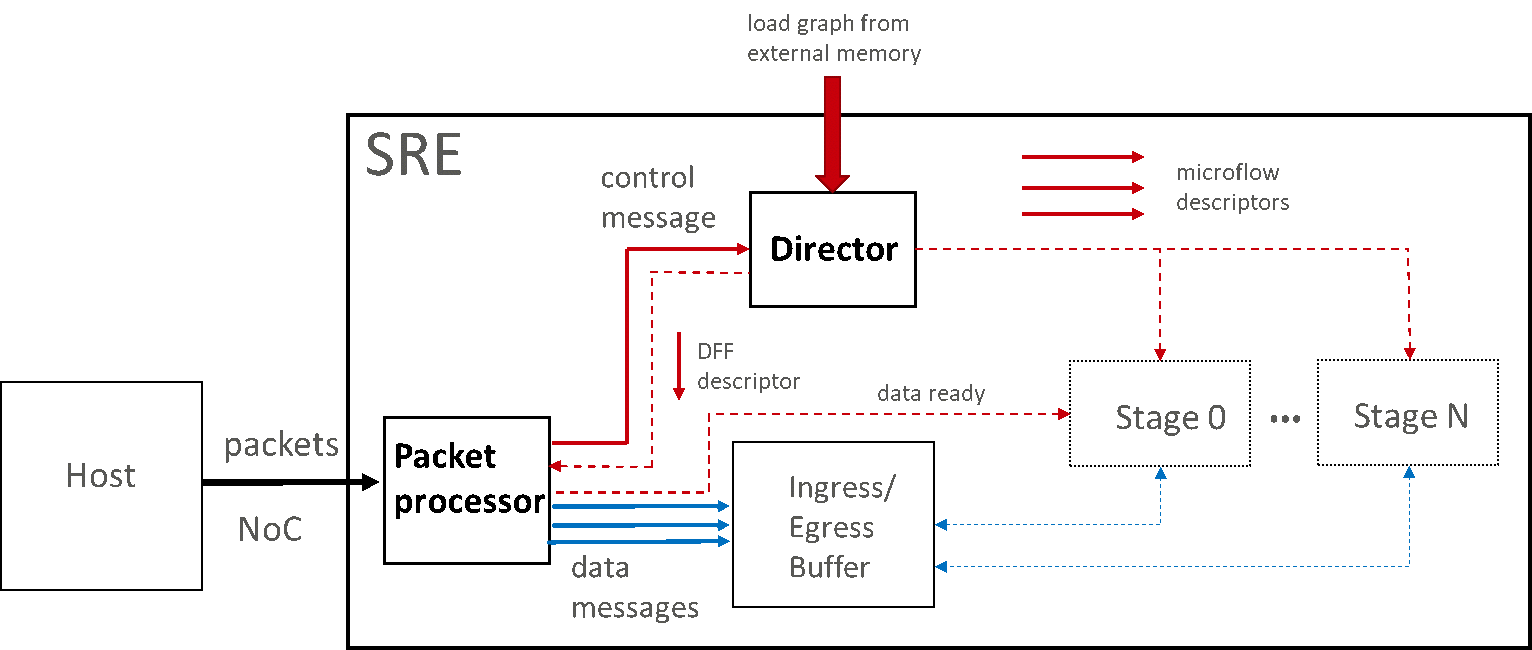}
	\centering
	\caption{Director and Packet Processor}
	\label{fig:director_packet_subsystem} 
\end{figure}

The director and packet processor are two different cores that work in tandem to provide top level control for the SRE (Figure \ref{fig:director_packet_subsystem}) The packet processor's primary responsibility is to keep track of control and data messages, while the director handles policy implementation, descriptor creation, and supervisory functions. Both processors are expected to be RISC V based cores with instruction set extensions and specialized coprocessors.

The general steps for executing a DFF on an SRE are:

\begin{enumerate}
	\item Request DFF resources
	\item Wait for input data to arrive
	\item Execute DFF
	\item Send back output data
\end{enumerate}

\begin{figure}[!h]
	\includegraphics[width=\linewidth, height=6cm]{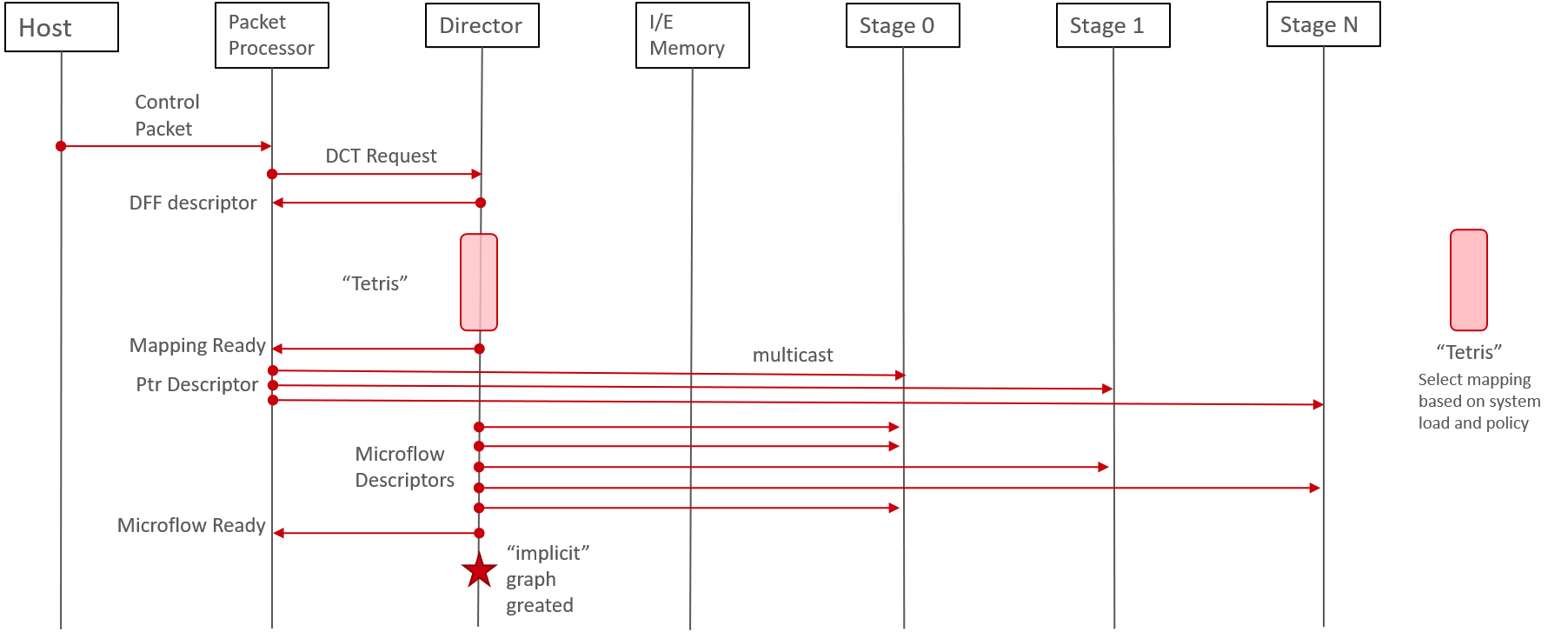}
	\centering
	\caption{Requesting DFF Resources}
	\label{fig:request_dff_resources} 
\end{figure}

When the host wants to process a DFF, it first sends the SRE a control packet, which tells the SRE which resources it requires (Figure \ref{fig:request_dff_resources}). The control packet contains a global tag, which identifies the unique instance of a DFF in the system, and a container id that specifies the type of graph that best matches the DFF to be executed. The control packet is forwarded to the director as a DCT (container type) request. If the particular container type is supported by the SRE, then the director sends a DFF descriptor back to the packet processor so that it can begin receiving data packets. However, at this point, the director has not yet accepted the request as it has not yet determined resource requirements.

The director applies a policy to the request to determine if it can be accepted by the SRE. In order to avoid being a computational bottleneck, resource requirements of a container are pre-computed in one or more "feature vectors". The director examines several feature vectors associated with a DFF container to see if any mapping will fit inside the machine given the current system load. We refer to this step as playing "Tetris". If no mapping can be found, the director reports an error and instructs the packet processor to discard the DFF descriptor and any corresponding data. However, if a mapping can be found, the director sends a mapping ready signal back to the packet processor and begins to load the graph from external memory. 

Note that the graph is not loaded explicitly, in the form of nodes and edges, but implicitly as microflow descriptors. Each microflow descriptor contains information about its inputs, outputs, kernel(s) to be executed, and the stage on which the microflow executes. Microflows within a DFF are identified using a local tag, which is statically determined. When the director reads a microflow descriptor during run time, it appends the DFF global tag to the microflow descriptor. This allows a microflow instance to be uniquely identified using a (global\_tag, local\_tag) tuple. The complete microflow descriptor is then sent to the appropriate stage where it will be executed. Once all microflow descriptors in a container have been sent to their appropriate stages, the director sends a microflow ready signal back to the packet processor.

\begin{figure}[!h]
	\includegraphics[width=\linewidth, height=6cm]{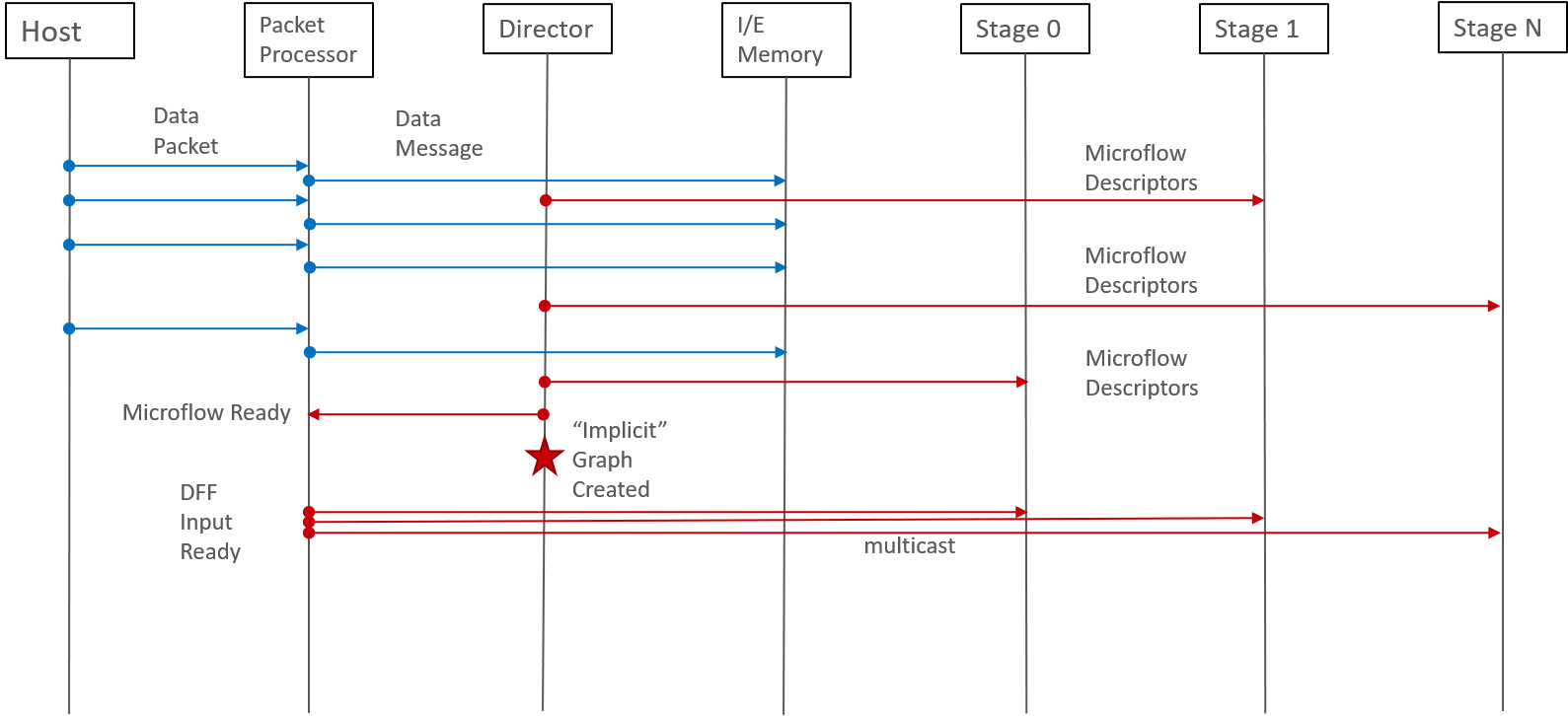}
	\centering
	\caption{Receiving Input Data}
	\label{fig:receive_input_data} 
\end{figure}

While the director reads microflow descriptors and dispatches them to the appropriate stages, the packet processor is collecting data in its ingress buffer (Figure \ref{fig:receive_input_data}). The packet processor sends a pointer descriptor to the appropriate stages indicating where the DFF inputs are arriving. Once all input data for a dataflow fragment has arrived, and all microflow descriptors have been dispatched by the director, the packet processor sends a data ready message to the appropriate stages, indicating that the DFF is ready to execute.

\begin{figure}[!h]
	\includegraphics[width=\linewidth, height=6cm]{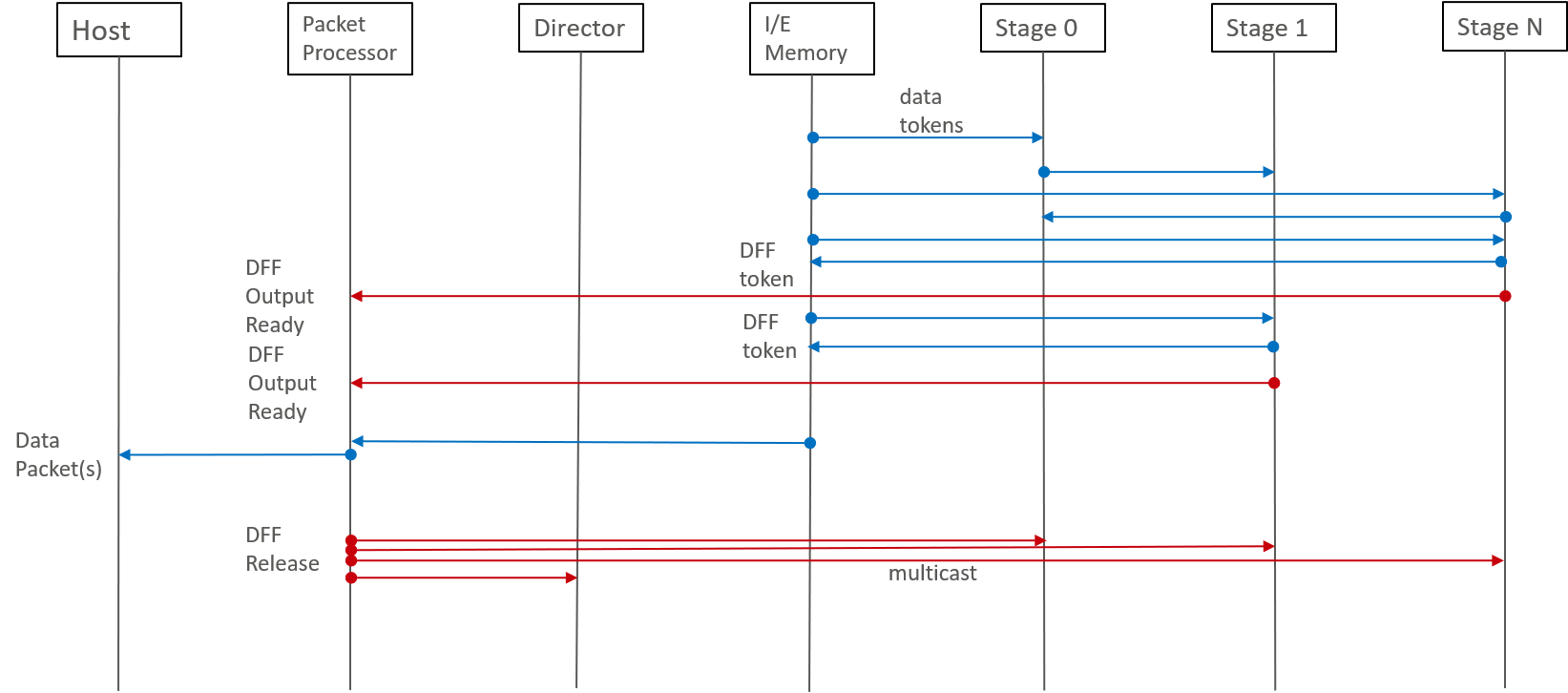}
	\centering
	\caption{Executing the DFF}
	\label{fig:execute_dff} 
\end{figure}

Once the input data is ready, all awaiting microflows are activated and execute the DFF. Microflows within a DFF continue to pass intermediate tokens within a stage or between stages until a DFF output token is produced (Figure \ref{fig:execute_dff}). When a DFF output token is written back to the egress buffer, a DFF output ready signal is also sent back to the packet processor so it can keep track of the output tokens. Once the packet processor receives all output tokens for a DFF, it sends the output data back to the host in the form of data packets. The packet processor also sends a DFF release message to the director and all stages indicating that the DFF has completed execution.

\subsubsection{\label{subsec:Stages} The Flexible Stage Subsystem}
\begin{figure}[!h]
	\includegraphics[width=0.6\linewidth,height=0.6\linewidth]{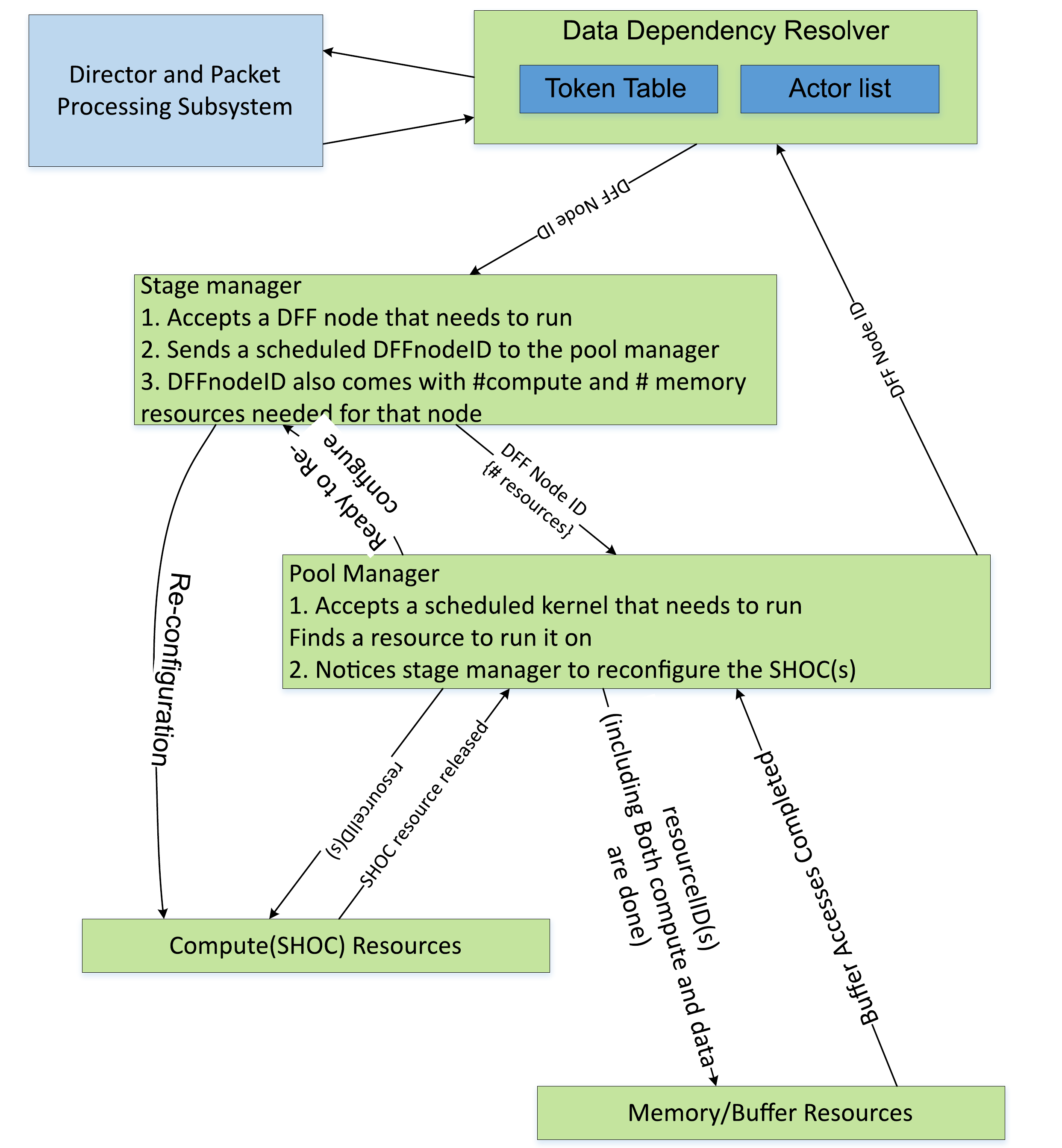}
	\centering
	\caption{Interactions Between the Director and Some Key Components Inside a Stage}
	\label{fig:SRE-stages-funcs} 
\end{figure}

Multiple homogeneous stages can be instantiated during the integration time. As shown in Fig. \ref{fig:SRE-stages-funcs}, every stage is activated and triggered by the input token(data) and works on the granularity of a microflow/kernel. It is passively driven by the microflow/kernel assignments deployed by the director. Once the director has finished to assign the miroflow/kernels of a DFF to a particular stage via control messages. The token table and actor list for a stage are populated to capture the connection information of data flow graphs, microflow/kernel trigger conditions and dependency information among them. The data dependency resolver is triggered by any control message sent from the director and other stages or data triggering events generated inside the same stage. \\

\begin{figure}[!h]
	\includegraphics[width=\linewidth]{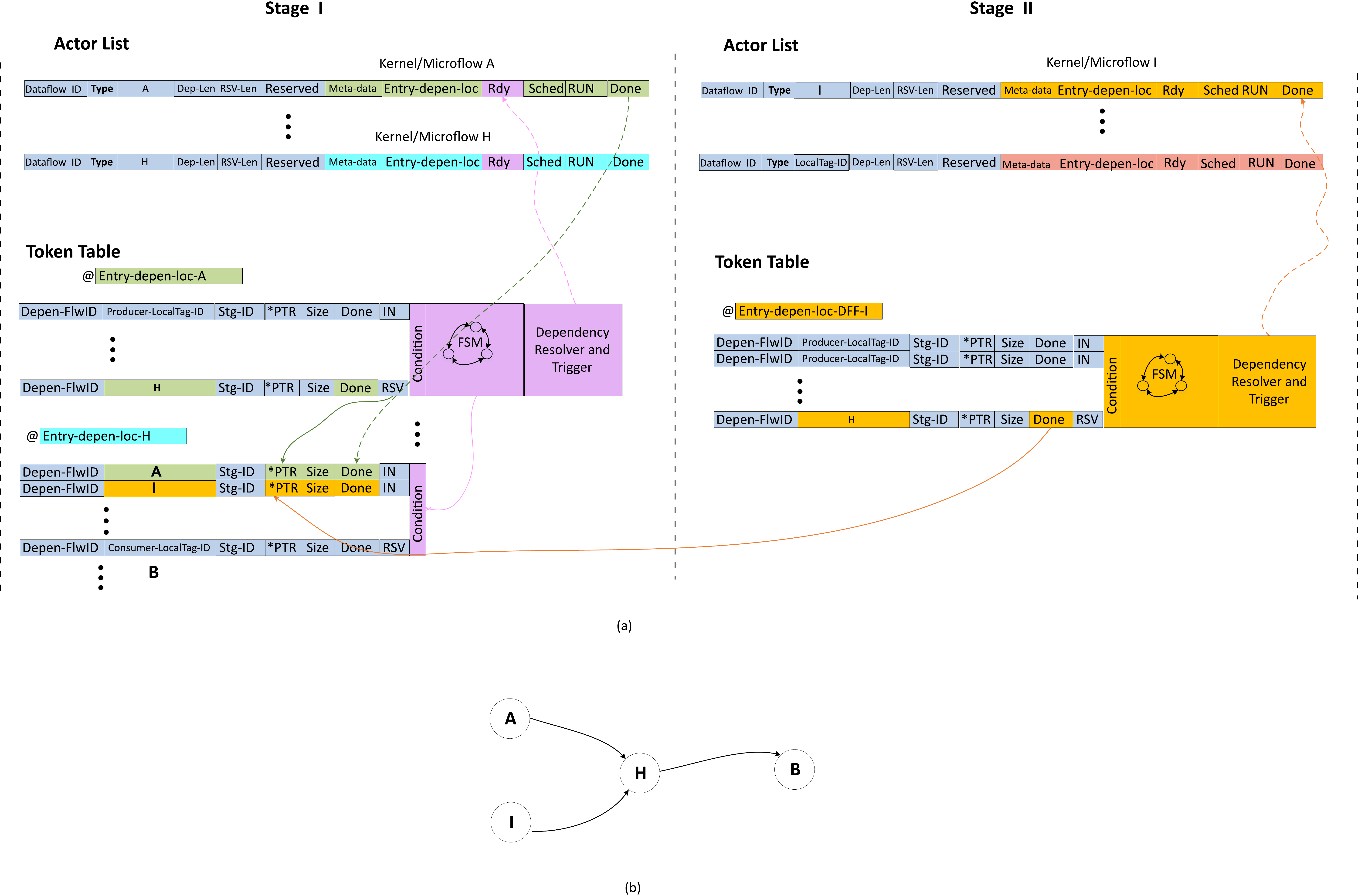}
	\centering
	\caption{Micro-architecture of Token Tables and Actor Lists and How They Work Together}
	\label{fig:SRE-stages-tokenActor} 
\end{figure}

The data dependency resolver is responsible for interpreting the control messages from outside a stage and any relevant events generated inside the same stage. It can be implemented by a group of finite state machines to fulfill table searching and condition evaluations as shown in Fig. \ref{fig:SRE-stages-tokenActor}. \\

Once one or multiple DFFs are triggered and ready to run after the evaluation of the data dependency resolver, it(they) will be presented to the stage manager via the ready list of the DFFs, which is normally implemented by a queue. The stage manager is responsible for the following tasks inside a stage even though its functionalities are captured in two separated boxed in Fig. \ref{fig:SRE-stages-funcs} \\
\begin{itemize}
	\item \textbf {Miscellaneous Managements Tasks Inside a Stage}: 
	\begin{itemize}
		\item{Taking the relevant information from the input arcs  such as the size of input tokens of a  microflow/kernel that is ready to run }
		\item{Preparing a DMA transfer parameter list if a DMA is required to move the input tokens from other stages.}
		\item{Checking the available buffer space for a stage and presenting the information to the scheduler and pool manager for further decision making}
		\item{Maintaining and updating the ready queue list based on a microflow/kernel completion status}
		\item{Sending necessary control message to the other stage(s) or generating the necessary event for the same stage once a microflow/kernel is completed successfully}		
	\end{itemize}
	\item \textbf {Run-time Scheduling}: It is a very important role for the stage manager to apply different scheduling algorithms such as Early Deadline First, round robin based on the timing information carried in the meta-data sub-field in the actor list as shown in Fig. \ref{fig:SRE-stages-tokenActor}.
	\item \textbf {Resource Pool Management and Allocation }: Once a micro-flow is scheduled, the stage manager acts as a pool manager to reserve output tokens from the shared buffer memory inside a stage. It also reserves the computing resource (SHOC in this case) so the scheduled microflow/kernel has the necessary resources to finish on time.
	\item \textbf {SHOC Reconfiguration}: It is responsible for re-configuring a SHOC(s) during run-time if and when a new SHOC image is required to perform a different function than the one it is currently assigned. This task is called upon once a DFF is scheduled and the pool manager has reserved the necessary resources for the scheduled microflow/kernel.	
\end{itemize}

\begin{figure}[!h]
	\includegraphics[width=0.9\linewidth,height=0.9\linewidth]{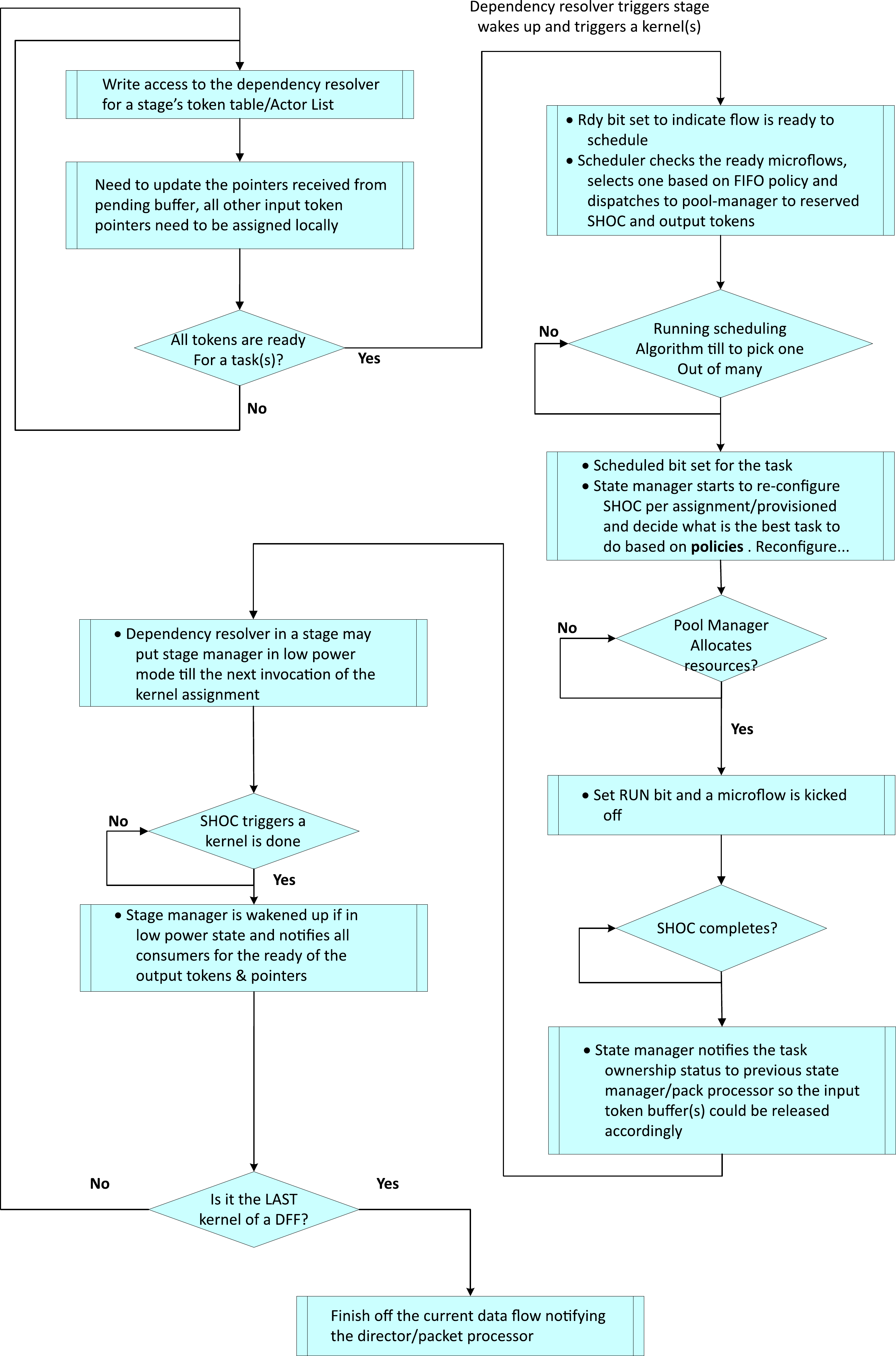}
	\centering
	\caption{Operation Flow Chart of the Stage Manager Inside a Stage}
	\label{fig:SRE-stages-flowchart} 
\end{figure}
Fig. \ref{fig:SRE-stages-flowchart} shows the operation flow of a typical stage. We need to cross-reference the definitions of Actor List and Token Table shown in Fig. \ref{fig:SRE-stages-tokenActor} to understand how it works. Fig. \ref{fig:SRE-stages-tokenActor}(b) shows an example snippet of a DFF. Microflow A and I are input arcs to trigger the microflow H, and microflow B is triggered by the completion of the microflow H afterwards.\\
\\
As shown in Fig.  \ref{fig:SRE-stages-tokenActor}(a), there are multiple entries in the actor list inside Stage I and II. The actor list consists of multiple uniform-sized entries, each entry is employed to describe one microflow/kernel out of many assigned to a stage. Because each microflow/kernel might have a different number of input and output arcs, we employ a second table called token table to to handle the irregularity. In each actor list entry, there is a subfield called Entry-depen-loc, this is the pointer to locate the full definition of all the input and output arcs of a microflow/kernel defined in the corresponding actor list. \\
\\
As mentioned previously, the data dependency resolver evaluates the readiness of all microflow/kernels assigned in a stage once a control message is received from other stages or an internal data event is generated.  Please note the definition of ready is the availability of tokens for all input arcs of a specific microflow/kernel. With that, we called the microflow H is ready \textbf{only} when both the input tokens of input arc A and I are available for microflow H in stage Stage I. The data dependency resolver sets the Rdy bit in entry of microflow H and pushes relevant information to the ready queue in a stage. \\

As described in Fig. \ref{fig:SRE-stages-flowchart}, the stage manager takes over beyond this point. It prepares all the necessary information such as the timing information carried in meta-data of the actor list, and the loading of a stage and the configurations of a SHOC(s) etc. to the scheduler so that it can make an optimal decision. Once a microflow/kernel is scheduled, the Sched bit in the corresponding entry of the actor list will be set. At the same time, the scheduled task is passed to the pool manager so it can allocate the buffers/tokens for all the output arcs. The pool manager also needs to reserve the computing resource (SHOC in this case) so the microflow/kernel can be accommodated. The pool manager notifies the stage manager if a re-configuration to the allocated/reserved SHOC(s) is(are) required. \\
\\
The last step before a microflow/kernel runs is to wait till the re-configuration to a SHOC(s) is(are) completed. The stage manager monitors the re-configuration and it also sets up the necessary parameters to SHOC(s) so it(they) can run properly once it triggers the SHOC(s) to run. The corresponding RUN bit for a scheduled entry in the actor list will be set once the corresponding SHOC(s) starts to run. \\
\\
Once the SHOC has completed an assigned microflow/kernel, it triggers a data dependency resolver and stage manager in the same stage so that the data dependency resolver(s) located in other stages can start to evaluate any other microflow/kernel dependencies based on the completed microflow/kernel. For example, for microflow A and H, they both belong to stage I. Therefore, once microflow A is completed, an internal event will be created and it triggers the data dependency resolver in stage I to evaluate whether microflow H can be ready to run or not. In another scenario, microflow/kernel I belongs to stage II as shown in Fig. \ref{fig:SRE-stages-tokenActor}, the stage manager located in stage II is responsible for sending a control message to stage I once microflow I is completed. The corresponding control message routed from stage II to stage I via the control and management crossbar triggers the data dependency resolver located in stage I to evaluate all microflows/kernels that depend on the microflow I. Then the operation flow chart described in Fig. \ref{fig:SRE-stages-flowchart} starts. \\
\\
As the very last step of the operation flow, and after a microflow/kernel is completed, the stage manager removes the corresponding entry from the ready queue list, resets the corresponding Rdy, Sched, RUN bits for a completed microflow/kernel in the actor list. The same microflow/kernel now is back to the normal waiting state and waits for further data triggering. The stage manager also needs to release the input buffers reserved for the completed task and be ready to give the ownership of all the output buffers for the relevant triggered microflow/kernel(s). \\  
\\
If the running microflow/kernel is the last segment within a complete DFF, it needs to send a control message to the director/packet processor to notify the completeness of the whole DFF. This way, they can take further actions with regard to the whole DFF for things include but not limited to the following:
\begin{itemize}
	\item{Set up DMA parameters to move the final results of a DFF to the rest of SREs on the SoC. Release the output token buffer spaces once the data is moved out completely}	
	\item{Retire the completed DFF entry deployed across multiple stages if the same DFF will not be run beyond this point}			
\end{itemize} 
\subsubsection{\label{subpara:shared-buffer} The Shared Buffer and Memory Subsystem}
The SRE is a native dataflow architecture, thus it works on the granularity of a token, which is directly mapped to a buffer or FIFO in the SoC micro-architecture. During the run-time, buffers that hold tokens are created and released along with a dataflows that are activated or completed on the fly. The size of a token varies a lot and it could have multiple producers and consumers to access it at any time. Also many producers and consumers spread across multiple stages, spatially all need to access the shared buffer(s).\\
\\ 
This dictates that the memory architecture be shared and distributed in nature. It also needs to accommodate multiple shared buffers with different sizes that can be accessed from many ports. This brings up a difficult challenge to the overall memory subsystem inside an SRE. Therefore, we need to apply our latest shared memory techniques described in \cite{9241708} and aided with the window-access technique described in \cite{4429955} as the starting point for the next level of micro-architecture. \\
\subsubsection{\label{subpara:Interface} The Data and Control Infrastructures}
Inside of an SRE, there are three switching networks in Fig. \ref{fig:SRE-top-level-view} to support the following needs:
\begin{itemize}
	\item \textbf {The SRE Internal Control and Management Crossbar}: It is responsible for the control message communications among the director, packet processor and stage managers inside the SRE. The interconnect carries short control messages and needs to have very short reactive time to make sure all important messages can be routed from source to destination as quick as possible. It also needs to have properly acknowledgment sent from the recipient to make sure no important information gets dropped.  
	\item \textbf {The NoC Interface Crossbar}: It is responsible for the data exchanges between the rest of SoC and the shared buffers inside of an SRE. It often carries big and long data packets, however, any connected route has a much longer live time than that of the control and management crossbar because director/packet processor sets up the connections on a granularity of a dataflow. It can be treated as a slower switching network inside of an SRE. 
	\item \textbf {The Dataflow Mapped Distributed Buffer Network}: It can be considered as part of the overall shared buffer memory subsystem. Once again, it works at the granularity of a dataflow and it carries the bursty and windowed traffics to a portion of a shared buffer(s) most of the time. The access protocol is a very simple since it directly interfaces with single or two port SRAM.  			
\end{itemize} 
We have considered TileLink \cite{Risc-v/Tile-Link} as the basis to implement all of the three switching networks described because it offers multiple protocols with the same foundation but at different level of implementation complexity. With that, the SRE internal control and management and NoC interface crossbar are based on TileLink Uncached Heavyweight (TL-UH) protocol because it provides robust handshake scheme to guarantee that no information gets lost; and the dataflow mapped distributed buffer network is based on TileLink Uncached Lightweight (TL-UL) protocol for the purpose of easy implementation. One more reason why TileLink is considered is because a couple of open-source RISC-V cores are evaluated for the implementation of the director, packet processor and stage managers. They all support Tilelink interfaces so TileLink enables a fast prototype and smooth integration capability.\\
 
\subsubsection{\label{subsec:DMA} The DMA subsystem}
There isn't any special requirements to the DMA engine from the SRE perspective. The only requirement is that it can support multiple "data moving engines" that understand the nature of multiple dataflows moving tokens in and out any stage, or in and out of the SRE simultaneously. There are many choices for this sub-system.

\subsection{\label{subpara:SHOC} The Scalable, Hybrid and Organizational Computing (SHOC) Architecture for Multi-Application Domains}
\subsubsection{Introduction to Scalable Compute}
Until breakthroughs of computing technologies such as quantum computing and/or graphene/CNT processors are productized, CMOS based technology remains the only choice for the chip processor industry.  Therefore, effectively using existing technologies to build PPA efficient products, that need to meet exponentially increasing computing demands, remains a big challenge which is faced by the entire chip industry in the post-Moore’s law era.  Particularly in the wireless industry, as we evolve from 5G to 6G, not only will the computing demand increase dramatically, but the variety of application domains will no longer just be wireless modem applications, these would be extended to other domains like AI/ML and applications like sensing and positioning. This requires a "6G modem" that's more flexible and generic than conventional wireless modems, in addition to having much higher PPA efficiency and supporting hard real time requirements than previous generations (4G/5G) systems.  An efficient and flexible computing system is a must for future 6G computing.  \\

Since Dennard scaling ended in 2004, the clock rate could no longer be raised, and computing components can only be spatially expanded, from single core to multi-core, and then to many cores. Around 2008, the concept of “dark silicon” started getting popular, it claimed some parts of the chip would have to be powered off to avoid high power consumption and excess heat generation. However, one might ask the question, if silicon is partially used, why put it in in the first place? perhaps re-usability would be a better solution.  In the end, there is little clarity on how to deal with the post-Moore’s Law era and most chip companies continue to rely on scale rather than efficient use. \\

Scientist and engineers realized long time ago that a single technology cannot meet computing demands, and that different computing technologies should be combined to meet the challenges for low power and area designs, particularly in post Moore’s Law era.  For many years, the GPP industry has been trying to solve these problems by adding more programmable and customized accelerators to the main CPU/DSP cores; the FPGA industry has been taking the approach of embedding more pre hard-wired DSP/CPU blocks with fixed bit precision, which sacrifices the FPGA’s fine grained flexibility that supposedly could generate arbitrary types and precision computing elements; GPUs, originally designed for video processing, have been modified to support more generic applications and integrate TPUs to improve the compute efficiency for neural networks.  The recent NVIDIA plan to acquire ARM confirmed this direction.  Since Google’s TPUv2 published stunning energy and performance numbers over CPU and GPU in 2017, TPUs have been evolving from a programmable ASIC processor to a more flexible and higher performance processor to cover ML training tasks, (not just inferencing as in TPUv2) and hoping to be used for an expanding set of applications.  In the recent two years, the top two processor manufacturers, Intel and AMD each acquired the two largest FPGA companies, hoping to increase the computational efficiency of GPPs through the flexibility of FPGA.  The decisions seem largely impacted by the research results of Microsoft’s Catapult project, which claimed that at a cost of 10\% power increase brought by FPGA boards they could gain 90\% throughput improvement for their search infrastructures  \cite{MS_Catapult_2016}. \\

We have observed in past several years that there isn't a single technology, or multiple technologies for that matter, that can use a brute force approach in single or multi die chiplet, to dominate the industry as a mainstream general purpose processor with wide acceptance. Particularly in the wireless industry, where energy and area are extremely sensitive, GPP processors do not fit the requirement of low power and area, let alone meet the requirements for hard real time and latency.  We have to consider a less general purpose architecture targeting the wireless application domain, enter DSA for wireless or WDSA.  And for future 6G, we have to also support additional domains as discussed earlier in this section.  In the rest of this section, we propose a computing architecture that can support multi-application domains but still meet the high PPA efficiency required by those applications.
\subsubsection{Design Considerations for SHOC}
The industry is moving towards more heterogeneous computing on a single die or multi dies to meet the challenges in post Moore’s Law era. Although different technologies can be packaged into a chip, there still exist large data movement, which not only increases latency, but also consumes too much energy not used for computing.\\

SHOC is hybrid and organizational, by which we mean there is no clear cut boundary between different technologies and/or computing structures, some part of the structure could change its functional roles and become a sub structure of another part in the structure at run time.  With this method, we could let the computing system minimize data move and keep the processing locally as much as possible, and make the entire computing system more harmonious among each sub-system to achieve high utilization of the actual computing operators.  After all, it is those operations that are useful for solving practical problems.
\begin{itemize}
	\item \textbf{Configurable coupling (tightly/loosely) co-processors }: By tightly coupled co-processor, we mean that the host processor is dedicated to the co-processor for control and/or data LD/ST; loosely coupled means that the co-processor and the host can execute independently of each other after the co-processor is configured.  In loosely coupled mode, the co-processors are acting like regular hardware accelerators.
	\item \textbf{Complementary of all aspects }: Real world problems are not so neat and tidy to fit in a pre-designed computing systems.  People must partition and formulate the problems into computable equations that may have all large mix of the algorithms, operator types, precision, sizes, etc.  On the other hand, the compute architecture should also provide sufficient flexibility and capability to efficiently solve the problems.  Therefore, we need to consider different complementary pairs in computing systems: time-space, big-little, complex-simple, tightly-loosely, local-global, and so on.  The complementary structure should also adhere to the 80-20 rule, which indicates the large part of a the computing problem can be solved with simpler and regular computing structures.  However, one shall not let the 20\% complex part stall the 80\% .  The SHOC architecture fully takes these considerations into account as part of its design.
	\item \textbf{X-architecture co-designs }: It is well known in the industry that 80\% energy optimization is from algorithms and architecture co-design. In their 2018 Turing award lecture, Hennessy and Patterson also mentioned the benefits of HW/SW co-design for high level languages, and compiler-architecture co-design for C-compiler and RISC architectures.  They also predicted that new compilers for DSA could raise 10X computing efficiency \cite{HP2019}. In addition to this, SHOC also takes an approach of scheduling-architecture co-design, the figure below illustrates an example of how the latency could be reduced by the flexibility of SHOC. 
\end{itemize} 
\begin{figure}
	\includegraphics[width=1.0\linewidth,height=0.3\linewidth]{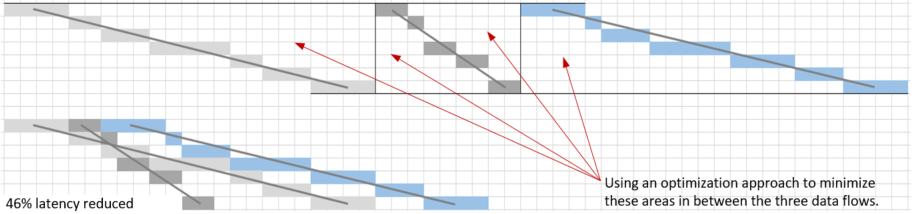}
	\centering
	\caption{HW assisted scheduling}
	\label{HW_assisted} 
\end{figure}
\subsubsection{\label{SHOC_arch} The SHOC architecture}
\begin{figure}
	\includegraphics[width=0.8\linewidth,height=0.5\linewidth]{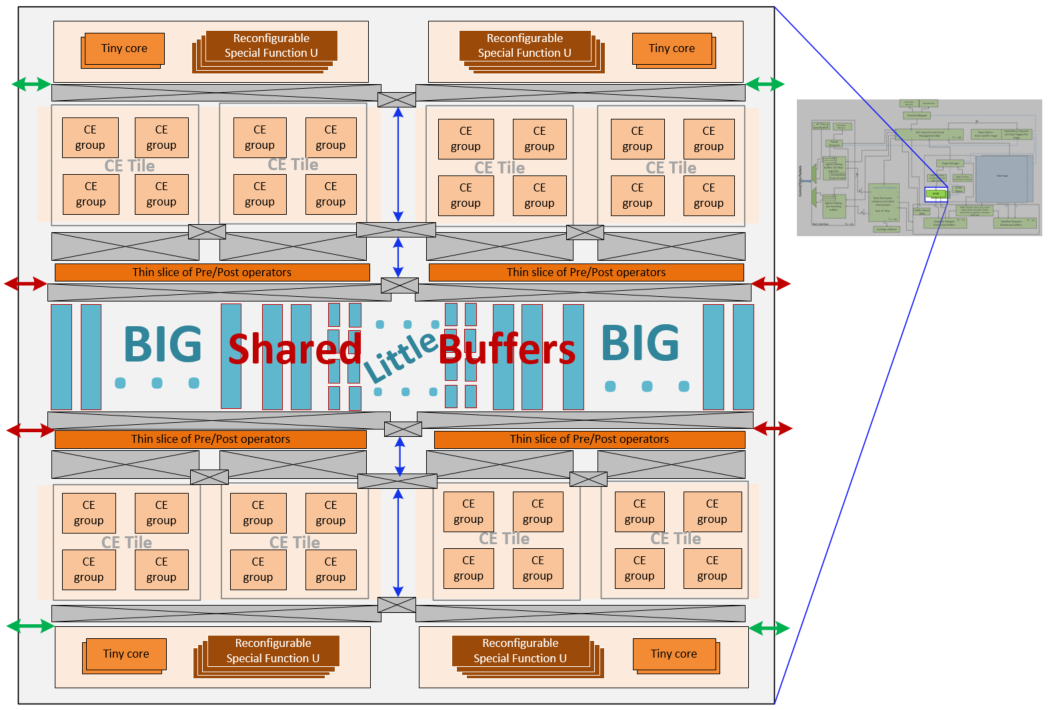}
	\centering
	\caption{SHOC top diagram}
	\label{SHOC_top} 
\end{figure}

It is physically difficult and area/power cost prohibiting to try to build a fully connected network for all components even in a medium sized compute unit, say one with 16 CMAC (Complex MAC) and 32 memory banks.  Therefore, a locally dense globally sparse interconnect scheme is used for SHOC.  It takes a hierarchical structure and uses a full cross bar interconnect for local small Compute Element (CE) group and uses a less dense cross bar for high level interconnect.  The gray cross boxes in Figure \ref{SHOC_top} show the interconnect top level.  \\

Each CE group contains 4 CMACs of FP16 and configurable for a 2x2 array or a 4x1 vector.  Four CE groups comprise a CE tile, and each tile is connected to the scratch buffers via a two stage cross bar.  In between, a sharable thin slice of simple operators that can, on the fly, perform operations like data scaling, shifting or add/sub etc. for the  data that passed through the two stages of interconnect, if being configured to do so.  This way, the expensive CMAC operators can avoid performing simple real number operations, while the computing stages are reduced and the latency and storage could be significantly decreased. \\
The tiny cores and some of the special functional units and CE groups can be tightly or loosely coupled depending on the algorithms and performance needs to improve computing resource utilization.  All the interconnects support neighboring sharing for computing and memory resources.  Most of the CE groups are semi statically configured and operated like ASIC, thus SHOC can achieve close to ASIC performance.  (A design comparison indicated SHOC is even better than ASIC for some applications. \\

In Figure \ref{SHOC_top}, there are 128 CMACs, and the SHOC can be scaled up to larger system horizontally and/or vertically.  The green arrow buses support control/configuration loading in addition to data from/to outside, and the red arrow buses support data transfer only from/to outside world.  The tiny cores take care of the control/configuration in addition to computing.
\subsubsection{\label{performance_analysis} Some performance analysis and comparisons}
Because the majority of computing elements in SHOC execute like ASIC after configurations, which is a significant difference from the conventional programmable CGRA \cite{CGRA_survey_2020},  we expect SHOC to achieve close to ASIC PPA performance.  It is all about improving the utilization rate of computing elements that are actually needed to solve algorithms in applications.  SHOC takes this as a key design criteria.

As SHOC is targeted for real time application domains, we mainly tackle single block data (batch 1) or small batch sized data blocks (batch 4).  The reader should note that, it is not fair to compare parallelism in small batch processing to that in large batch sized processing, often used in offline computing, as it is much more difficult to apply parallel processing on a single data block (batch 1) where the parallelism must come from within the block in order to reduce processing time for real time applications, rather than from an average over many blocks processing in parallel.  However, processing fast for a single data block  is important, and it can also significantly save areas and power, for instance in NN inferencing. In general, large batch sizes are easier to implement and can achieve higher utilization rate but cost more for data storage and data movement.  Even TPUs can only achieve less than 50\% utilization rate for training of NN with large batch sizes.  GPUs do not fair any better \cite{TPUGPUCPU_benchmark_2019}.  Here are some examples of SHOC performance.
\begin{itemize}
	\item \textbf{FFT/IFFT functions}: In SHOC, there does not exist a dedicated module for FFT Special Functional Unit (SFU).  Instead, we separate the butterfly structures and the twiddle MUL vectors such that only the butterflies are designed as configurable SFU (so the butterflies can be used for other functions if FFT is not needed) and the MUL vectors could be used as regular CMAC operations.  This way, we saved significant computing areas and memory buffers for dedicated FFT hardware accelerators.  Furthermore, the scalability of SHOC can combine smaller FFT components to support larger radix FFTs to achieve much better PPA performance. 	
	Based on a paper from Microsoft \cite{MS_FFT_2008}, the performance for SHOC is more than two orders of magnitude that of the GTX280 for a single 4096 point FFT, since SHOC as a DSA has much less area and power, the energy and area efficiency for SHOC would be much higher.  Also, in the cases of smaller than 4096 FFT sizes,  GTX280 would even be worse.  (see the left curve diagram in Figure \ref{GTX280_FFT}) From Figure \ref{GTX280_FFT}, GTX280 utilization rate is less than 0.72\% for FFT4096 and smaller (close to 0 for FFT16), and those size FFTs are the one required in 5G infrastructure applications.  In the best case of large batch and large sized FFTs, the utilization rate is still less than 11\%, while SHOC could reach above 90\% utilization rate.  This result is mainly achieved by algorithm-architecture co-designs.  In SHOC, its FFT configurations runs like an ASIC, and use large radices as much as possible to improve parallelism and performance.  If radix 3 and radix 5 need to be supported, we would expect GPU performance to be even worse.  (For an early version of SHOC compared to a CGRA FFT configuration, SHOC achieved 100X energy efficiency for a 256 point FFT.) 
	\begin{figure}
		\includegraphics[width=1.0\linewidth,height=0.5\linewidth]{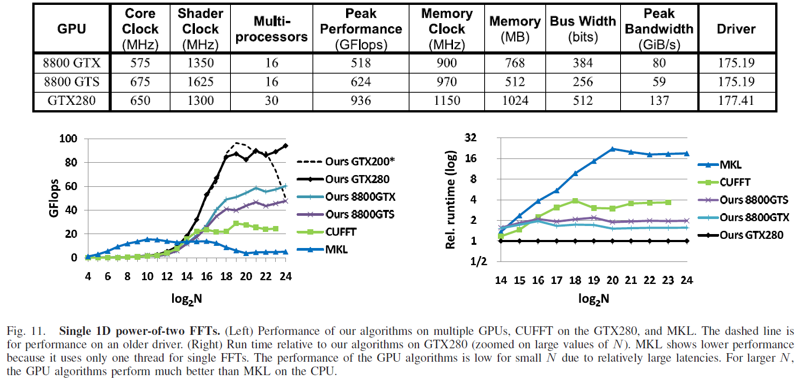}
		\centering
		\caption{GTX280 FFT batch 1 performance (Source: \cite{TPUGPUCPU_benchmark_2019})}
		\label{GTX280_FFT} 
	\end{figure}	
	\item \textbf{Cholesky decomposition}: Cholesky decomposition is a widely used algorithm in matrix linear algebra, and heavily used in wireless baseband processing.  It is highly spatially structured and highly data dependent.   And its irregular matrix computation usually causes low utilization rate for many parallel processing systems on a single matrix (batch 1), which is preferred for our targeted application domain (here again the idea of DSA comes into play).  Again, we applied algorithm-architecture co-design to take the advantages of SHOC’s spatial and temporal flexibility.  We experimented on SHOC architecture for different matrix sizes ([8x8], [16x16] and [32x32]), and different parallel levels (CMAC vector [4x1], [8x1] and [16x1]).  With the same number of parallel levels, SHOC achieved $5X \sim 8X$ performance efficiency on internal data comparisons and the utilization rate ranged from 38\% to 98\% for this data set.  Based on these numbers from different combinations of cycle-CMAC number pairs, the scheduler would choose the right parallel level based on available CMAC resources and load balancing considerations, thanks to the capability of SHOC’s variable length SIMD support.

	\item \textbf{Scheduling-architecture co-design}: A problem is usually partitioned into multi data flows (or micro flows), and then multi data blocks are fed in for processing.  It is hoped that those data blocks could be processed in a way that they would start at the same time and finish at the same time, such that computing resources assigned to these data flows can be all ready for the next set of data flows.  Otherwise, the processing latency would be dependent on the largest data flow processing time and those resources of early completed data flows would be wasted.   Unfortunately, this is real world applications, and nothing is ever perfect.  To deal with this issue, SHOC provides a switching mechanism that can facilitate the scheduler to find the best job sequencing to reduce the total latency.  
	
	Figure \ref{3D_tess} illustrates this scheme, which is an example of temporal tessellation.  The four colored data flows started at the same time from the bottom.  They completed at different times (along the vertical axis).  If all the four flows do not change their CE group assignments, as the new data blocks continue to be fed in, the green flow would complete first, and would be sitting there until the red flow completed last, and the idle time would be accumulated if many data blocks need to be processed.  By switching their resource assignment, the idle time would not accumulate, the total latency will be significantly reduced, and utilization rate improved.  The cost is more reconfiguration times, but it is acceptable because this is not cycle based dynamic reconfiguration, and it is only switching, no new different control bits need be loaded.
	
	In a real modem algorithm of 3 matrices MUL case, this scheme can achieve 100\% utilization rate and ideal latency, while a system without this method resulted in 75\% utilization rate and longer processing time,  given the condition that both cases have the same amount of resources.
	\begin{figure}[!h]
		\includegraphics[width=0.2\linewidth,height=0.5\linewidth]{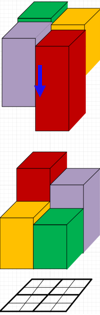}
		\centering
		\caption{3D tessellations for a CE tile }
		\label{3D_tess} 
	\end{figure}
	\item \textbf{Image processing examples}: The SHOC design considerations and architecture were also tested for the ISP (Image Signal Processing) domain, where the area and energy are very sensitive.  We tested two algorithms and both comparisons showed SHOC is even better than ASIC designs on both area and energy thanks to the algorithm-architecture codesign and flexibility for area reuse (~50\% and 30\% smaller areas respectively).
\end{itemize} 
\subsubsection{\label{challenges} Challenges of next stages}
We still have much work ahead.  We need continue to work on the compiler-architecture co-design such that an automatic compiler tool can ease the use of SHOC.  We will leverage existing high level languages and tools to speed up the development.  Also, more features like variable precisions will be considered to support AI/ML domain, which is expected to have a high demand in 6G modems.
\subsection{\label{subsec:SRE-modeling-results} SRE Architectural Modeling and Explorations}
\subsubsection{\label{subsubsec:SRE-modeling-overview} Overview of the Mirabilis SRE Model}
\begin{figure}[!h]
	\includegraphics[width=1.0\linewidth]{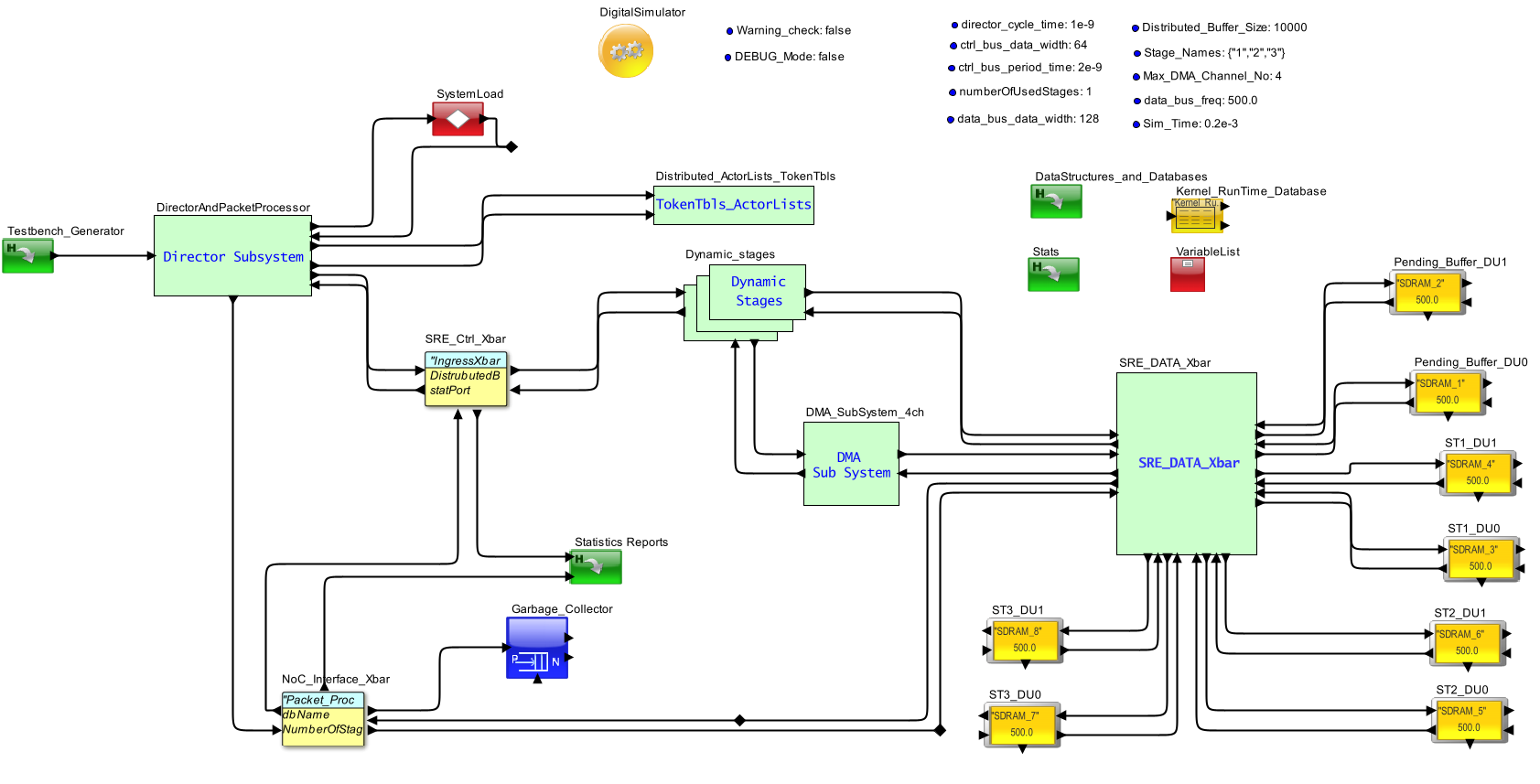}
	\centering
	\caption{Top Level Mirabilis Model for An SRE}
	\label{Mirabilis-SRE-top} 
\end{figure}

Mirabilis VirsualSim Architect \cite{Mirabilis/VirsualSim} is deployed to do the SRE architectural modeling and explorations. Fig. \ref{Mirabilis-SRE-top} is the top level view of the model and it has one-to-one mapping to the key components shown in Fig. \ref{fig:SRE-top-level-view}. 
The very leftmost submodule in Fig. \ref{Mirabilis-SRE-top} is the stimulus generator that feeds in necessary control and data packets to the SRE. The rest is the full comprehension of an SRE. The model is realized to follow the operation flow described in Fig. \ref{fig:SRE-stages-flowchart}. The details of the overall SRE model are described in Appendix \ref{appendix-Mirabilis-Model}. \\
\subsubsection{\label{subsubsec:SRE-modeling-overview} Early Results of the Architectural Modeling and Explorations}
\begin{figure}[!h]
	\includegraphics[width=0.8\linewidth,height=5cm]{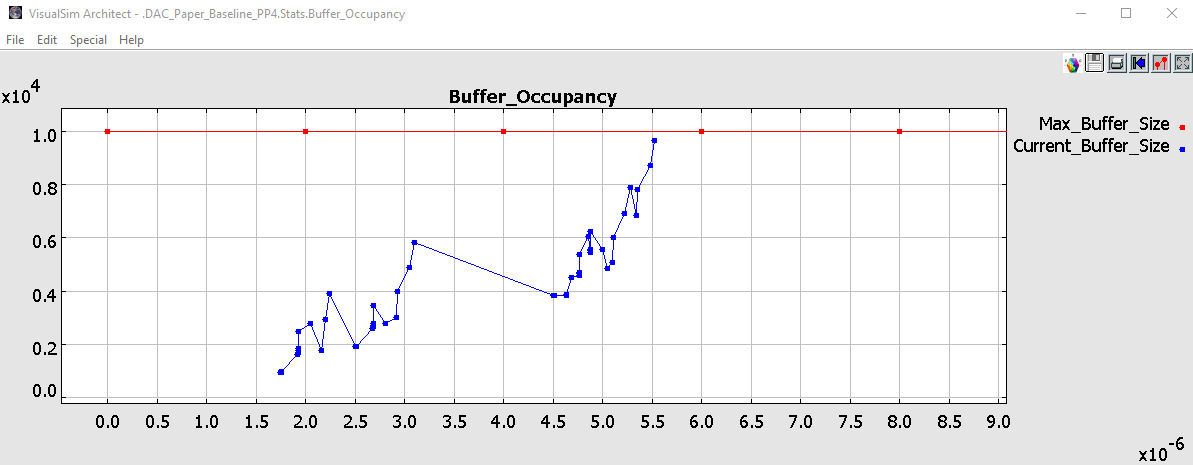}
	\centering
	\caption{Memory Leakage Issue Observed with Four DFFs Running Together}
	\label{Mirabilis-Simulation-Memory-Leakage-4} 
\end{figure}

The SRE Mirabilis model is ready to produce some earlier results based on the Channel Estimation Dataflow Fragment described in Fig. \ref{fig:chest_dff}. This DFF is used to bring-up the whole integration of the SRE model and is able to identify memory leakage issue exists earlier in the architectural exploration as shown in Fig. \ref{Mirabilis-Simulation-Memory-Leakage-4}.  

The x-axis represents the execution time of four DFFs, they are input to the SRE one by one with some overlaps among them. The y-axis logs the buffer usages along the dataflow execution against the time. The correct memory usage is that the buffer is being allocated after a microflow/kernel is triggered and should be released once a kernel is completed. And the memory usage should be back to zero once all DFFs are completed. However, one can observe that the memory usage keeps going up even though all of the four DFFs are completed in Fig. \ref{Mirabilis-Simulation-Memory-Leakage-4}.
\begin{figure}[!h]
	\includegraphics[width=0.8\linewidth,height=5cm]{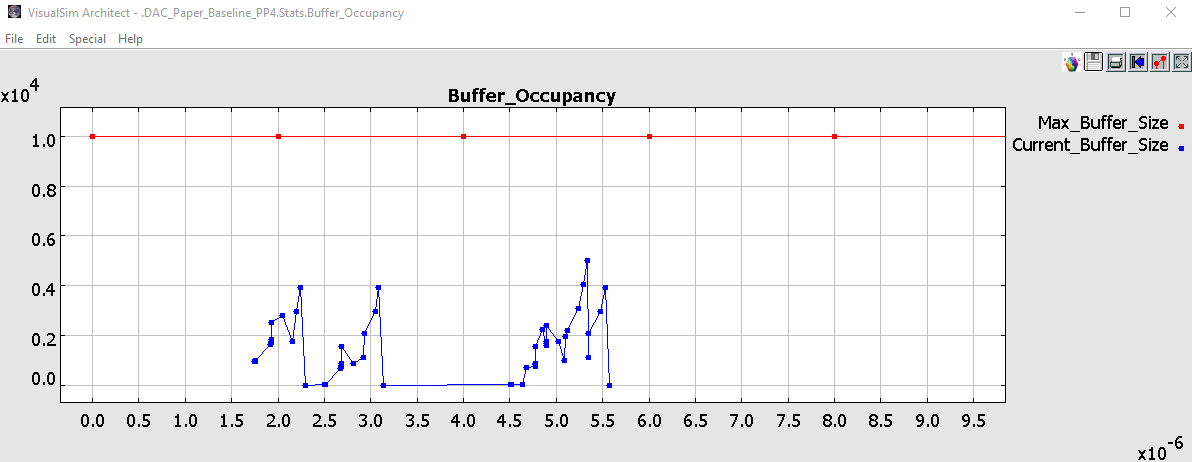}
	\centering
	\caption{Memory Leakage Issue Corrected with Four DFFs Running Together}
	\label{Mirabilis-Simulation-Memory-Leakage-4-Corrected} 
\end{figure}
The memory leakage issue is resolved after the relevant stage and memory subsystem is re-architected. As observed from Fig. \ref{Mirabilis-Simulation-Memory-Leakage-4-Corrected}, the memory usage returns back to zero once all DFF flow are completed so we can conclude the memory leakage issue is cleanly resolved.\\

\begin{figure}[!h]
	\includegraphics[width=0.8\linewidth,height=5cm]{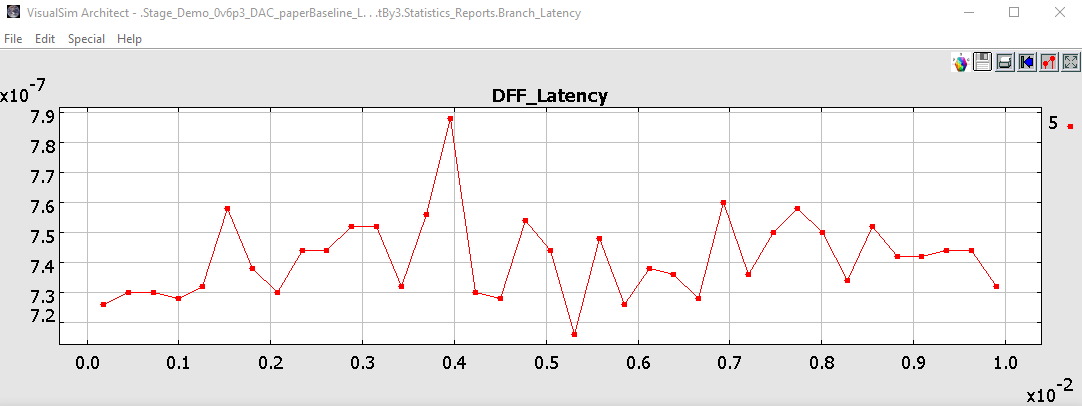}
	\centering
	\caption{Channel Estimation DFF Execution Time Observations with the SRE Mirabilis Model}
	\label{Mirabilis-Simulation-Latency} 
\end{figure}

Other useful information such as DFF execution time/latency as shown in Fig. \ref{Mirabilis-Simulation-Latency} can be observed to help fine tune the overall SRE architecture. This is the simulation result where the same DFF has been executed over twenty times. The x-axis indicates the run time from start to finish, the y-axis logs the individual DFF execution time for every iteration. We can observe that the DFF execution time is quite consistent with a small range of variation, which hints we have a solid start point for further architectural explorations.\\

\section{\label{sec:FV}Automated Formal Checking of 5G and next generation Base-station Modem SoC Functionality}
 In wireless infrastructure, modem requirements are regularly updated in software, and simulation is not sufficient to ensure fault tolerance for the wide variety of scenarios. As a result, leakage of Heisenbugs through integration testing and into field deployment is a real concern. As mentioned in earlier chapters, we outline an architecture that is designed to allow practical formal checking for Heisenbugs and performance guarantees, and demonstrate how formal checking of blocks can be automated at the dataflow specification level. 
 
 Figure \ref{formal_toolchain} shows an highlevel overview of our Formal Toolchain methodology. In this chapter we describe that methodology in detail starting from how to automate the process from algorithm and requirements specification thru mapping to the SoC with formal checking of performance and correctness. 
 \begin{itemize}
 \item We start by describing the algorithms as Directed Acyclic Graphs (DAGs) using a Parameterized and Interfaced Meta-Model (PiMM) Model of Computation (MoC) \cite{PIMM} using the PRESSM \cite{PREESM} tool. This tool allows us to manipulate the complicated dataflows into a series of loosely connected, generic DAGs, which we call Data Flow Fragments (DFFs), to run on the SoC, that are mapped at runtime to achieve the desired flow. 
   
 \item We defined a template for the DFF representation of any application (including Wireless Baseband Applications) in PREESM,and developed Python scripts to translate those templates into UPPAAL's Timed Automata (UTA) meta-model format. We call these generated UTA meta-models as untimed Behavioral Automata (BA) meta-model since they are not timed. Generated BA meta-model can then be automatically manipulated within the UPPAAL tool \cite{Behrmann06atutorial} to provide guarantees of performance, correctness and other desired properties. Section \ref{behavior_TA} describes the use of PREESM to generate BA models of DFFs that can be integrated with fixed hardware models in UPPAAL. Section \ref{FV:5GChestBA} describe the generation of an untimed BA for one of the 5G Dataflow Graphs, called Channel Estimation DFF.
  
 \item We developed an UTA model of SoC architecture model based on an array of \lq\lq DFF acceleration" IP, called a Service Resource Element (SRE \ref{sec:SRE-Arch}, each receiving a stream of DFFs for processing. Section \ref{SRE_models} present the target architecture and the Hardware Models that represent it in UTA metamodel.  
 Inspired by \cite{Lampka2010} we reduce the complexity of the formal check by checking the operation of an SRE in isolation with models for DFF arrival patterns formally defined as TA. We found that the TA models must be carefully designed to prevent an explosion of the search space resulting in hours/days to finish the formal verification and we share our philosophy for efficient model design in section \ref{model_design} as well as formal verification runtime results in section \ref{Results}. 
 
 \item   Section \ref{classes_of_queries_section} outlines how to test the resulting TA for robustness and correctness and in section \ref{Results} we demonstrate the performance of the automated formal tooling.
\end{itemize}  
 
 \subsection{Why formal methods based checking \& verification approach?}
 Mapping of many DFF simultaneously onto 5G SoC must occur at runtime because, each symbol period, different DFFs are used to achieve the goals of the chosen users and rates. Static mapping strategies do not allow for such flexibility. Our tool permits runtime flexiblity while maintaining  high reliability by applying formal methods designed specifically for flexible FRT operation to verify that the recombination of DFFs for an assumed pattern of DFF arrival can achieve the desired QoS for a given SoC.  
 
 Using a single, DFF, or  \lq\lq generic DAG" with runtime, data buffer size and resource requirement ranges to represent DFFs that have a similar structure, we can keep the number of DFF to be analyzed formally to a manageable number.  
 
 Though there are numerous papers on scheduling multiple DAGs of different shape/size onto multi-core SoC \cite{5546233} \cite{Xie_DAGs} \cite{8743275}, these papers address only one or two cases, usually worst case scenarios, and do not check for correctness and robustness (from effects such as Heisenbugs, Livelock, buffer overflow and so on) at the architecture level. Correctness and robustness are usually checked via extensive simulation during integration testing but the complexity of 5G operations and the regularity of feature additions and modifications makes this an increasingly impractical methodology that often leads to Heisenbugs in the field. 
 
 Our goal is to demonstrate a path toward automating and formalizing the correctness and robustness checking as well as the mapping of the algorithms to the SoC such that a high degree of flexibility and continuous integration of new features is possible.
 
\subsection{Overview of the UPPAAL Tool}\label{uppaal_tool}
Uppaal is a toolbox for verification of real-time systems jointly developed by Uppsala University and Aalborg University. It has been applied successfully in case studies ranging from communication protocols to multimedia applications. The tool is designed to verify systems that can be modelled as "networks of timed automata" extended with integer variables, structured data types, and channel synchronisation. These systems are called Uppaal Timed Automata (UTA) metamodel in Uppaal toolbox.

A timed automaton is a finite-state machine extended with clock variables. It uses a dense-time model where a clock variable evaluates to a real number. All the clocks progress synchronously. In Uppaal, a system is modelled as a network of several such timed automata in parallel. The model is further extended with bounded discrete variables that are part of the state. These variables are used as in programming languages: they are read, written, and are subject to common arithmetic operations. A state of the system is defined by the locations of all automata, the clock constraints, and the values of the discrete variables. Every automaton may fire an edge (sometimes misleadingly called a transition) separately or synchronise with another automaton, which leads to a new state.

The model-checker Uppaal is based on the theory of timed automata and its modelling language offers additional features such as bounded integer variables and urgency. The query language of Uppaal, used to specify properties to be checked, is a subset of CTL (computation tree logic).

Purpose of this chapter is to explain the novel techniques we developed to represent dataflow models and SoC architecture (with FRT) as Networks of Timed Automata in Uppaal toolbox and formally verified to provide guarantees of performance, correctness and other desired properties.

Disclaimer: This section does not explain the concept of Time Automata and the usage of Uppaal toolbox. Refer \cite{6979532}.

\subsection{Overview of the auto-generation of behavioral TA models from LTE/5G algorithms}\label{behavior_TA}

\begin{figure}[h]
	\includegraphics[width=1.0\linewidth]{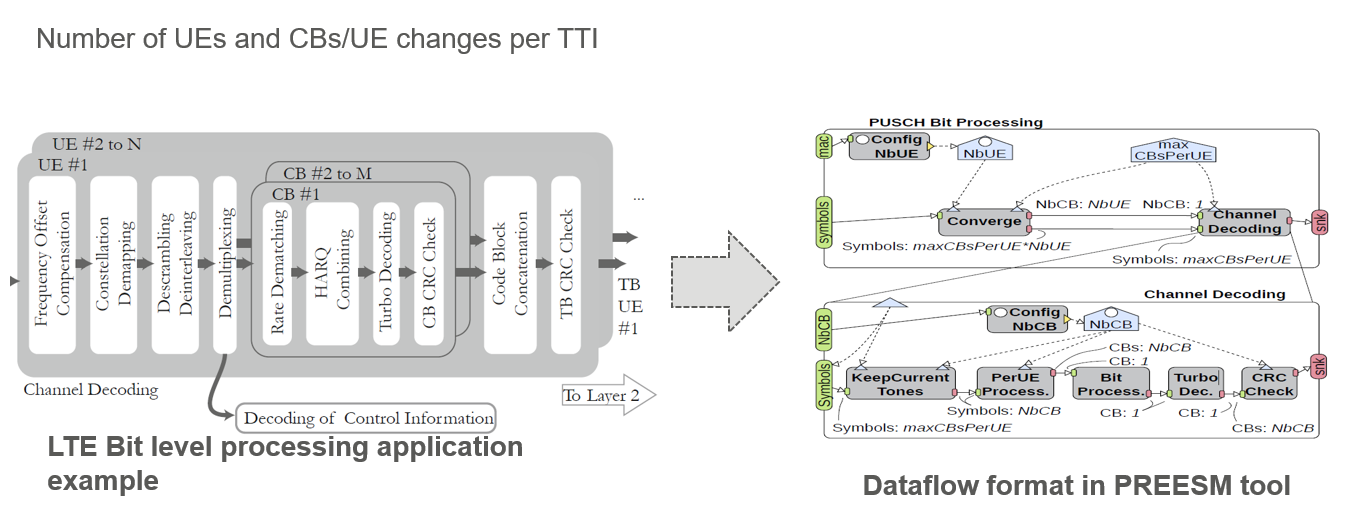}
	\centering
	\caption{LTE's Bit Level Processing representation as Dataflow models (PiMM MoC) in PREESM tool (inspired by source:\cite{PIMM})}
	\label{BLP_in_PREESM} 	
\end{figure}

Translating LTE applications into Dataflow models using the PREESM tool is covered in \cite{PIMM}. Figure \ref{BLP_in_PREESM} shows an example of representing LTE's PUSCH's Bit Level processing (BLP) applications as Dataflow models. For details on algorithms mentioned in Figure \ref{BLP_in_PREESM} refer \cite{5G_book}. 

\begin{figure}[h]
	\includegraphics[width=1.0\linewidth]{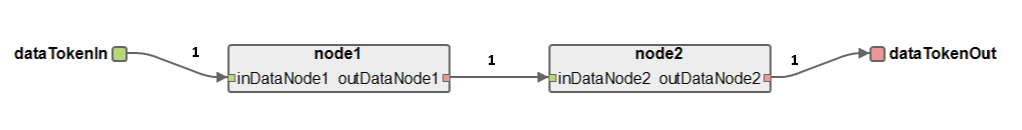}
	\centering
	\caption{A simple DFF example with two Nodes}
	\label{DAG_2NODE} 	
\end{figure}

 Similarly 5G physical layer algorithms can be translated into Dataflow models and appropriate DAGs can be generated. Therefore how to develop DFFs of LTE/5G physical layer algorithms is not covered and a simple DFF example is used to explain the TA auto-generation mechanism followed by an overview of auto-generated TA model for Channel Estimation dataflow shown in figure \ref{fig:chest_dff}.
Figure \ref{DAG_2NODE}, shows a PREESM's PiMM MoC representation of a DFF with two nodes, connected by an edge. Note that the PREESM tool was designed for analyzing PiMM MoCs \cite{PIMM} but we are using its graphical interface to draw DFF.

\begin{figure}[h]
	\includegraphics[width=1.0\linewidth]{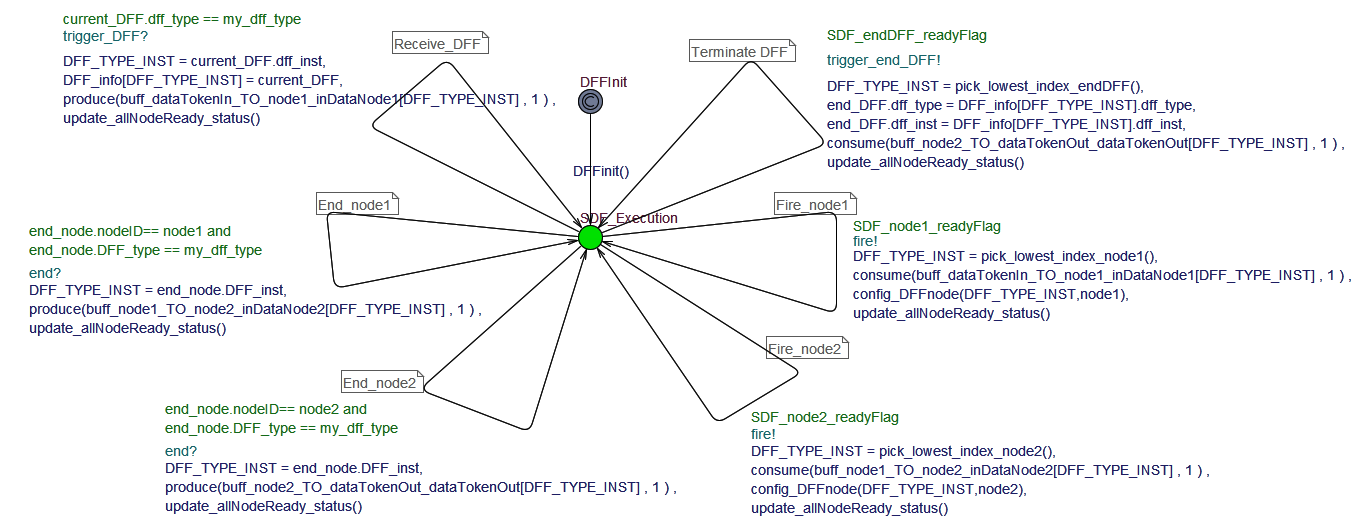}
	\centering
	\caption{Behavioral Timed Automata of DFF (shown in figure \ref{DAG_2NODE})}
	\label{DAG_2NODE_TA} 
\end{figure}

Figure \ref{DAG_2NODE_TA} shows the auto-generated (via python script) untimed Behavioral Automata (BA) model of the example 2-node DFF shown in figure \ref{DAG_2NODE}. We take the approach laid out in \cite{wandeler2006}, separating the system into hardware TA (which are referenced in this parapraph and described in section \ref{SRE_models}) and untimed BA models , connected by well defined interfaces. This allows us to generate the behavior of the modem algorithm as DFF without having to touch the timed models of the hardware it is running on. All behavioral DFF models use a one to many \textit{Receive\_DFF} edge to handshake a DFF descriptor from the Director. The DFF model uses internal token passing to create a pattern of kernel descriptors in the correct order for the DAG and finally the DFF model handshakes a return DFF descriptor using a many to one \textit{Terminate\_DFF} edge. This allows many DFF to be connected to a single Director with the correct DFF being chosen based on the guard pattern of the edge.

All other edges in the DFF behavioral model either \textit{fire} or \textit{end} the running of a kernel inside the DFF. In the example this adds four edges for two nodes. The \textit{fire} edges create a kernel descriptor and handshake it into the Scheduler. The \textit{end} edges handshake a kernel descriptor back from the Pool manager.  Any DFF developed in PREESM is automatically converted into the BA format and is included in the list of DFF in the UPPAAL model. As hook up is directly to the Director and Pool manager interfaces it is also be easily automated.

\subsubsection{\label{FV:5GChestBA} 5G Channel Estimation auto-generated untimed Behavioral Automata}
\begin{figure}[h]
	\includegraphics[width=1.0\linewidth]{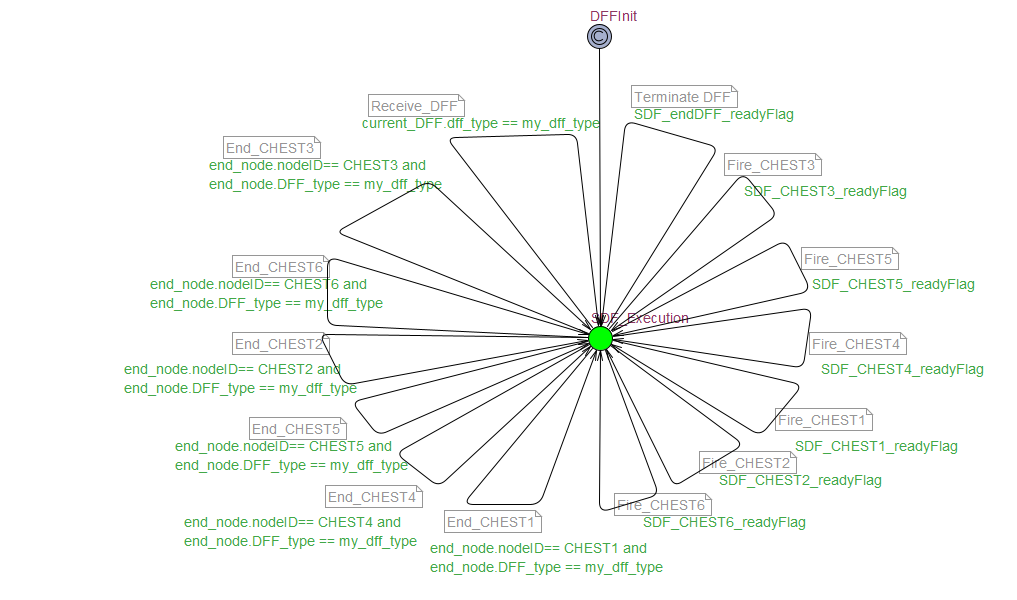}
	\centering
	\caption{Behavioral Timed Automata of Channel Estimation DAG DFF (shown in figure \ref{fig:ChEst_DAG_TA})}
	\label{fig:ChEst_DAG_TA} 
\end{figure}

Figure \ref{fig:ChEst_DAG_TA} shows the auto-generated untimed BA of the Channel Estimation DAG dataflow shown in figure \ref{fig:chest_dff} in the form of PREESM's PiMM MoC representation with six nodes (aka actors). The auto-generated Channel Estimation BA preserves the unique name of each actor from the PREESM model, for example, \textit{CHEST1\_L1metaDataExtractor} actor name in figure \ref{fig:chest_dff} is preserved as \textit{CHEST1} in generated BA in figure \ref{fig:ChEst_DAG_TA}.

\subsection{Breakdown of the SRE Architectural Model and Timing}\label{SRE_models}
The architectural model used in this paper is based on an array of SRE each receiving a stream of DFFs for processing. The top level SRE architecture is shown in Figure \ref{SRE_toplevel}. For this paper we focus on automated analysis of the ability of an SRE to support streams of DFFs from multiple sources. Each source sends a control packet to the Director, to be described, and the pattern of arrival of the DFFs is modeled by a TA that represents a jittered periodic arrival in the same way that \cite{Lampka2010} used a real time calculus source to model the input stream to the IP under analysis. We can use any arrival model that can be represented as a TA, though some will lead to state space explosions and therefore careful thought as to the correct model is needed. Note that the model of arrival can be a superset of the actual arrival pattern as the formal checker will check all possible patterns within the model. This allows for TA models to be used even if the actual arrival cannot be modeled as a TA. However a very generic arrival model will cause an increase in the state of the system so more accurate is generally better.

As DFFs are large elements of compute, consisting of several dependent kernels and running for tens of thousands of clock cycles (cc), modeling only their arrival pattern limits the complexity of the IO model for the SRE and allows the formal checker to evaluate complex and highly parallel flows on the SRE. A DFF is processed in the Director using a received control packet, setting up the control structures to run the DFF. The SRE does not trigger the DFF until all the required data packets arrive. Currently we only model the arrival of the complete, ready to run, DFF, but we will extend this to comprehend the arrival of control and data packets separately in the future. The DFF is broken down into kernels by the control structure which may be spread across multiple stages of the SRE, each of which has multiple pools of resources (for this paper we model memory and compute pools) which are assigned when the kernel is triggered to run. The stages control the triggering of kernels in the correct order and this is modeled by the DFF BA described in section \ref{behavior_TA}.

The hardware model described in this report models the Director and a single stage of the SRE. Its goal is to formally check the policy of the Director and the scheduling strategy of the Stage when presented with time jittered, quasi periodic DFFs of multiple types arriving at the SRE. It consists of the following elements
\begin{itemize}
\item The Director which receives a trigger when a DFF is ready to run, along with details on the type of DFF. The Director runs a policy to determine if the DFF should be run and also how it should be staged. The Director can maintain several different staging strategies (modelled by different DFF BA) and can also reject the DFF entirely, reporting it as a dropped DFF to the higher layers. As the policy is still under development the Director currently runs the DFF in a FCFS manner.
\item A Stage Scheduler that receives kernels when they are ready to run and time schedules their initiation in the Stage. This scheduler can prioritize more urgent DFFs for instance. Currently the Stage also runs a FCFS schedule.
\item a pool manager that receives a scheduled kernel and implements its resource use on a stage, reporting completion to the DFF behavioral model. Currently each pool is a "pure" pool with equal and homogeneous access to all resource in the pool. More complicated pool structures are easily modeled.
\end{itemize}

\subsection{A brief Review of the Formal Models}\label{Formal_Model_Review}

Without going into the details of how these formal models work we point out the main features of the models that together make up the SRE formal model. Some knowledge of the input formal for UPPAAL woyld be helpful when reading this section, though not necessary.

\subsubsection{Model design philosophy}\label{model_design}
This is a model of a processing system and as such events commence as soon as the conditions are satisfied. Therefore urgent channels are extensively used to align edges in TA. In a TA there is generally no requirement that an edge is taken once it is valid unless it is explicitly required to do so. Urgent channels cannot remain valid if time increments and must either become invalid immediately (due to an action as the result of the choice of another path) or must be taken. This models he a real time system works in practice and we therefore use urgent channels as much as possible. We also use multiple "its\_urgent" automata that are a single loop connected to a single state with an urgent channel to force edges to complete without any more time being consumed. This allows us to accurately model a program flow which proceeds without timing uncertainty once it is triggered. It also reduced the state space size of the model by removing behavior that simply doesnt reflect the real system. Another modeling technique we use extensively is the committed node. This node type must be left on the next transition, before anything else happens. It allows us to force sequences of events to happen in an atomic manner. 

Data passing between processing elements is generally passed using a handshake where an urgent channel represents a semaphore. The transmitting model places the data in a global variable once the edges are synchronized (so the semaphore is locked) and the receiving model picks up the data on the next edge using a committed node and an its\_urgent channel to ensure this happens immediately. If multiple writers are trying to lock the semaphore then the checker will try all orderings of locking.  We use queues to enforce ordering of handshakes whenever possible as this reduces the number of choices for the checker leading to significant speed up of the formal checking. Generally we use urgent channels and committed nodes as often as possible to minimize the number of path choices available to the checker while maintaining correct behavior. Non determinism in the model is created due to uncertainty in runtime of functions, which happens because parameters in the data and input buffer sizes can change processing runtime in a real system, and because of uncertainty in the arrival time of the DFF due to uncertainty in runtime in the rest of the SoC.

We track DFF through the system by assigning a unique instance number to each DFF as it arrives. This number is picked determinisitically as the lowest currently available instance number to minimize the number of paths that need to be traversed by the checker. We set a maximum number of DFF that can be active in the system so there is a  DFF\_MAX\_PAR constant that defines the maximum index number. In a DFF behavior each kernel is given a unique, hardwired number and for any system we calculate and set a constant MAX\_NUMBER\_KERNELS that is the maximum number of kernels in any DFF. We then use 2D arrays of size DFF\_MAX\_PAR by MAX\_NUMBER\_KERNELS to keep track of booleans and clocks and descriptors associated with each kernel. This use of large, sparsely used and redundant arrays is also used successfully in \cite{Jaghoori2008}\cite{Mikucionis2010} and reduces the formal checking time by providing a deterministic location for each kernel.
\subsubsection{The Director Model}
\begin{figure}
	\includegraphics[width=1.0\linewidth]{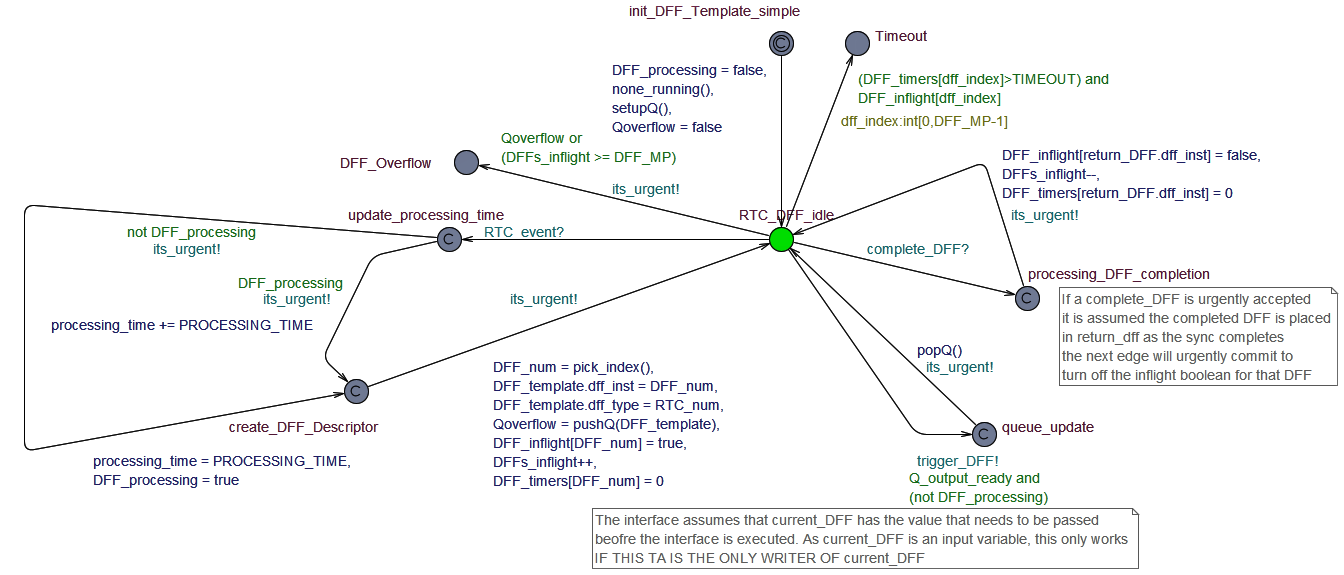}
	\centering
	\caption{Director Model}
		\label{Director_Model} 
\end{figure}

\begin{figure}
	\includegraphics[width=0.8\linewidth]{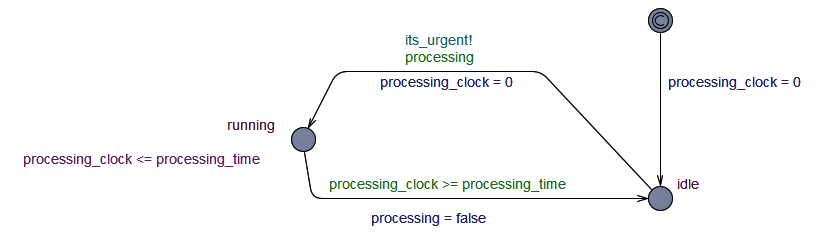}
	\centering
	\caption{Director Processor Model}
		\label{Director_Processor_Model} 
\end{figure}

The Director Model is shown in Figure \ref{Director_Model}. It consists of three main loops. The loop on the left receives a DFF event and adds time to a variable that monitors how much time the Director has left to complete the processing of all outstanding DFFs. The Director will not trigger the functional behavior of a DFF until it has completed all outstanding processing of incoming DFFs. The Director Processing Model shown in Figure \ref{Director_Processor_Model} is used to model this processing time. The Director pushes the DFF request onto a queue and this is done by a function that can choose to reorder DFFs if there is a backlog, to prioritize certain DFF over others. More generally the Director can run a policy to accept or reject DFFs if there are too many arriving to meet its policy. This allows a distributed management of the modem functionality that reduces the control overhead of the modem. The Director never back pressures the DFF source and this simplifies the formal checking process as well orthogonalizing the formal checking of one SRE from its neighbors. If the SRE receives more DFF than it can handle the DFF queue will overflow and there is a formal check for this error.

The lower loop in Figure \ref{Director_Model} pushes the chosen DFF into the appropriate DFF BA that then breaks it down into individual kernels in the correct order and these kernels are passsed on to the scheduler model. The right loop in the Director model is used to receive completed DFF events from the Pool Manager and removes them from the model. Note that each DFF is assigned a unique instance number by the director and that a timer is started for this instance number to track the lifetime of this DFF. This timer is switched off when the completed loop cancels the DFF. As pointed out in \cite{Mikucionis2010} this translates the schedulability analysis problem into a reachability problem for TA. The Director model has a terminal state that can be reached if any of the DFF violates its' timeout. This terminal state is tested for reachability to test the correct real time functionality of the system. The Director only allows a certain number of DFF in flight at the same time and the model also has a terminal state that is reached if a DFF arrives when the maximum number of DFF are still in flight.
\subsubsection{The Scheduler Model}
\begin{figure}
	\includegraphics[width=1.0\linewidth]{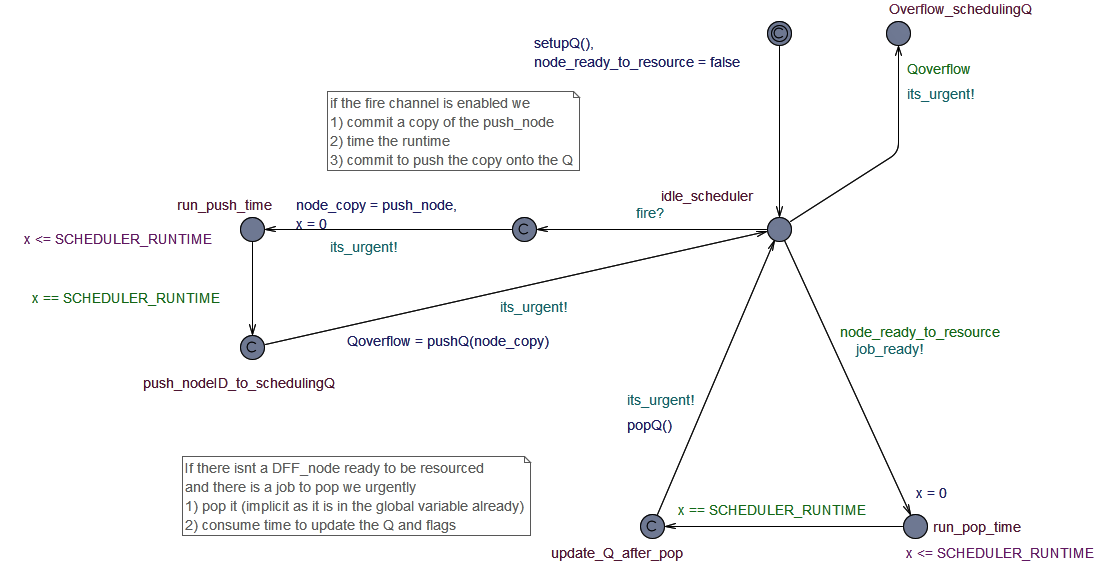}
	\centering
	\caption{Scheduler Model}
		\label{Scheduler_Model} 
\end{figure}

The Scheduler Model is shown in Figure \ref{Scheduler_Model}. The scheduler model simply receives kernels descriptors and pushes them onto a queue. It also pops kernels off the queue and pushes them at the pool manager as the next kernel to allocate. The function that pops the kernels can be used to implement a scheduling policy based on deadline information in the kernel descriptor. The scheduler model has a terminal state that is reached if the queue overflows.
\subsubsection{The Pool Manager Model}
\begin{figure}[h]
	\includegraphics[width=1.0\linewidth]{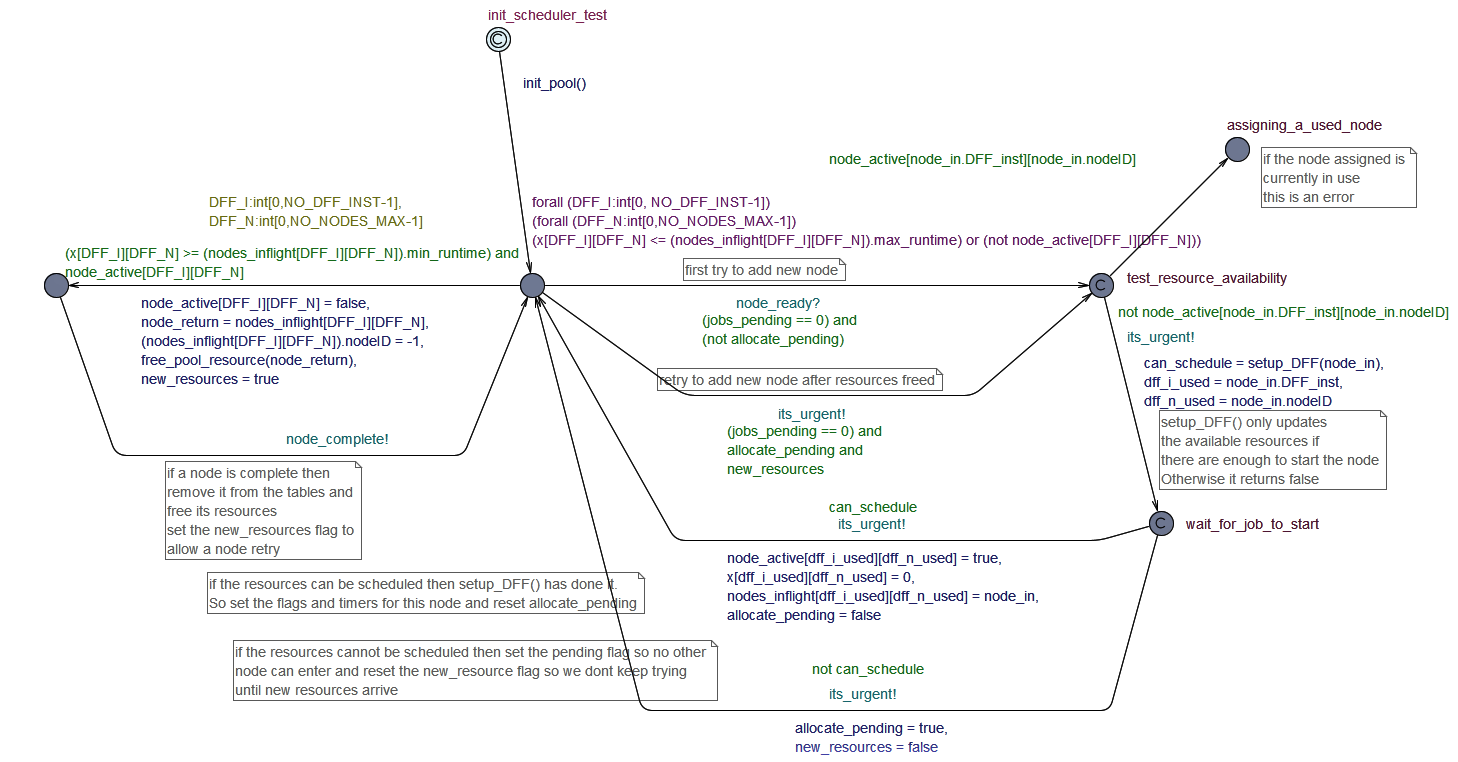}
	\centering
	\caption{Pool Manager Model}
		\label{Pool_Model} 
\end{figure}

The Pool Manager Model is shown in Figure \ref{Pool_Model}. It has two main loops to allocate and free resources. The allocation loop, the three nested loops on the right, reacts to a kernel arrival event and tries to allocate the resources in a pool. If it fails to find the resources it needs it will wait for another kernel to complete to free up some resources and try again. Until it has allocated this kernel no other kernel arrival event can occur. Once a kernel is allocated, and the remaining resource count adjusted, the kernel is assigned a unique timer with a minimum and maximum runtime that is taken from the kernel descriptor. The free resources loop on the left is triggered by a complex invariant on the central node in the graph which kicks the model out of the central node if at least one kernel timer reaches its maximum runtime. The loop also becomes enabled once any kernel has reached its minimum runtime. This pattern means that any active kernel must leave the central node and complete sometime between its minimum and maximum runtimes. If the loop is traversed the kernel in question is deleted from the model and a completion event is activated that allows the DFF BA to progress to the next kernel.
\subsection{The Complete SRE Formal Model}
\begin{figure}[h]
	\includegraphics[width=1.0\linewidth]{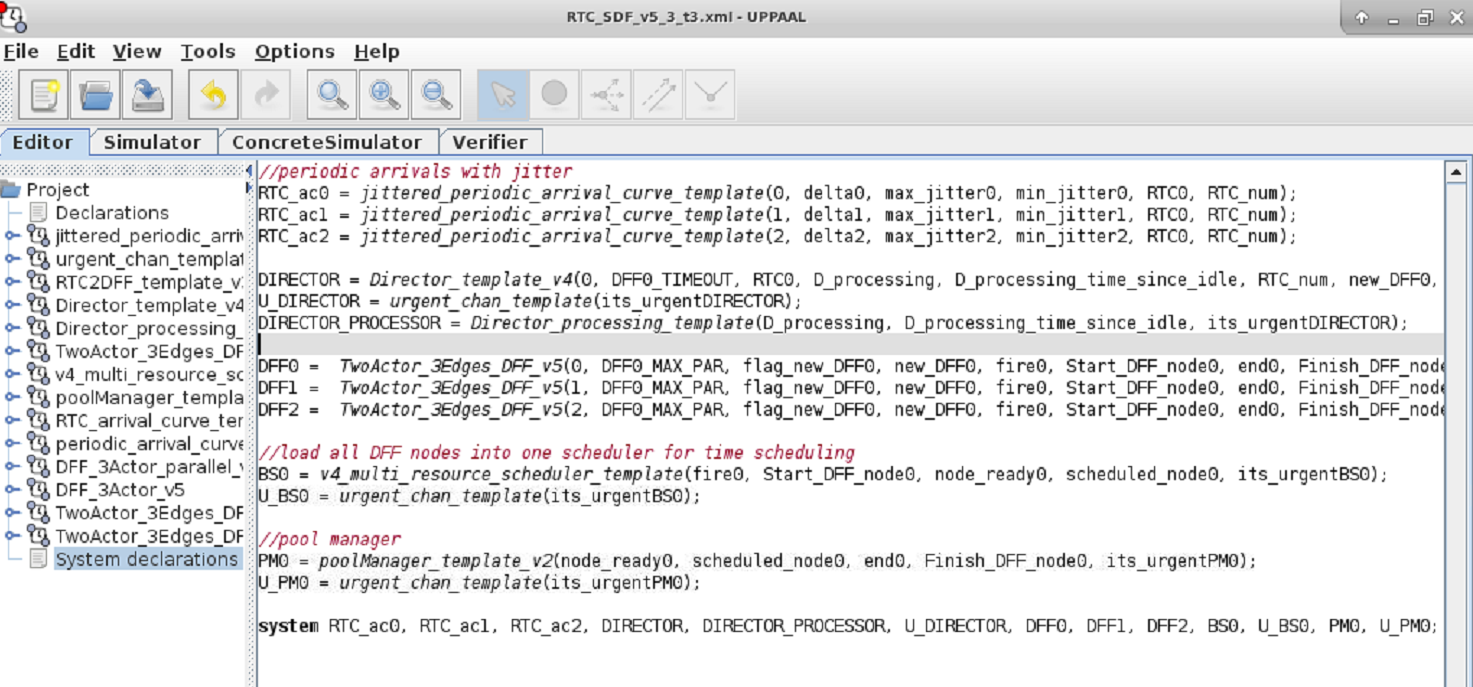}
	\centering
	\caption{SRE Model Hook Up in UPPAAL}
	\label{SRE_Model} 
\end{figure}

The formal model used in this report is made up of a single director (there can only be one director in the current model) and a single stage (there could be multiple stages hooked up) that consists of a scheduler and pool manager. Any number of DFF models can be added to the Director. Finally, for each DFF model there is a arrival pattern generator. When the formal model runs the checker tries all combinations of arrival patterns possible from all the DFF for all possible runtime variations of all compute elements. An example of this hook up from the main UPPAAL window is shown in Figure \ref{SRE_Model}. In this example there are three sources of DFF, a Director (with its urgent channel TA and its processor TA), three DFF behavioral models, one for each source (each the same model in this case but they can be different), the stage scheduler (with its urgent channel TA) and the pool manager (with its urgent channel TA). At the bottom of the screenshot you see the system deployment of these individual TA into a single connected model.
When we run the model we use queries to check that the model behaves correctly under all conditions.  
\subsection{Classes of Queries for Formal Checking}\label{classes_of_queries_section}
Any formal checker is only as good as the queries developed to extract information from the design. This model is focused on finding Heisenbugs that are difficult to identify using simulation and are likely to survive through unit test and even integration test into the product. For our system level checking there are three classes of queries we need to consider:
\begin{itemize}
\item Functional correctness. These queries ensure that the hardware accesses resources in a certain order at all times or that correct timing is maintained for operations. Such queries are developed when checking basic hardware modules, the most obvious being a check for Deadlock or Livelock in the system. An example of this is in the Pool manager where we formally check that a kernel that is currently active is not activated again.
\item Runtime Errors. These queries check for buffer overflow, parameter initialization errors and system overload. They are a side effect of "operator error" in setting up and running the system. Usually they are hard to find with static checks. Runtime should not cause system error but the generation of error message feedback. Runtime errors in our system occur when we overflow queues. We cannot back pressure queues as it breaks our system analysis model.
\item Performance checks. These queries check that performance timing is met for the real time system so that deadlines are not missed in the SoC.
\end{itemize}
For the examples in this report we perform the following checks. The true system list is considerably larger:
\begin{itemize}
\item Run time check for each DFF (Performance check)
\item Deadlock (Functional correctness)
\item Director queue overflow (Runtime Error)
\item Scheduler queue overflow (Runtime Error)
\item Pool Manager Node Assignment (Functional Correctness)
\item Livelock of node assignment (Functional Corectness)
\item DFF activity level (Performance check)
\end{itemize}

\subsection{Results}\label{Results}
\begin{figure}
	\includegraphics[width=1.0\linewidth]{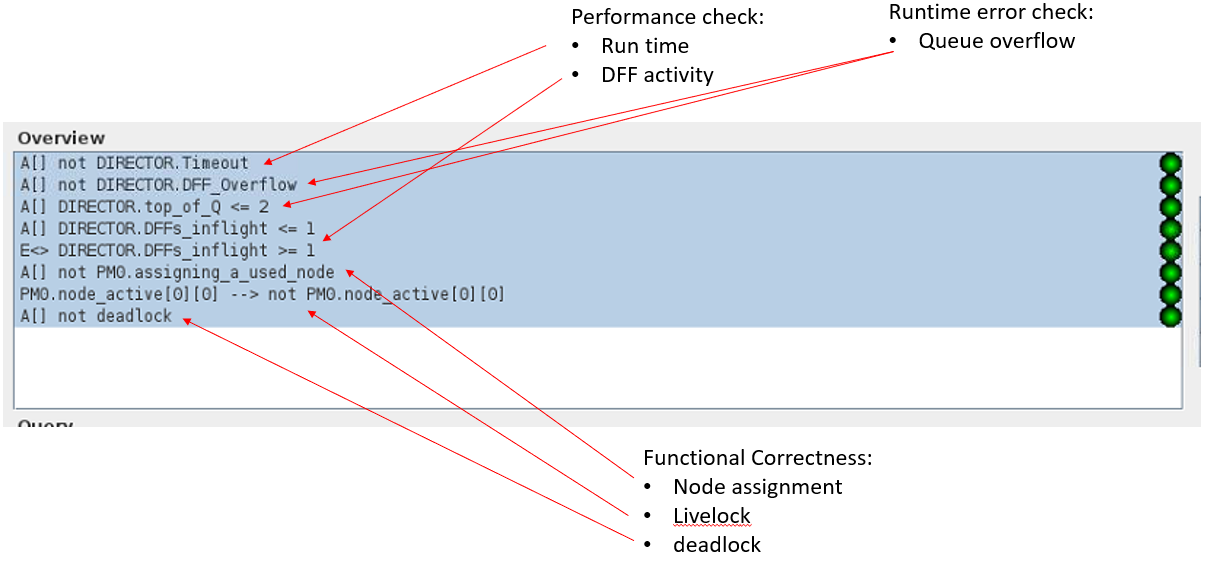}
	\centering
	\caption{UPPAAL query list}
		\label{UPPAAL_query_list} 
\end{figure}

For the model checking we used uppaal64-4.1.24 running on an Intel(R) Xeon(R) CPU E5-2695 v3 @ 2.30GHz, supported by Ubuntu 18.04.4 LTS and 94G of usable memory. As UPPAAL cannot fork multiple queries in parallel, the timing we quote is for the complete query list running serially. Based on the query classification in Section \ref{classes_of_queries_section} we developed a basic query list shown in Figure \ref{UPPAAL_query_list}.
The Performance checker tested to make sure each DFF achieved its runtime and monitored the number of DFF that could be in flight at any moment. The Runtime error checking monitored the maximum queue size for the director to make sure the DFF did not back up into the system. The Functional Correctness tested that a node was not assigned while still in process, a kernel once activated will eventually complete, and that there is no deadlock across the whole model. Obviously many more checks would be included for a full analysis of the model, but these form a representative set of the kind of checks we would like to perform. 

Refering to Figure \ref{SRE_toplevel}, the hardware is modeled so that the Director takes 100cc to process a DFF into the stage manager. This processing includes running the behavioral model of the DFF and setting up the kernels to be run in the stage manager. The stage manager consists of a scheduler to order the kernels and a pool manager to allocate the resources in the pool as described in Section \ref{Formal_Model_Review}. The scheduler takes 123cc to schedule each kernel into the pool manager. The input DFFs arrive as several sources from other parts of the SoC as a control packet followed by some data packets. We currently model the arrival of the control packet at the director and assume that the data can arrive before the DFF is triggered in the pool manager. In future work we will add a packet processor model. The arrival of the control packet at the Director is modeled by a quasi periodic source that has a specific average period about which it is jittered up to a maximum number of clock cycles. We model a single pool manager in this paper.
\begin{itemize}
\item \textbf{Simple Tests:} We ran some initial burn tests to make sure the model functioned correctly and that no state space explosion was slowing down the checker. We set the time out for all DFFs to be a healthy 100000 cc. We set up three DFF sources with period of 5000cc and no jitter and spaced the arrival of the DFF by 1000cc so that they did not overlap. The DFF flow was the same for each source and consisted of two kernels in a chain, each taking exactly 200cc per kernel. The checking was near instantaneous due to the lack of resource contention and any runtime variation in timing. Adding a runtime range to the kernels of [200,300] did not change the checking time as it does not add to the number of states in the TA of this very mildly loaded system.
\item \textbf{Overlapping DFF of different types:} Changing this model so that the sources were offset by 200cc allows the sources to bunch up and overlap periodically. This produces a more complex resource usage pattern increasing the total checking time to about one minute. The maximum number of DFF in flight changed from 1 (as shown in lines 4 and 5 of the query list in Figure \ref{UPPAAL_query_list}) to 3, as would be expected as each of the three DFF run their kernels one at a time. Different types of DFF were then tested with 2 or 3 kernels in different configurations and we found that with runtime variation of 50\% and three kernels, the checking time might increase to closer to 1 hour. The worst case backlog of the Director queue remained small for all of these DFF combinations showing that the checking time was due to the combinatorial number of ways that a larger number of overlapping kernels with significant runtime jitter can combine. 
\item \textbf{DFF Arrival Jitter} Adding jitter of 10\% of the DFF start time separation (so 20cc of jitter to a 200cc arrival gap between DFF types) to the overall arrival time of the DFF didn't not change the checking time,  adding jitter of 25\% of the separation took the checking time to about 30 minutes, and increasing this jitter to 50\% did not allow even a single query to complete in 24 hours. Clearly the uncertainty in arrival time of the DFF is adding a lot of new potential orderings of operations within the hardware and this is having a dramatic impact on the search space. Note that most time is spent in the Livelock and deadlock testing.
\item \textbf{Running close to runtime requirements} We simplified the model back to 3 sources of 2 kernel DFFs at 200cc spacing with 10\% jitter in arrival time and a runtime range for the kernels of [200, 300]. This example checks in about 3 minutes with most of the checking time in deadlock and livelock. Counting the time consumed in the path of a single DFF, the time to complete a DFF is best case in the range [623, 823]. Focusing on the timeout check only we see that the system actually fails at 1750 and the trace of this timeout involves 3 DFFs and 6 kernels which, due to runtime range, complete in a different order to the one in which they were submitted, but also partially in parallel with each other so that it is hard to break down the 176 state change steps in the formal checker that led to this state into a series of time updates. The timeout check takes longer as we change the timeout value to be close to the edge of failing, but it remains less than 15 seconds.
\item \textbf{Using all the Pool resources} We kept the same simulation as in the previous step but increased the timeout check to 1800cc. We then reduced the number of compute and memory resources in the pool. In our simulation we require 1 compute element and 3 internal memory resources from the pool for each kernel in this example. If we allocate 4 compute resources and 13 memory resources to the pool we successfully achieve our timeout goal. But a reduction to 12 memory elements causes a timeout failure which requires an increase in timeout to 1807 to fix. The formal checking again shows the subtlty of the failure in resource allocation. It might be expected that a lack of resources would stall a DFF by a time somewhere close to the runtime of a kernels (so 100s of cc). But in this case the stall was less than 10cc. These subtle corner cases are practically impossible to analyze by hand.
\end{itemize}

\subsection{\label{FVConclusions} Conclusions and Future Work}
\subsubsection{What have we demonstrated so far?}
The above results show that we can develop a formal model of a large processing engine, the SRE, and check its behavior for critical heisenbug generating properties with reasonable compute time. This is true because the SRE is being developed in a hierarchical structure where lower level functional blocks, such as the Director and the SHOC have specific interfaces that can be modeled formally. The SRE processes in a RESTful manner and this keeps the total state space down to a minimum. Essentially, the SRE has been designed with one of its goals to be formally checked at the system level. This is critical in the 5G wireless space (and even more so in future 5.5 to 6G modems) which is a Firm Real Time problem requiring high availability of real time flows.
\subsubsection{How do we improve the formal tooling?}
There remains plenty of room to scale this result to more complicated cases as most checking times were relatively small. But there is also much research to be completed in order to develop a formal checking strategy that can deal with state space explosions caused by the setting of parameters in a way that the complexity of operation of the SRE expands dramatically. In such cases the UPPAAL tool will just sit and process for a long time and may run out of memory. Rather we need new techniques that recognize state space explosion and either report it back to the operator or fall back to a simpler, if more pessimistic, model of formal checking. Improvement to the SRE architecture may also allow us to control the state space under most parameter settings.

The speed of the formal check can be improved using better compute platforms. Queries can be run in parallel with either multiple UPPAAL licenses or by using the multi-core OPAAL+LTSMIN tool \cite{opaal}.
\subsubsection{A next generation of formal checking for SRE}
Eventually a TA checker can be written specifically for this class of architecture speeding up checking enormously because there are invariants in the architecture that are hard to communicate to the general UPPAAL checker. By using flows of DFFs without back pressure we can orthogonalize the checking, vastly simplifying the formal checking of the SoC. 

The formal check of the runtime of a complete dataflow, spanning multiple SRE is not addressed in this report. We use the same technique as in \cite{Lampka2010} to check that the output DFF flow of an SRE matches its intended quasi periodic pattern and this allows us to check formal properties at the SoC level and SRE level separately.

In the model presented in this paper we model time consumed by the Director and the Scheduler and assume it is constant, though this could be easily changed to be a range of timing. As in the PRET processing philosophy \cite{Liu2012_PRET} it is important that bounds on the runtime of each component can be enforced, in our case to formally guarantee the correct operation of the SRE (for instance that it is deadlock free) as well as to ascertain that it meets its performance requirements for a given mixture of DFF arrivals.

Some further study is required on this system level formal checking and this may also impact the architecture. We have seen that the formal check can cope with complicated overlapping patterns of DFF processing in an SRE, even with significant runtime uncertainty in the kernels, which is a natural effect in parameterized processing of blocks of data that may vary in size.  But if there is significant jitter in the arrival time of the DFFs at the SRE there is an explosion in the number of possible state sequences that have to be checked. This implies we may want to control the arrival jitter by delaying and buffering  the DFF before releasing them into the SRE. Such a strategy would add to the total latency of a flow but would allow for higher utilization of the SRE and therefore better overall performance and is for further study.

\section{\label{sec:Conclude_NextSteps} Conclusions and Next Steps}
In the course of this research we have developed a new, dataflow based architecture for efficient implementation of the diverse use cases that mix together in 5G modems while setting the foundation for future generations of wireless modems. The architecture was based on a hierarchical decomposition of the large dataflow problem into
\begin{enumerate}
	\item a top level SoC view where DFF flow through a network of SREs, the DFFs being processed and triggering each other. The network maintains a leaky bucket type of flow control managing the rate at which DFF arrive at each SRE
	\item SRE Director/Packet Processor control and policy management of nodes from the DFF onto Stages
	\item Stage level scheduling and management of multiple nodes in flight 
	\item SHOC level computation of nodes
\end{enumerate}
\subsection{SRE architecture progress and next steps}
We focus on the design of the SRE and describe the packet protocol that allows all levels of the hierarchy within the SRE to communicate with each other in a logical manner. There is no sense of a global memory map, or a global scheduling and real time management strategy. For the SRE architecture we present a detailed breakdown of the packet protocol as well as the operation of the Director, Packet Processor and Stages. All of this is simulated at an event driven level in VisualSim.\\
In the coming months we plan to continue to test the SRE at the event driven level using more complicated and realistic DFF mixtures as well as experiment with new policies to maximize the capacity of the SRE. We will also examine the complexity of the SRE in critical sections of the architecture.\\
\subsection{Toolchain progress and next steps}
We have developed a prototype formal checking tool for the SRE that we hope can be a model and an inspiration for others to do the same. Because of the hierarchical architecture, the SRE can be formally checked separately from the rest of the SoC provided the flow patterns are agreed upon. 

However there is more work to be done to make the formal tool robust to a wide range of DFF flow parameters. We also have to align the model with the latest verison of the SRE detailed specification. There are some features that are not presently modeled and these need to be added. We expect that we will eventually move to an in house formal checking program and away from UPPAAL, as the design get more specific and to allow increased robustness and speed. In particular we need a methodology for coping with operator error, for instance an incorrect or infeasible set of patterns being input. 
Though most checking times were relatively small, some scenarios exhibited state space explosion and failed to complete formal checking. Understanding and managing this problem for the SRE is clearly an important medium term goal.

More effort needs to be put on developing a comprehensive set of queries to maximize the value of the formal checking and minimize any Heisenbugs.

We did not go into detail in this report on our efforts to use Spectral graph theory and SAT and ILP to provide options on how to map map DFF nodes to stages and also map DFF to SREs but we will report on this in the next technical report. Our strategy is to provide several different ways of mapping a DFF on an SRE (which we call "tetris blocks") and then use a policy running on the Director to choose the best one of these tetris blocks, with appropriate time offset, given the current loading of the SRE. The Policy can be formally checked in the toolchain but development of the policy is a subject for further study.

\bibliographystyle{IEEEtranS}
\bibliography{WDSA_refs_Dec3}
\pagebreak

\appendix
\section {Mirabilis Model Details \label{appendix-Mirabilis-Model}}
The top level SRE Mirabilis model is shown in \ref{Mirabilis-SRE-top}. The below describes the details of each important component. \\
\begin{figure}[!h]
	\includegraphics[width=1.0\linewidth,height=0.5\linewidth]{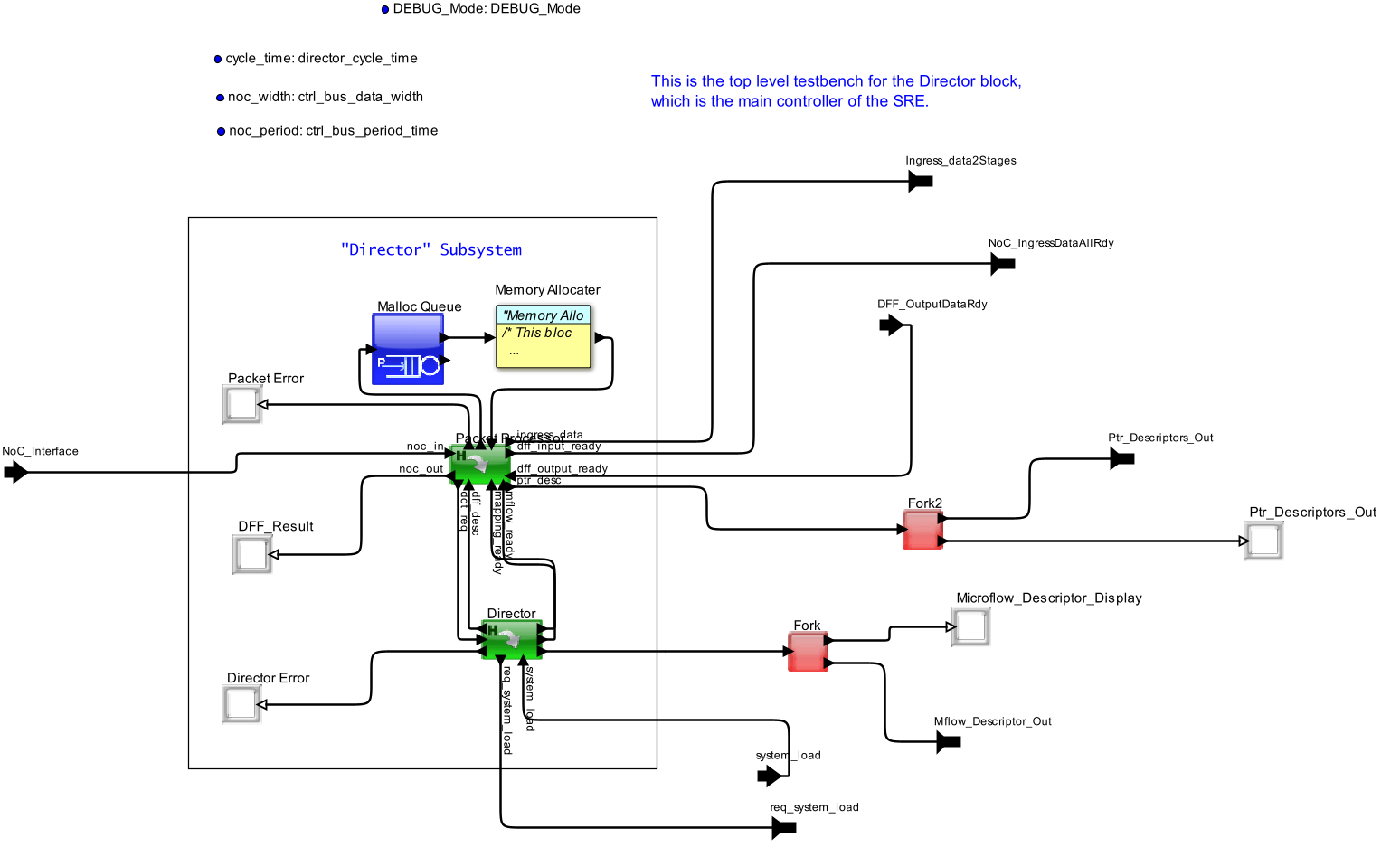}
	\centering
	\caption{Mirabilis Model of the Director and Packet Processing Subsystem}
	\label{Mirabilis-Director} 
\end{figure}

Fig. \ref{Mirabilis-Director} shows the model of the director and packet processing subsystem, it interprets the control and data packets from the NoC interface, generates and dispatches the microflow descriptors to different stages. Microflow descriptors not only describe a complete data flow graph, but also data dependencies among multiple segments of a data flow. This way, the token table and actor list can be built properly to run a data flow seamlessly once this process is completed. \\

\begin{figure}[!h]
	\includegraphics[width=0.9\linewidth,height=0.5\linewidth]{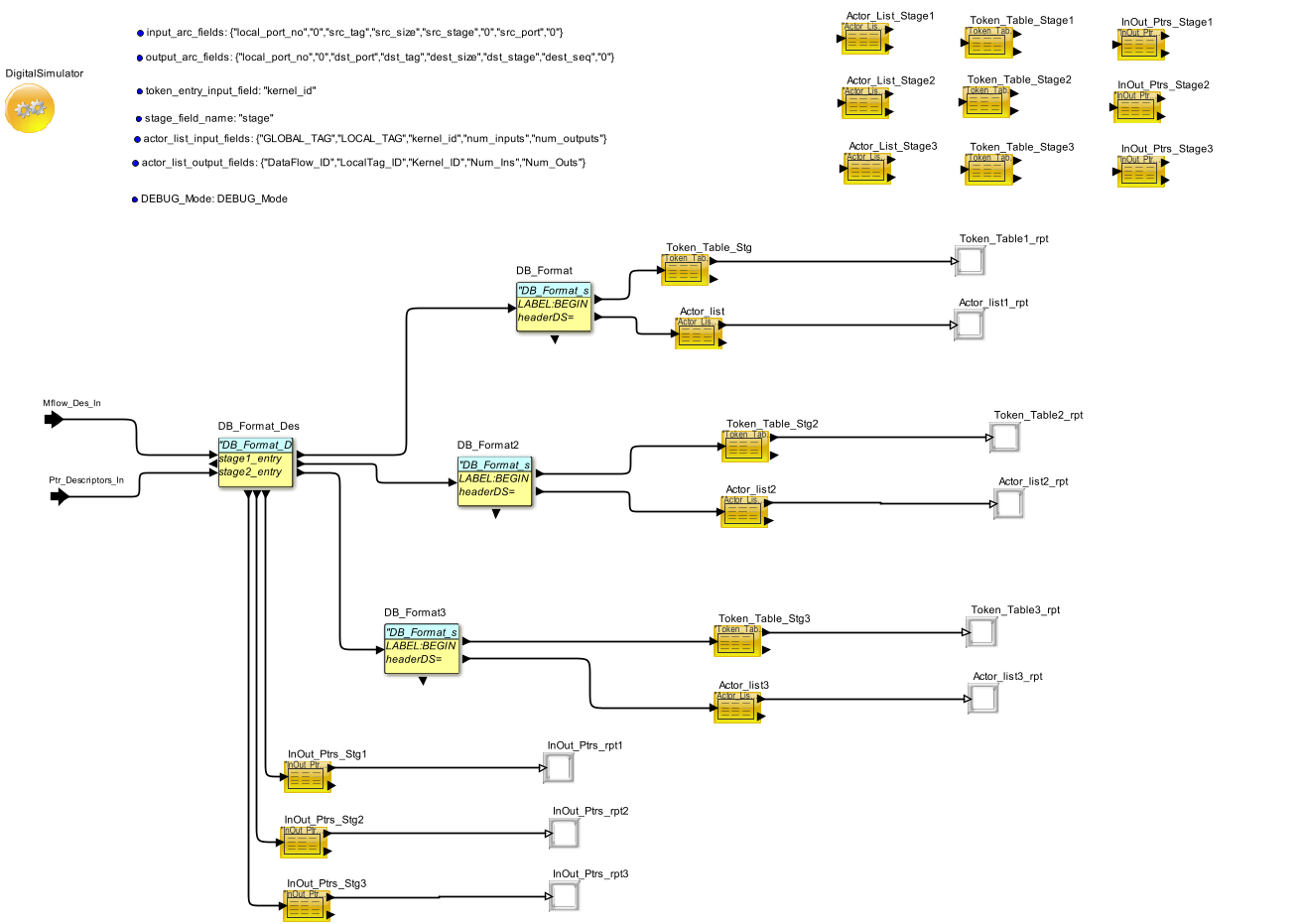}
	\centering
	\caption{Model of Token Tables and Actor Lists for All Stages }
	\label{Mirabilis-Tokentable-ActorLists} 
\end{figure}

\begin{figure}[!h]
	\includegraphics[width=1.0\linewidth]{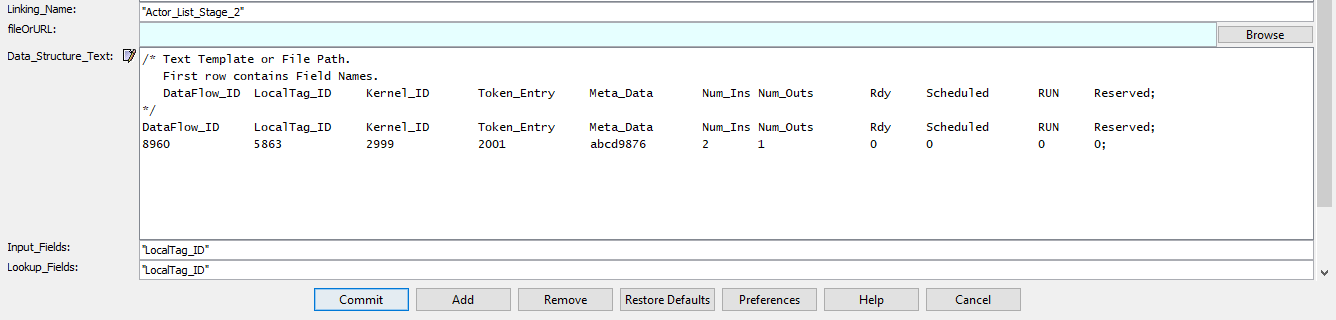}
	\centering
	\caption{Mirabilis Model of the Actor List}
	\label{Mirabilise-ActorLists} 
\end{figure}

\begin{figure}[!h]
	\includegraphics[width=1.0\linewidth]{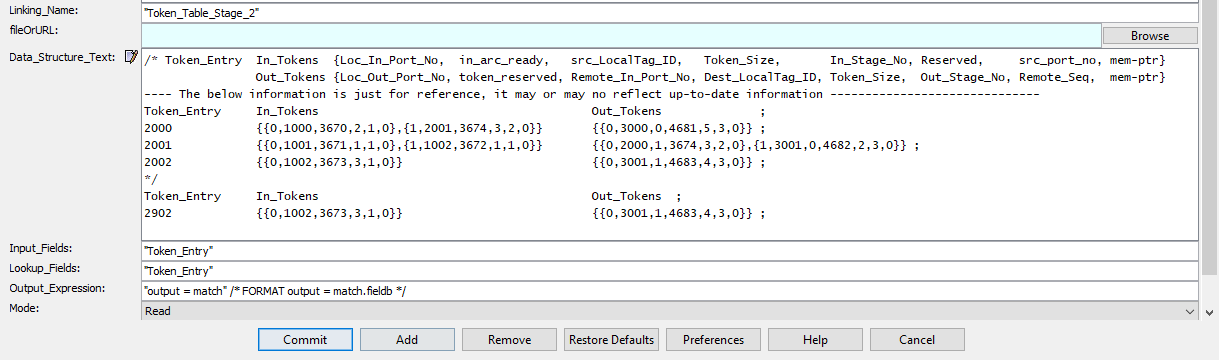}
	\centering
	\caption{Mirabilis Model of the Token Table }
	\label{Mirabilis-Tokentable} 
\end{figure}

Each stage has its own actor list and token that are built upon the microflow descriptors received from the director and packet processing subsystem as shown in Fig. \ref{Mirabilis-Tokentable-ActorLists}. They are modeled as databases in Mirabilis VirtualSim to hold multiple entries of microflow/kernels as shown in Fig. \ref{Mirabilis-Tokentable} and Fig. \ref{Mirabilise-ActorLists}. \\

Please note that, the director and packet processing also generates and dispatches pointer descriptor to describe those segments that have ingress and egress interfaces of an SRE. They are modeled as separate databases for the easy bring-up purpose now, however the necessary pointers information can be comprehended as mem-ptr subfield in the token table as shown in Fig. \ref{Mirabilis-Tokentable} later on. The Token-Entry that exists in the actor list in  Fig. \ref{Mirabilise-ActorLists} and token table Fig. \ref{Mirabilis-Tokentable} is the link that connects the both tables as described as the Entry-depen-loc in Fig. \ref{fig:SRE-stages-tokenActor}. \\

\begin{figure}[!h]
	\includegraphics[width=1.0\linewidth]{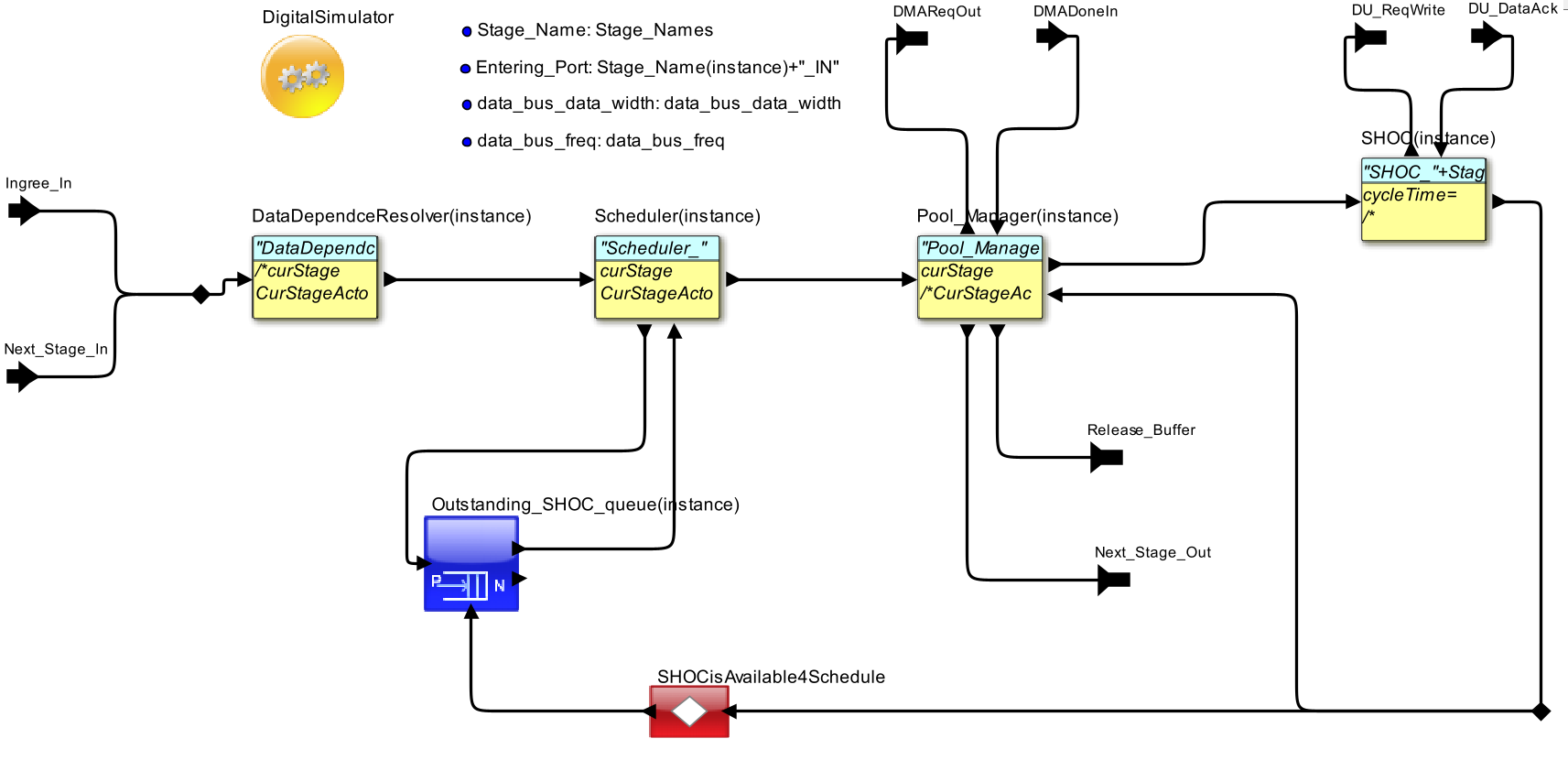}
	\centering
	\caption{Mirabilis Model of the Dynamic Stage Subsystem}
	\label{Mirabilis-Stages} 
\end{figure}

The current model supports maximum three dynamic stages. The director can assign up to three stages depends on the overall loading and other QoS requirement to conduct architectural exploration and analysis. As shown in Fig. \ref{Mirabilis-Stages}, each stage has separate submodule for the scheduler and the pool manager to comply the architectural considerations described in Fig. \ref{fig:SRE-stages-funcs}. The data dependency resolver automatically selects and searches its own set of token table and actor list whenever a relevant microflow/flow in the same stage or from other stage is completed. The evaluation result is passed to the scheduler and the pool manager for further actions by following the flow chart illustrated in Fig. \ref{fig:SRE-stages-flowchart}. \\

\begin{figure}[!h]
	\includegraphics[width=1.0\linewidth]{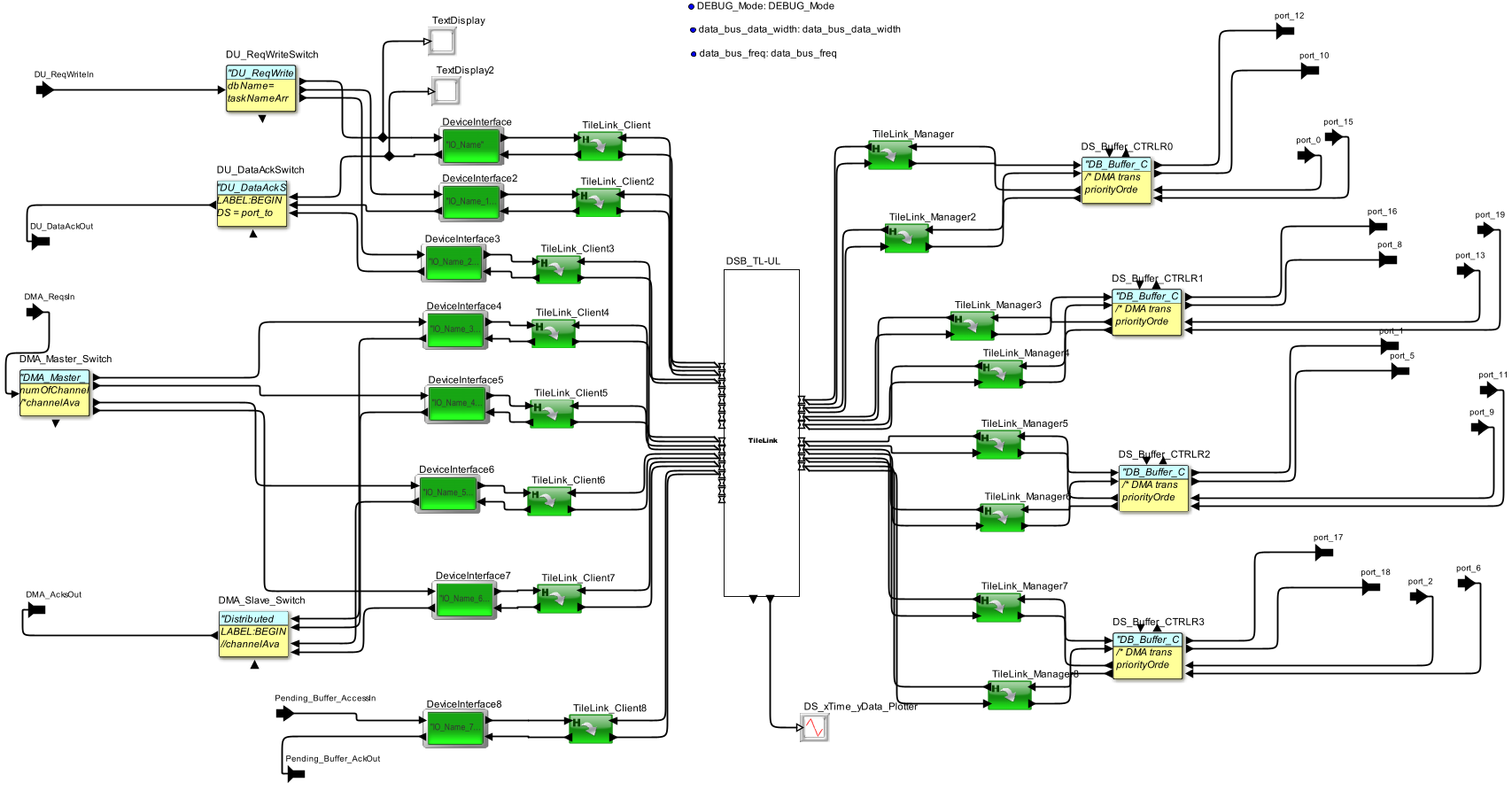}
	\centering
	\caption{Mirabilis Model of the Dataflow Mapped Distributed Buffer Network}
	\label{Mirabilis-Data-Xbar} 
\end{figure}

The Mirabilis TileLink module is instantiated to model the dataflow mapped distributed buffer network as shown in Fig. \ref{Mirabilis-Data-Xbar} for further architectural trade-off and data-traffic analysis. \\

\begin{figure}[!h]
	\includegraphics[width=0.7\linewidth,height=0.3\linewidth]{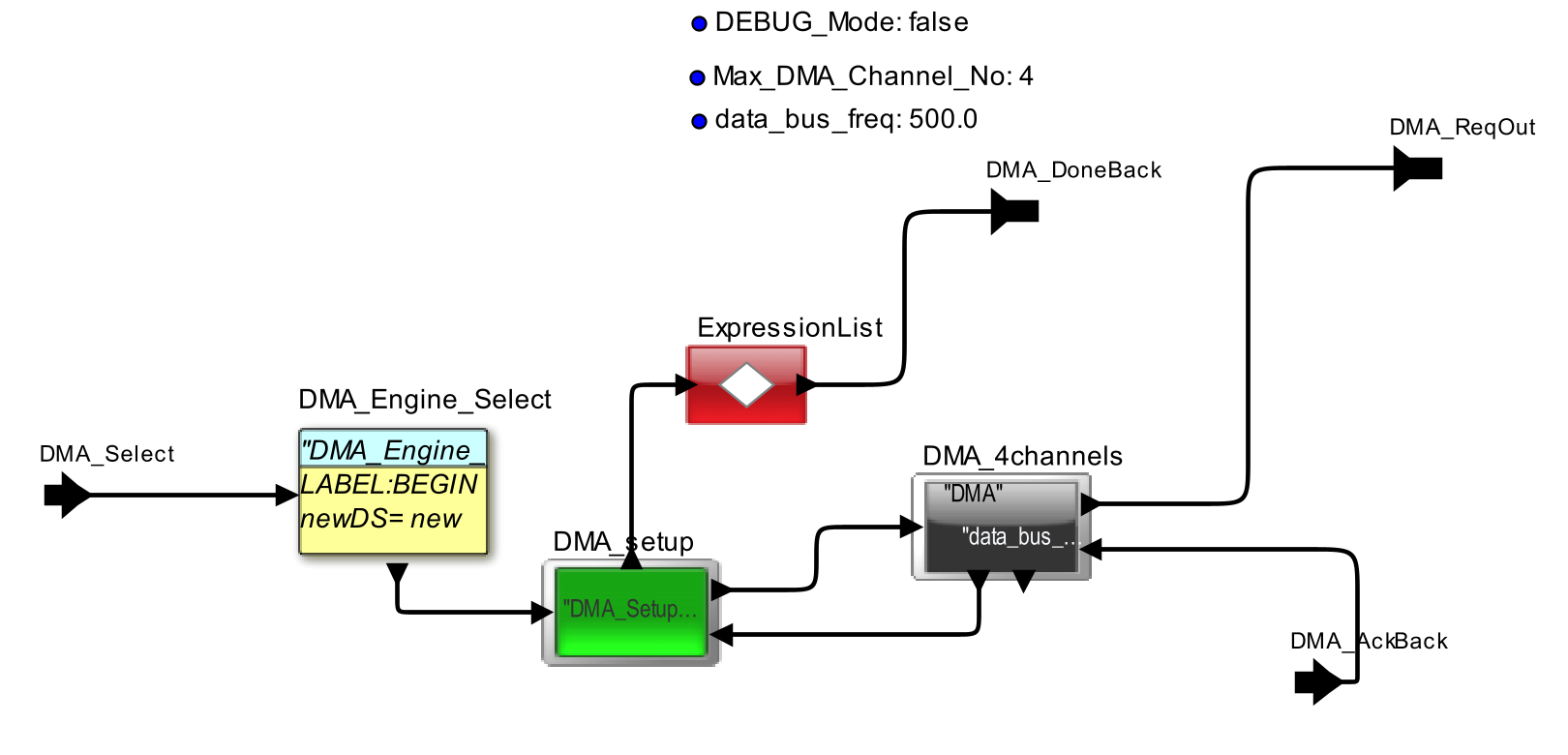}
	\centering
	\caption{Mirabilis Model of the DMA Subsystem}
	\label{Mirabilis-DMA} 
\end{figure}

As mentioned previously, any conventional DMA can be leveraged for the quick SRE architectural exploration, the Mirabilis DMA engine with simple configuration of four channels shown in Fig. \ref{Mirabilis-DMA} is instantiated to complete the SRE architectural model. 

\section{\label{sec:appendix}Message Formats and Error Codes}

\subsection{\label{sec:message_formats}Message Formats}

The following section describes internal and external message formats used by the SRE.

\includegraphics[width=0.8\linewidth, height=7cm]{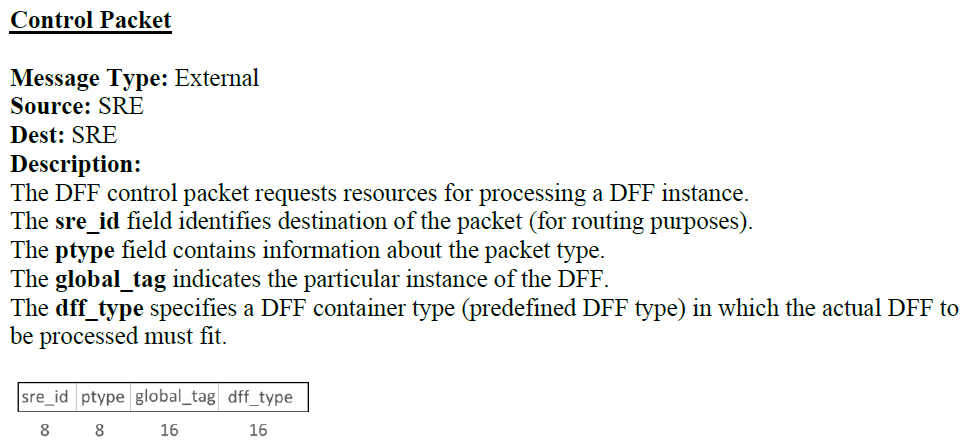}

\includegraphics[width=0.8\linewidth, height=10cm]{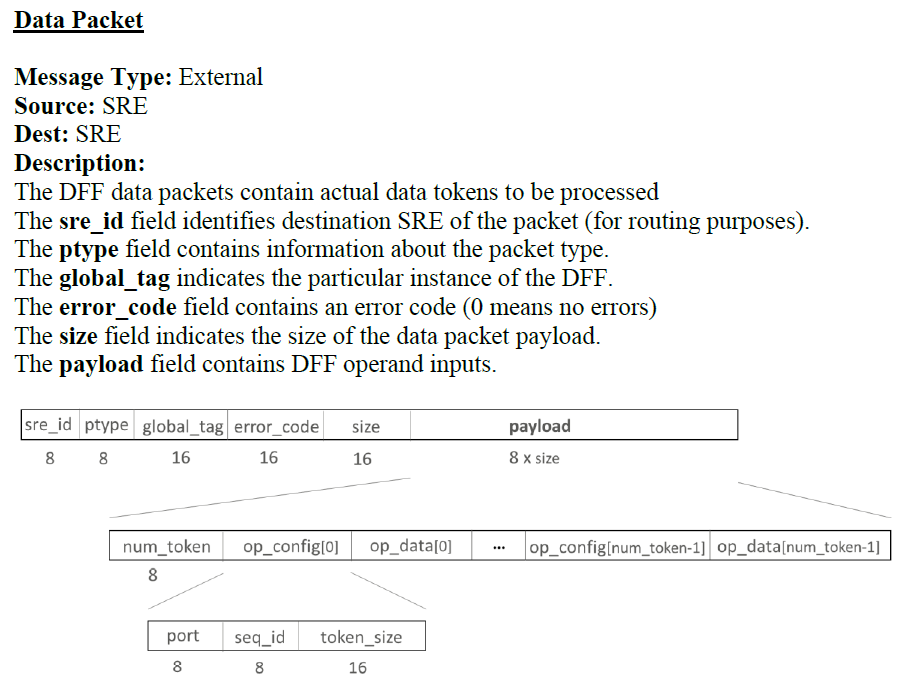}

\includegraphics[width=0.7\linewidth, height=7cm]{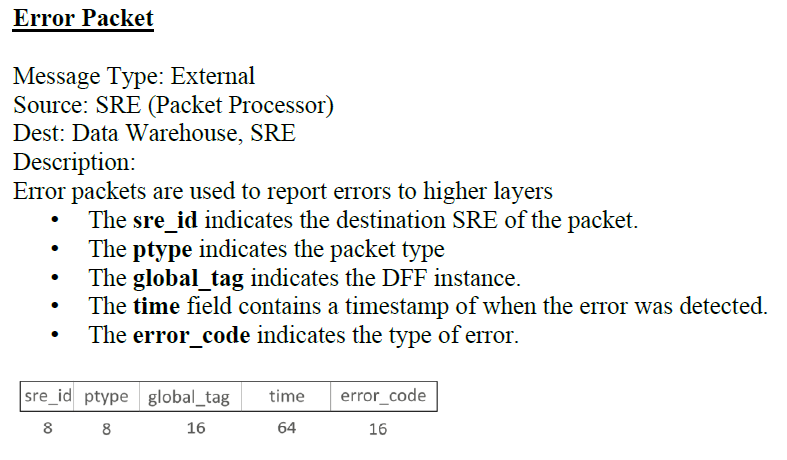}

\includegraphics[width=0.8\linewidth, height=7cm]{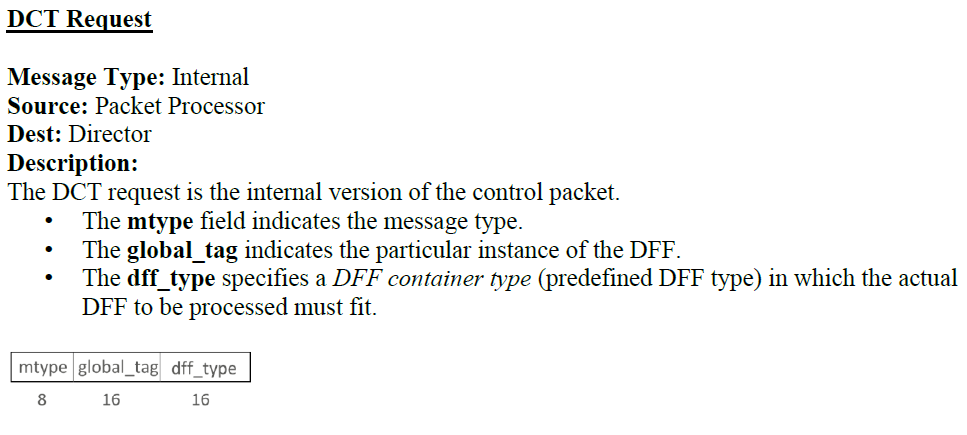}

\includegraphics[width=0.8\linewidth, height=12cm]{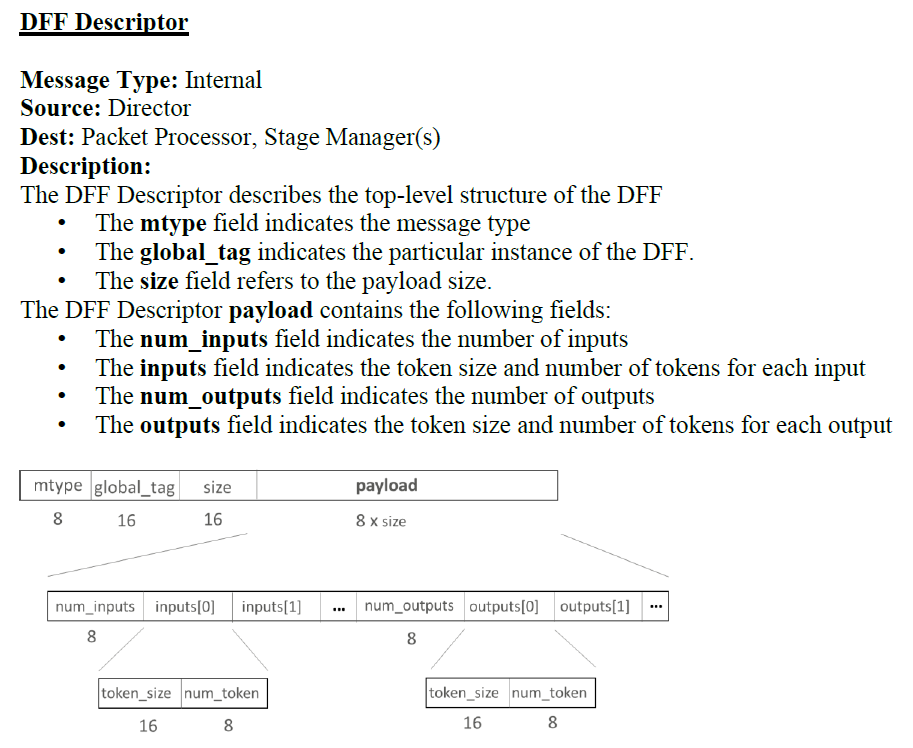}

\includegraphics[width=0.6\linewidth, height=6cm]{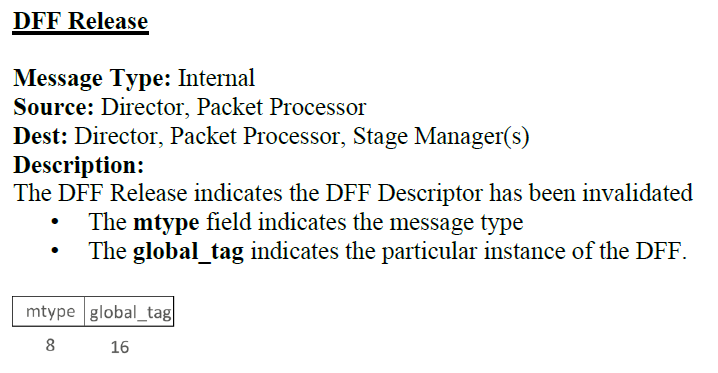}

\includegraphics[width=0.6\linewidth, height=10cm]{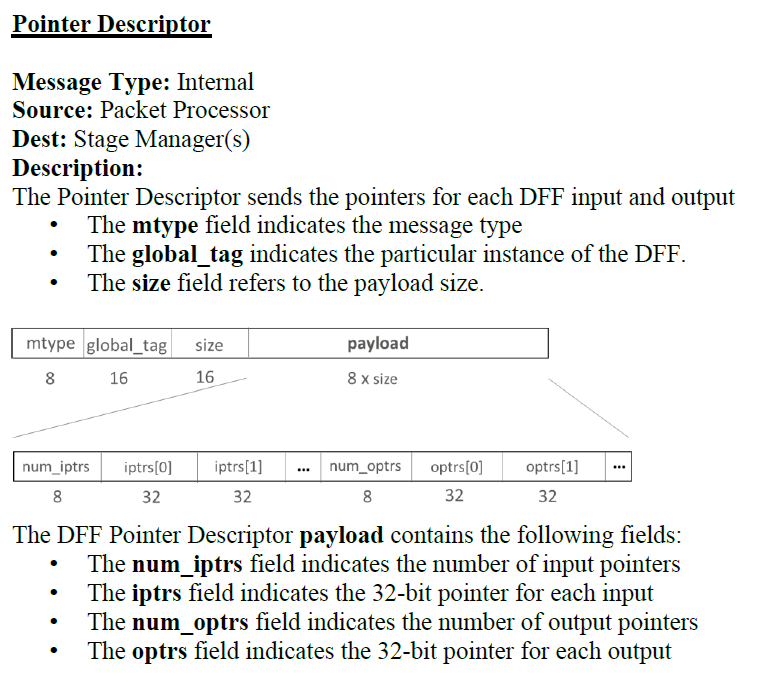}

\includegraphics[width=0.8\linewidth, height=12cm]{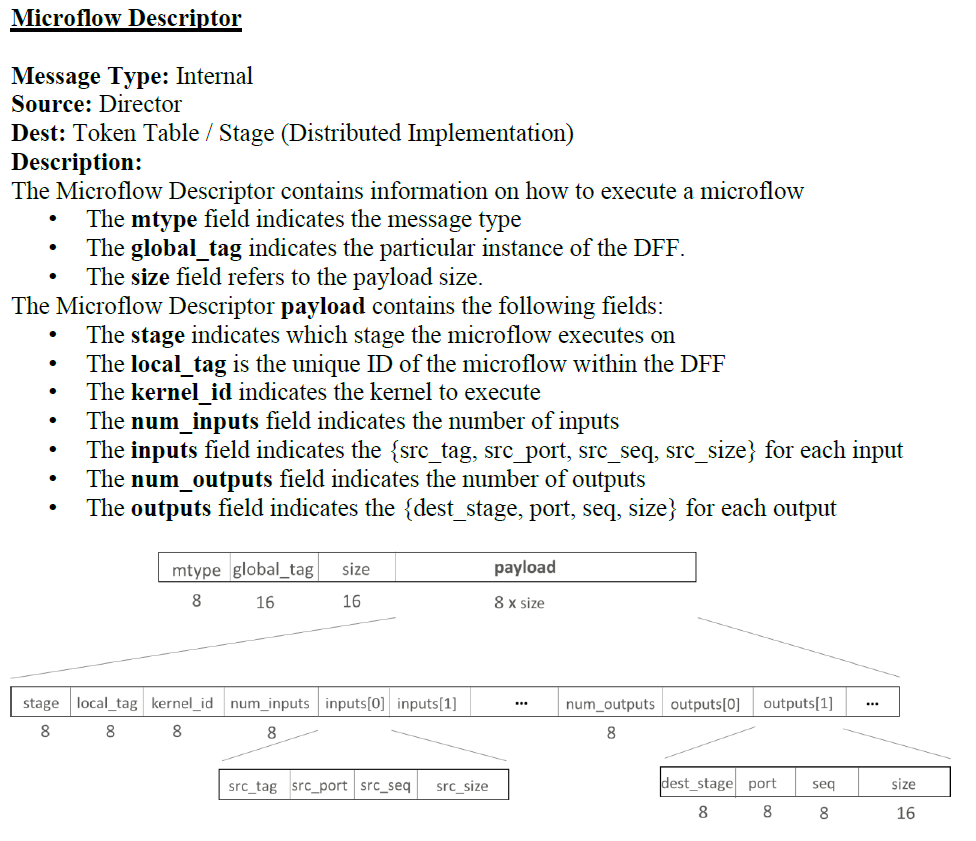}

\includegraphics[width=0.6\linewidth, height=6cm]{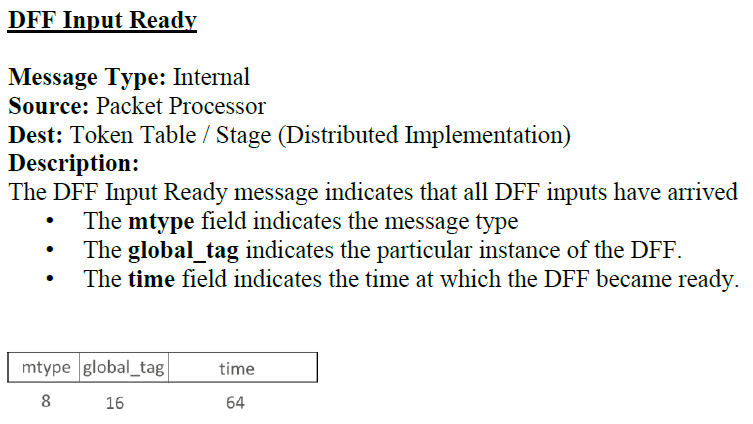}

\includegraphics[width=0.6\linewidth, height=6cm]{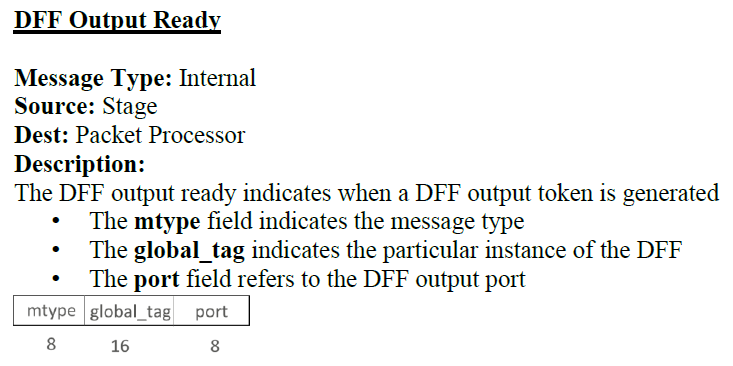}

\includegraphics[width=0.6\linewidth, height=6cm]{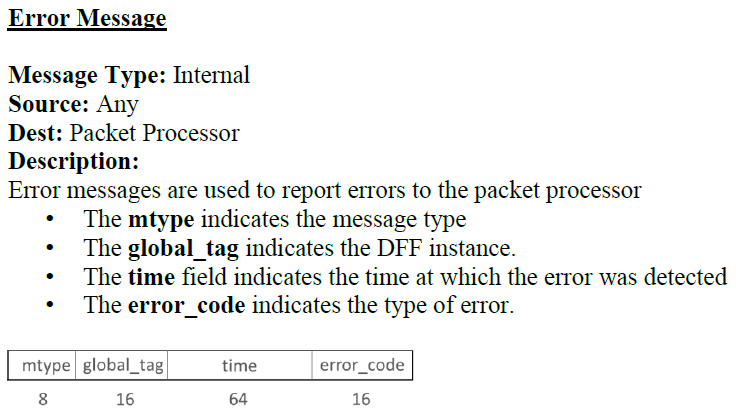}

\subsection{\label{sec:error_codes}Error Codes}

This section describes a list of basic error codes. All consumers inherit the error code of their producer in order to facilitate debug. A time stamp is also appended to the error message (see Error Message format in subsection \ref{sec:message_formats}).

\includegraphics[width=0.6\linewidth, height=3cm]{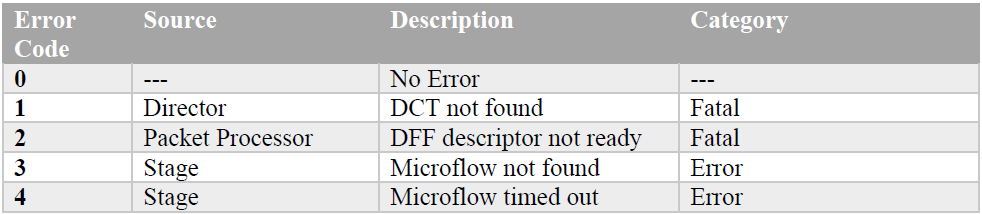}

\end{document}